\pdfoutput=1

\documentclass[12pt]{article}%
\usepackage[nosort]{cite}
\usepackage{graphicx}
\usepackage{multicol}
\usepackage{amsfonts}
\usepackage{amssymb}
\usepackage{amsmath}
\usepackage{heck}
\usepackage[all]{xy}
\usepackage{setspace}
\usepackage{verbatim}
\usepackage{color}
\usepackage{epsfig}
\usepackage[margin=1in]{geometry}
\usepackage{titletoc}
\usepackage{lscape}%
\providecommand{\U}[1]{\protect\rule{.1in}{.1in}}
\numberwithin{equation}{section}

\newcommand{\cI}{\mathcal{I}}
\newcommand{\cJ}{\mathcal{J}}

\usepackage{tikz}

\usepackage[debug,pageanchor=false]{hyperref}
\definecolor{link}{rgb}{.8,.15,.1}
\hypersetup{colorlinks=true,linkcolor=link,citecolor=link,urlcolor=link,linktocpage}

\begin{document}

\date{July 2014}

\title{6d Conformal Matter}

\institution{HARVARD}{\centerline{${}^{1}$ Jefferson Physical Laboratory, Harvard University, Cambridge, MA 02138, USA}}

\institution{UNC}{\centerline{${}^{2}$ Department of Physics, University of North Carolina, Chapel Hill, NC 27599, USA}}

\institution{BICOCCA}{\centerline{${}^{3}$ Dipartimento di Fisica, Universit\`a di Milano Bicocca, Milan, Italy}}

\authors{Michele Del Zotto\worksat{\HARVARD}\footnote{e-mail: {\tt eledelz@gmail.com}},
Jonathan J. Heckman\worksat{\HARVARD, \UNC}\footnote{e-mail: {\tt jheckman@email.unc.edu}},
\\[4mm] Alessandro Tomasiello\worksat{\BICOCCA}\footnote{e-mail: {\tt alessandro.tomasiello@unimib.it}},
and Cumrun Vafa\worksat{\HARVARD}\footnote{e-mail: {\tt vafa@physics.harvard.edu}}}

\abstract{A single M5-brane probing $G$, an ADE-type singularity, leads to a system which has
$G\times G$ global symmetry and can be viewed as ``bifundamental'' $(G, G)$ matter.  For the $A_N$ series,
this leads to the usual notion of bifundamental matter.  For the other cases it corresponds to a strongly interacting
$(1,0)$ superconformal system in six dimensions.  Similarly, an ADE singularity intersecting
the Ho\v{r}ava-Witten wall leads to a superconformal matter system with $E_8\times G$ global symmetry.
Using the F-theory realization of these theories, we elucidate the Coulomb/tensor branch of $(G,G')$ conformal
matter. This leads to the notion of fractionalization of an M5-brane on
an ADE singularity as well as fractionalization of the intersection point of the ADE singularity with the Ho\v{r}ava-Witten wall.
Partial Higgsing of these theories leads to new 6d SCFTs in the infrared, which we also characterize. This
generalizes the class of $(1,0)$ theories which can be perturbatively realized by
suspended branes in IIA string theory.  By reducing on a circle, we arrive at novel duals for
5d affine quiver theories.  Introducing many M5-branes leads to large $N$ gravity duals.}

\maketitle

\tableofcontents

\enlargethispage{\baselineskip}

\setcounter{tocdepth}{2}

\newpage

\section{Introduction \label{sec:INTRO}}

One of the remarkable developments in string theory is the close interplay
between the geometry of its extra dimensions, and the resulting low energy
theories. In the case of geometries with singularities, such methods have led
to a host of tools in the construction and study of conformal field theories
in diverse dimensions.
  Of particular significance are conformal field theories in six dimensions,
which resist a UV Lagrangian description. The key ingredients of these
theories are tensionless strings coupled to dynamical tensor modes.

Notable examples of such theories include the ADE$\ (2,0)$ theories
\cite{Witten:1995ex}. For the A-type series, this is realized by a coincident
stack of M5-branes \cite{Strominger:1995ac}. Alternatively, all of the ADE
theories can be realized by type IIB\ strings compactified on an ADE\ orbifold
singularity \cite{Witten:1995ex}. Comparatively less is known about $(1,0)$ theories; some examples were found in the past in
\cite{WittenSmall, Ganor:1996mu, Seiberg:1996vs, Bershadsky:1996nu, Brunner:1997gf, BlumINT, Blum:1997fw, intriliaddme, Hanany:1997gh}. Recent work \cite{Heckman:2013pva}
gave a complete classification of $(1,0)$ theories without a Higgs branch. Those results also
give a systematic starting point for pursuing a full classification of theories which have a Higgs branch.

The CFTs in the classification in \cite{Heckman:2013pva} do not have a weakly coupled UV\ Lagrangian.
However, one can always go to the Coulomb/tensor branch of these theories, which corresponds to giving vevs to scalars in tensor multiplets.  In such cases one
can find an effective Lagrangian description for $(1,0)$ theories in terms of a weakly coupled quiver gauge theory, where the scalar in the tensor multiplet controls the coupling constant of the corresponding gauge groups (i.e. the multiplet containing the gauge coupling) and is promoted to a dynamical collection of fields, ending up with a quiver-type theory. Moving to the origin of the tensor branch typically leads to a 6d\ SCFT with $(1,0)$ supersymmetry. Given the ubiquity of quivers in string theory, it is perhaps not surprising that some of these theories have a straightforward realization in string theory \cite{BlumINT, Hanany:1997gh, Gaiotto:2014lca}.


There are, however, some seemingly obvious quiver gauge theories which do not
have a realization in perturbative string theory. For example, the structure
of the orientifold projection forbids a bifundamental between $SO(2p)$ and
$SO(2k)$, but instead leads to bifundamentals between $SO(2p)$ and $Sp(k)$.
Perhaps even more conspicuous is the absence of E-type gauge theories, let
alone an understanding of what a bifundamental between two such nodes would mean.

In this paper we point out that such generalized quivers do exist in
string theory, but their matter sector is itself a strongly coupled 6d\ SCFT.
We focus on two primary examples.  One of them involves the realization
of such 6d\ SCFTs by treating M5-branes as domain walls in a higher
dimensional theory.  This case is realized by the theory of M5-branes probing
an ADE\ singularity $\mathbb{C}^{2}/\Gamma_{ADE}$ in M-theory.
The other type involves intersecting an ADE singularity with
a Ho\v{r}ava-Witten wall \cite{Horava:1995qa,Horava:1996ma}.

ADE singularities define a seven-dimensional super Yang-Mills theory; being
one dimension lower, M5-branes correspond to domain walls in this theory. Because
it cuts the space in two, each such domain wall contributes additional light states
to the low energy theory. Each subsequent parallel M5-brane introduces another
domain wall, and for each finite segment between adjacent M5-branes we get a
dynamical gauge symmetry whose inverse squared coupling constant is
proportional to the length of the segment. We thus end up with a linear
quiver consisting of gauge groups $G$, where the ``bifundamental''
between each adjacent group is interpreted as the associated superconformal
matter with $G\times G$ symmetry. For $N$ M5-branes, we therefore get theories
$\mathcal{T}(G, N)$.

The question is thus reduced to understanding this matter sector,
i.e. the theory living on such a domain wall. To determine this, we use
a dual description of conformal matter in F-theory. We take F-theory
on a non-compact elliptically fibered Calabi-Yau threefold. In
F-theory, the conformal matter degrees of freedom on
the domain wall are associated with the collision of two seven-branes, each
supporting a gauge group $G$ and wrapping a non-compact curve. Conformal
matter is located at the intersection of two such curves, where the associated
elliptic fibration can become more singular. In the case of a collision of two
A-type gauge groups, there is a single hypermultiplet in the bifundamental of
$G_{L}\times G_{R}$. For D- and E-type gauge groups, such a collision leads
to a theory of tensionless strings which can be studied by introducing a
minimal resolution in the base of the F-theory geometry.
The existence of additional tensor multiplets suggests that the M5-brane
fractionates on a singularity, leading to new gauge symmetries between
the fractional M5-branes.   We suggest an interpretation of fractional M5-branes as domain
walls separating loci of M-theory singularities with different fractional discrete three-form flux of the type
proposed in \cite{deBoer:2001px}.

For example, the strongly coupled conformal matter produced by the collision
of two $\mathfrak{so}_{2p+8}$ factors has a non-trivial tensor branch,
consisting of a single tensor multiplet, an $\mathfrak{sp}_{p}$ gauge theory,
and a half hypermultiplet in the $(\mathbf{2p+8},\mathbf{2p},\mathbf{1}%
)\oplus(\mathbf{1},\mathbf{2p},\mathbf{2p+8})$ of $\mathfrak{so}_{2p+8}%
\times\mathfrak{sp}_{p}\times\mathfrak{so}_{2p+8}$. In this case, we
can view the M5-brane as fractionating to two 1/2 M5-branes between which the
gauge symmetry has changed from the $\mathfrak{so}$ type to the $\mathfrak{sp}$ type.
As another example, conformal matter between $\mathfrak{e}_{8}$ and $\mathfrak{e}_{8}$
leads to a strongly coupled CFT with an eleven-dimensional tensor branch and gauge
algebra $\left(  \mathfrak{sp}_{1}\times\mathfrak{g}_{2}\right)  _{L}%
\times\mathfrak{f}_{4}\times\left(  \mathfrak{g}_{2}\times\mathfrak{sp}%
_{1}\right)  _{R}$ with a half hypermultiplet in the $(\mathbf{2}%
,\mathbf{7}+\mathbf{1})$ for the $\left(  \mathfrak{sp}_{1}\times
\mathfrak{g}_{2}\right)  _{L}$ factor, and a half hypermultiplet in the
$(\mathbf{7}+\mathbf{1},\mathbf{2})$ for the $\left(  \mathfrak{g}_{2}%
\times\mathfrak{sp}_{1}\right)  _{R}$ factor.   In this case the M5-brane
fractionates to 12 fractional M5-branes, and the gauge groups
arise from the finite intervals between the fractional M5-branes.\footnote{Between
some fractional M5-brane pairs there are no gauge groups.}
Such considerations show that even in the case of a \textit{single} M5-brane,
the resulting probe theory of a D- or E-type singularity leads to a
non-trivial fixed point which is the reflection of the existence of fractional M5-branes.
This is in line with the expectation that additional degrees of freedom enter
the low energy theory near the singular point of the moduli space.
Upon compactification on a circle, these lead to novel duals of the
well studied affine quiver gauge theories.

As the second main example, we consider the M-theory background $\mathbb{R}/\mathbb{Z}_{2}\times\mathbb{C}^{2}%
/\Gamma_{ADE}$, i.e. ADE singularities intersecting the Ho\v{r}ava-Witten wall.
The $\mathbb{Z}_{2}$ fixed point gives an $E_{8}$ nine-brane which wraps $\mathbb{C}^2 / \Gamma_{ADE}$.
This leads to a conformal system with $E_{8}\times G_{ADE}$ global symmetry.
In this case we again find the phenomenon of fractionating, but now the intersection
point of the ADE singularity and the wall fractionate. As before, we can also
introduce M5-branes along the line of the ADE singularity.  In heterotic terms, this is
the theory of small $E_{8}$ instantons \cite{WittenSmall, Ganor:1996mu, Seiberg:1996vs} probing an ADE singularity.
Some aspects of this system have been analyzed using F-theory in \cite{Aspinwall:1997ye}. We find
$G$-type gauge symmetries with $(G,G)$ conformal matter system for all of them,
except the one adjacent to the wall, which gauges the $G$ symmetry of the
$(E_{8},G)$ conformal matter system at the wall. We label these theories
as $\mathcal{T}(E_8 , G , N)$.

These theories also have partial Higgs branches where operators develop vevs which
break some of the flavor symmetry, leading to {\it new}
conformal fixed points. By studying the vacua of the 7d SYM theory, or equivalently
the vacua of the flavor seven-branes, we show that partial Higgs branches of the $\mathcal{T}(G,N)$ theories
are classified by the orbits of nilpotent elements for the flavor symmetry factors.
In F-theory, such configurations are examples of T-brane configurations \cite{TBRANES,
Chiou:2011js,glueI,glueII,Anderson:2013rka}. These are non-abelian configurations of
intersecting seven-branes which can remain hidden from the complex structure
moduli of the Calabi-Yau geometry. We label these theories as $\mathcal{T}(G,\mu_{L},\mu_{R},N)$, which consists of
$N$ M5-branes, and a flavor symmetry $G_L\times G_R$ which can be broken, as dictated by the orbits in $\mathfrak{g}$ of
nilpotent elements $\mu_L\in$  $\mathfrak{g}_L$ and $\mu_{R}\in\mathfrak{g}_R$ for
the two Lie algebras.

There are also new conformal theories coming from the partial Higgsing of theories involving M5-brane
probes of the ADE singularities intersecting the Ho\v{r}ava-Witten wall.  These theories are
classified as $\mathcal{T}(E_{8},G_{R}$, $\gamma_{L},\mu_{R},N)$: We have a theory of $N$ M5-branes, and flavor
symmetry $E_{8}\times G_{R}$ which can be broken, as dictated by a
nilpotent element $\mu_{R}\in\mathfrak{g}$ (and its associated orbit)
for the right Lie algebra, as well as a homomorphism
$\gamma_{L}:\Gamma_G\rightarrow E_8$ corresponding to the choice of a
flat $E_{8}$ connection on $S^{3}/\Gamma_{G}$.

Taking the limit of a large number of M5-branes also leads us to a collection
of gravity duals in both M-theory and IIA string theory. An interesting
feature of our analysis is that we can see how certain features of IIA duals
with a Romans mass show up in our construction.

One can also study, from the perspective of F-theory, the more general case
of colliding $G_{ADE} \times G_{ADE}^{\prime}$ singularities and the associated conformal matter. For completeness,
we also include this analysis.

The rest of this paper is organized as follows. To set the stage, we first
show in section \ref{sec:CBI} how to understand conformal matter sectors in F-theory.
We use this analysis in section \ref{sec:Mprobe} to study M5-branes probing an ADE
singularity. In section \ref{sec:5D} we show how reduction of these
theories on a circle leads to novel 5d dualities.
In section \ref{sec:HIGGS} we show how to characterize the additional SCFTs generated by moving onto
the partial Higgs branches of such theories.
Next, in section \ref{sec:SMALL} we turn to the theory of heterotic small instantons probing an ADE\ singularity,
determining both the associated generalized quivers, and their partial Higgs
branches. In section \ref{sec:HOLO} we turn to scaling limits of our
solutions, and characterize the corresponding holographic dual descriptions.
We present our conclusions in section \ref{sec:CONC}. Some additional
background, as well as examples of generalized quiver theories in F-theory are
presented in a set of Appendices.

\section{Conformal Matter \label{sec:CBI}}

One of the aims of our paper will be to show how conformal matter appears in
various contexts. In this section we show how to derive properties of
these theories via F-theory.

In F-theory, conformal matter arises from the collision of seven-branes where
the localized matter is not a weakly coupled hypermultiplet. A convenient way
to deduce properties of the matter sector is to formulate the collision of
seven-branes in terms of the geometry of an elliptically fibered Calabi-Yau
threefold. In minimal Weierstrass form, this is given by:%
\begin{equation}
y^{2}=x^{3}+fx+g,
\end{equation}
where $f$ and $g$ are sections of $\mathcal{O}(-4K_{B})$ and $\mathcal{O}%
(-6K_{B})$, with $B$ the base of the elliptic fibration. Seven-branes are
associated with irreducible components of the discriminant locus, i.e. the
zero set of $4f^{3}+27g^{2}=0$. The corresponding gauge symmetry on such a
seven-brane is dictated by the order of vanishing for $f$ and $g$, which in
turn determines the Kodaira-Tate type of the singular fiber. This, in
combination with additional geometric data can be used to read off the gauge
group on a seven-brane (see e.g. \cite{Aspinwall:1998xj}).

Localized matter is associated with the collision of two such irreducible
components of the discriminant locus. At these collisions, the Kodaira-Tate
singularity type of the elliptic fiber can become more singular, thus leading
to the phenomenon of trapped matter. In fact, the fiber can
sometimes become so singular that additional blowups in the base $B$ become
necessary to understand the resulting matter content. When such blowups are
introduced, there are additional exceptional curves in the base. Each such
curve can be wrapped by a D3-brane, contributing a string in the
six-dimensional effective theory. As these curves shrink to zero size, the
tension of this string also tends to zero, yielding a six-dimensional SCFT.
The total number of tensor multiplets for such a theory is simply the number
of independent curves which simultaneously contract to zero size.

In this section, our primary interest is in the collision of two seven-branes
which support the same ADE\ gauge group, and the corresponding conformal
matter. The result of this analysis has been performed in various places, for
example in \cite{Bershadsky:1996nu, Heckman:2013pva, MorrisonVafaII, BershadskyPLUS}.
Rather than launch into a detailed discussion of the
necessary blowup structure, we shall use the algorithmic procedure developed
and automated in \cite{Heckman:2013pva}, which involves stating some minimal
combinatorial data about intersections of curves in the base $B$.

Using this procedure, we can determine the corresponding degrees of freedom
trapped along each collision of singularities. For example, in F-theory, an
A-type $\mathfrak{su}_{k}$ gauge symmetry is realized by a Kodaira-Tate fiber
of split $I_{k}$ type.\footnote{Split means there is no monodromy by an outer automorphism
of the algebra.} At the collision of two split $I_{k}$ and
$I_{p}$ singularities, the singularity becomes $I_{k+p}$, so we have a
Higgsing of $\mathfrak{su}_{k+p}$ down to the product $\mathfrak{su}_{k}%
\times\mathfrak{su}_{p}$, with a corresponding hypermultiplet in the
bifundamental $(\mathbf{k},\overline{\mathbf{p}})$ of $\mathfrak{su}_{k}\times\mathfrak{su}_{p}$.

In the remaining cases we consider, the collision of two seven-branes will
lead to a strongly coupled conformal sector. Consider next the collision of
two D-type singularities. In F-theory, a D-type $\mathfrak{so}_{2p+8}$ gauge
symmetry is realized by a Kodaira-Tate fiber of split $I_{p}^{\ast}$ type. The
non-split case would realize an $\mathfrak{so}_{2p+7}$ gauge symmetry. At the
intersection point, the collision of $I_{k}^{\ast}$ and $I_{p}^{\ast}$ leads
to an order of vanishing for $f$ and $g$ which does not yield a standard
Kodaira-Tate fiber. Thus, a blowup at this point is required. This yields a
$-1$ curve which itself supports a non-split $I_{k+p}$ type fiber \cite{Aspinwall:1997ye}. The resulting gauge symmetry from
such a non-split singularity is $\mathfrak{sp}_{r}$ with $r = [(k + p) / 2]_{+}$, that is, the smallest
integer obtained from rounding up \cite{Bershadsky:1996nu, BershadskyPLUS, Aspinwall:2000kf}.\footnote{In
the case where $k+p$ is even, this can be understood by quotienting by the outer
automorphism of $\mathfrak{su}_{k+p}$, thus producing an $\mathfrak{sp}_{r}$
algebra. In the case where $k+p$ is odd, the analysis of roots in the
associated resolution of the fiber is more subtle, and only an $\mathfrak{sp}_{r-1}$ algebra can be
identified geometrically \cite{Aspinwall:2000kf}. However, the structure of 6d anomaly cancelation and consistent
Higgsing patterns in the field theory is such that the only self-consistent way to get a gauge symmetry
is to have $\mathfrak{sp}_{r}$ gauge symmetry, with some additional matter
fields attached to the $\mathfrak{sp}$ factor \cite{Bershadsky:1996nu}.} In addition to this
gauge symmetry, we also have a half hypermultiplet in the bifundamental
trapped at each collision of our $\mathfrak{sp}_{r}$ seven-brane with an
$\mathfrak{so}_{2k+8}$ and $\mathfrak{so}_{2p+8}$ seven-brane. In the case where $k + p$ is odd, we also have an
extra hypermultiplet in the $\mathbf{2r}$ of $\mathfrak{sp}_{r}$. Now, the key
point is that the \textquotedblleft matter sector\textquotedblright\ between
our two $\mathfrak{so}$ factors is really a conformal field theory, since the
$-1$ curve is shrunk to zero size in our geometry. This is our first example
of conformal matter.

Consider next the collision of two E-type singularities. The Kodaira-Tate
fiber for $E_{6}$, $E_{7}$ and $E_{8}$ is respectively a split $IV^{\ast}$
fiber, and a $III^{\ast}$ and $II^{\ast}$ fiber. The pairwise collisions can
be conveniently summarized by the minimal Weierstrass models:%
\begin{align}
(E_{6},E_{6}) &  :y^{2}=x^{3}+u^{4}v^{4}\\
(E_{7},E_{7}) &  :y^{2}=x^{3}+u^{3}v^{3}x\\
(E_{8},E_{8}) &  :y^{2}=x^{3}+u^{5}v^{5},
\end{align}
with conformal matter located in the base at the point $u=v=0$. Performing the
minimal blowups necessary to get all fibers into Kodaira-Tate form yields an
additional configuration of curves, which intersect pairwise at a single
point. For details of this resolution algorithm, see reference
\cite{Heckman:2013pva}. Letting a sequence of positive integers denote minus
the self-intersection number for these curves, we have the minimal
resolutions:%
\begin{align}
(E_{6},E_{6}) &  :%
\begin{tabular}
[c]{|c|ccc|}\hline
Gauge Symm: &  & $\mathfrak{su}_{3}$ & \\
Curve: & $1$ & $3$ & $1$\\\hline
\end{tabular}
\\
(E_{7},E_{7}) &  :%
\begin{tabular}
[c]{|c|cccccccc|}\hline
Gauge Symm: &  & $\mathfrak{su}_{2}$ & & $\mathfrak{so}_{7}$ & & $\mathfrak{su}%
_{2}$ &  & \\
Curve: & $1$ & $2$ & & $3$ & & $2$ & $1$ & \\
Hyper: &     &     & $\frac{1}{2}(\mathbf{2,8})$ &  & $\frac{1}{2}(\mathbf{8,2})$ & &  & \\\hline
\end{tabular}
\\
\label{eq:E8cm}
(E_{8},E_{8}) &  :%
\begin{tabular}
[c]{|c|ccccccccccccc|}\hline
Gauge Symm: &  &  & $\mathfrak{sp}_{1}$ & & $\mathfrak{g}_{2}$ &  &
$\mathfrak{f}_{4}$ &  & $\mathfrak{g}_{2}$ & & $\mathfrak{sp}_{1}$ &  & \\
Curve: & $1$ & $2$ & $2$ & & $3$ & $1$ & $5$ & $1$ & $3$ & & $2$ & $2$ &
$1$\\
Hyper: &  &  & & $\frac{1}{2}(\mathbf{2},\mathbf{7}+\mathbf{1})$ &   &   &
& & & $\frac{1}{2}(\mathbf{7}+\mathbf{1,2})$ &  &  & \\\hline
\end{tabular}
\end{align}
Here, the self-intersection of these curves also dictate the gauge symmetry
and matter content for this theory on the resolved branch, as we have indicated.
These repeating patterns were noted as basic building blocks of F-theory
compactifications in \cite{Morrison:2012np, Watiaddme1}. For earlier work
where these building blocks were also identified see \cite{Bershadsky:1996nu}.

Thus, what the F-theory realization gives us is a direct description of the
tensor branch of the conformal matter sector. By following a similar
procedure, other collisions with different singularity types $G\times G'$ lead to canonical notions of
conformal matter. We give a list of such conformal matter sectors in Appendix B.

\subsection{Higgsing and Brane Recombination}

A hallmark of bifundamental matter is that activating a vev breaks some of the
symmetries of the system. Even in our non-Lagrangian systems, this
characterization still carries over. As a warmup, consider again the collision
of two $A_{k-1}$-type singularities, with a bifundamental hypermultiplet in
the $(\mathbf{k},\overline{\mathbf{k}})$ of $SU(k)_{L}\times SU(k)_{R}$. The corresponding
geometric singularity is locally given by:%
\begin{equation}
y^{2}=x^{2}+u^{k}v^{k}.
\end{equation}
Activating a bifundamental corresponds to a brane recombination operation. As
explained in \cite{BHVI}, this can be viewed as the deformation $uv\longmapsto
uv+a$. So in other words, the flavor symmetry is broken to an
$SU(k)_{\text{diag}}$ stack supported at $uv+a=0$:%
\begin{equation}
y^{2}=x^{2}+\left(  uv+a\right)  ^{k}.
\end{equation}
Similar considerations hold for the strongly coupled conformal matter. For
example, in the collision of two $E_{8}$ singularities, we have the breaking pattern:
\begin{equation}
y^{2}=x^{3}+u^{5}v^{5}\longmapsto x^{3}+\left(  uv+a\right)  ^{5},
\end{equation}
that is, we break to the diagonal of $E_8 \times E_8$.

Following up on our discussion of collision of singularities given earlier, we
can see that a similar characterization holds for all of the other collisions.
In other words, if we have flavor symmetry $G_{L}$ supported on $u=0$ and
$G_{R}$ supported on $v=0$, then the brane recombination $uv\longmapsto uv+a$
breaks this to the diagonal subgroup.\footnote{For further details on these brane recombination operators see \cite{more6d}.}

\section{CFTs from Domain Walls \label{sec:Mprobe}}

In this section we introduce our first class of examples of 6d\ SCFTs with
conformal matter. In M-theory, these will be realized by M5-branes probing an
ADE\ singularity. In field theory terms, we introduce a class of $(1,0)$
superconformal field theories which are realized as domain wall solutions in
seven-dimensional gauge theory. This leads to theories
where the flavor symmetry of the CFT is a product $G_{L}\times
G_{R}$ with $G_{L}\simeq G_{R}$ an ADE\ group. The problem naturally reduces
to the study of a single M5-brane probing the singularity, leading to the conformal matter, from which one can
deduce the quiver theory associated with $N$ parallel M5-branes. We shall therefore label these theories
as $\mathcal{T}(G,N)$, in the obvious notation.

This section is organized as follows. First, we begin with some general
considerations in both M- and F-theory. Next, we consider in turn each type of orbifold
singularity. In the case of the A- and D-series, we also provide realizations
in IIA string theory.

\subsection{M5-brane probes of ADE\ Singularities}

To begin, we recall that seven-dimensional super Yang-Mills theory with 16
supercharges is realized by the M-theory background $\mathbb{R}^{6,1}%
\times\mathbb{C}^{2}/\Gamma_{G}$, where $\Gamma_{G}$ is an ADE discrete
subgroup of $SU(2)$. For additional properties of the group theory and
associated geometry of these singularities, see Appendix C. The bosonic field
content of this theory consists of a seven-dimensional gauge field, and three
real adjoint-valued scalars.

\begin{figure}[ht]
	\centering
		\includegraphics[scale=.5]{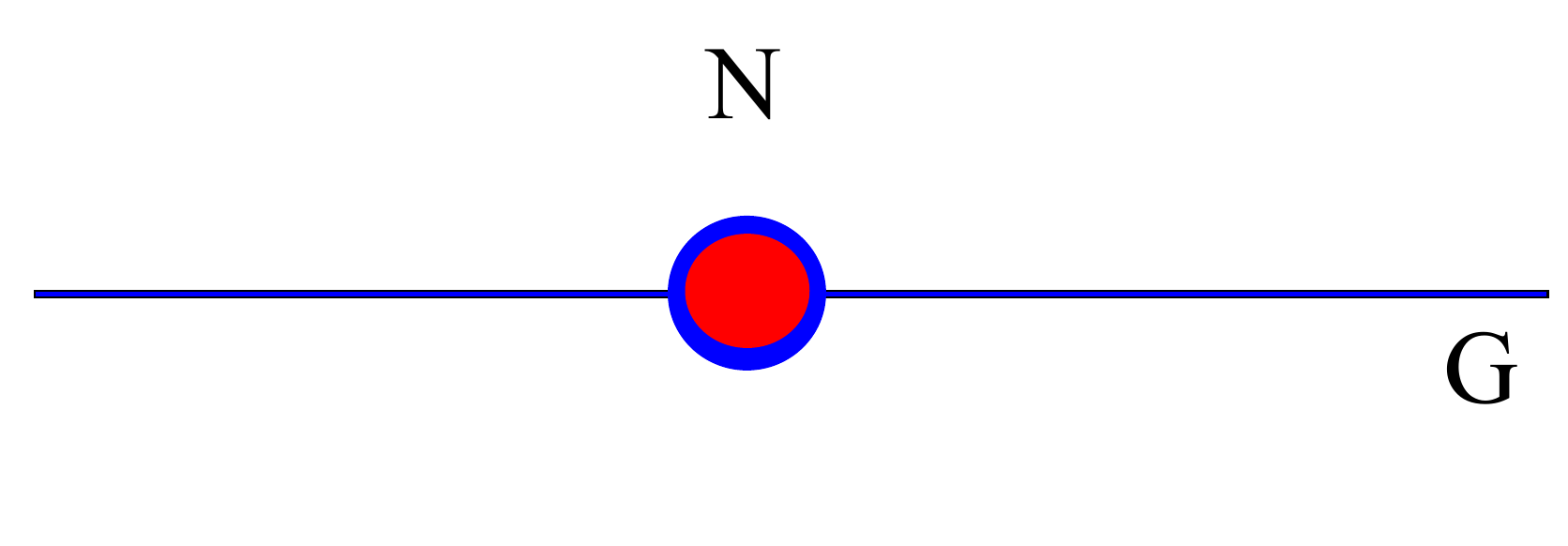}
		\includegraphics[scale=.5]{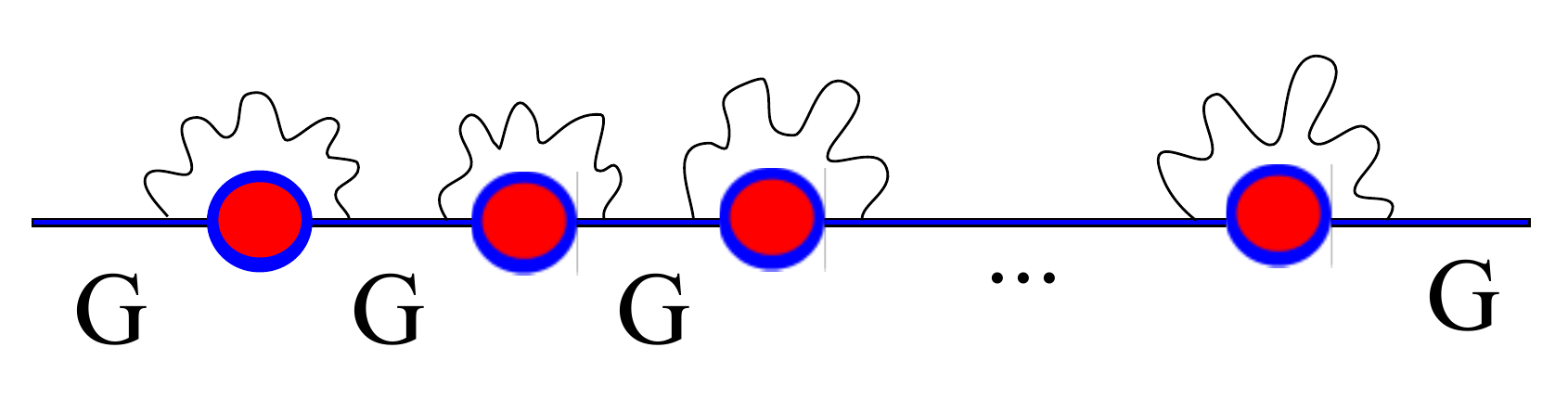}
	\caption{\textsc{up}: $N$ M5-branes at an orbifold singularity (SCFT point). \textsc{down}: The Coulomb/tensor branch deformation
	where the $N$ M5-branes are separated along the singularity locus and associated conformal matter which are symbolically represented by wavy lines between adjacent gauge factors.}
	\label{fig:Coulomb}
\end{figure}

Domain wall solutions of the M-theory realization correspond to M5-brane
probes which fill six spacetime dimensions and sit at the orbifold fixed
point. In more detail, the domain wall fills $\mathbb{R}^{5,1}$ and sits at a
point of the real line factor of $\mathbb{R}\subset \mathbb{R}\times\mathbb{C}^{2}/\Gamma_{G}$.
We are interested here precisely in the 6d theory living on the worldvolume of
this domain wall; see figure \ref{fig:Coulomb}. Being half-BPS, this 6d system has $(1,0)$ supersymmetry.
Moreover, being a domain wall for the 7d theory, the 7d gauge theory degrees
of freedom serve as flavor symmetry currents in the 6d\ system. Thus the
system has a flavor group $G_{L}\times G_{R}$. The 7d gauge symmetry may be
(partially) broken by suitable choices of boundary conditions for the 7d
fields at the domain wall \cite{Gaiotto:2014lca, DiacoNahm}.

By a similar token, we can introduce multiple domain walls and partition up
the real line factor in $\mathbb{R}\times\mathbb{C}^{2}/\Gamma_{G}$ into
finite size segments such that the the leftmost and rightmost segments are
still non-compact. Each such segment on the real line specifies a
six-dimensional gauge theory with gauge group $G$. The value of the gauge
coupling is in turn specified by the length of the interval. As an interval
segment becomes large, the corresponding gauge theory factor becomes weakly
coupled. In particular, we see that the leftmost and rightmost intervals are
non-compact and thus support flavor symmetries.

Summarizing then, we have arrived at a six-dimensional theory with $(1,0)$
supersymmetry, i.e. eight real supercharges. For each segment of the real
line, we have a corresponding gauge group:%
\begin{equation}
G_{\text{quiver}}=G_{1}\times...\times G_{N-1}%
\end{equation}
where we have partitioned up the real line into $N-1$ finite segments, and $G_{i}\simeq
G$ for all $i$. The 6d\ gauge coupling of each segment is proportional to the
length of the segment:%
\begin{equation}
\frac{1}{g_{i}^{2}}\sim L_{i},
\end{equation}
where $L_{i}$ is the length of the interval. Hence, the leftmost and rightmost
segments define flavor symmetries, while finite size intervals contribute
dynamical degrees of freedom. An additional feature of this construction is
that the length of each line segment is itself a dynamical mode, i.e. the scalar of a
tensor multiplet.

We reach a conformal fixed point by shrinking the distance between the domain
walls to zero size, that is, by passing to a strongly coupled fixed point of
the gauge theory. In other words, our description in terms of domain walls
partitioning up the real line characterizes the tensor branch of a
six-dimensional theory.

Each domain wall contributes additional degrees of freedom trapped along its
worldvolume. From this characterization, we see that we have a generalized
notion of a quiver gauge theory: We have a set of gauge groups and matter sectors
which act as links between them. Clearly then, it is important to know what
are the additional degrees of freedom living on each domain wall, i.e., the conformal matter.
The conformal matter sector corresponds to the special case of $N=1$ given by a single M5-brane which
in F-theory corresponds to the case of two non-compact $G$-type seven-branes
intersecting at a single point. The more general situation with $N$ M5-branes, translates
in F-theory to the case
where the seven-branes intersect at the $\mathbb{Z}_{N}$ fixed point of the
$A_{N-1}$ singularity.  The conformal matter will automatically have $G\times G$ symmetry
as discussed before.  In the F-theory setup this simply comes from the fact that the non-compact
seven-branes play the role of global symmetries.

As for what this conformal matter is, we know the answer, as it follows directly from the results reviewed
in section \ref{sec:CBI}. To see this, observe that in M-theory, the $A_{N-1}$
$(2,0)$ theory is realized by $N$ coincident M5-branes, while in F-theory,
it is realized by the geometry $\mathbb{C}^{2}/\mathbb{Z}_{N} \times T^2$. In
the resolution of the base, we have $N-1$ $\mathbb{P}^{1}$'s, each with
self-intersection $-2$, with neighboring intersections dictated by the
$A_{N-1}$ Dynkin diagram. Moreover, the volumes of the $\mathbb{P}^{1}$'s
control the relative positions of the M5-branes on the tensor branch. We are
interested in the case where there are some additional flavors, so we can also
introduce two stacks of non-compact seven-branes, with respective gauge groups
$G_{L}$ and $G_{R}$.\footnote{In fact, sometimes such a flavor symmetry is \textit{required} in order to
satisfy the condition that an elliptic fibration exists. In field theory, it is required to
satisfy 6d gauge anomaly cancelation.} In the resolution of the $\mathbb{C}^{2}/\mathbb{Z}_{N}$
singularity, $G_{L}$ intersects the leftmost $\mathbb{P}^{1}$ while $G_{R}$
intersects the rightmost $\mathbb{P}^{1}$. Following the analysis of
\cite{Heckman:2013pva}, we can see that the minimal singularity type over each
of the $-2$ curves enhances to an algebra $\mathfrak{g}$. So in other words, we have a configuration
of curves:%
\begin{equation}
\lbrack G_{L}]\,%
\begin{tabular}
[c]{cccccc}
$\mathfrak{g}$ &$\mathfrak{g}$& $...$ &$\mathfrak{g}$& $\mathfrak{g}$\\
$2$ & 2& $...$ &2& $2$\\
\end{tabular}
\,[G_{R}],
\end{equation}
that is, each $-2$ curve is wrapped by a seven-brane with gauge symmetry $\mathfrak{g}$.
At each intersection of a divisor in the base, we get a conformal matter sector.
for which the $G$-symmetries are gauged by the adjacent $G$ on the $-2$ curve.
In the following sections we discuss the $E$ case first, which
has no type IIA realization, and then turn to the $A$ and $D$ cases which
do have IIA realizations.

\subsection{M5-branes Probing E-type Singularity}

As discussed above, the problem reduces to finding the conformal matter which arises when
two $E_i$ singularities collide.  This was already discussed in section \ref{sec:CBI}.
For example, in the case $G=E_8$, the relevant conformal matter is given by line (\ref{eq:E8cm}); thus we end up
with a theory with gauge symmetries:
\begin{equation}\label{eq:E8cft}
	\mathfrak{sp}_{1} \times \mathfrak{g}_{2} \times \mathfrak{f}_4 \times \mathfrak{g}_2 \times \mathfrak{sp}_1  \ ,
\end{equation}
with half hypermultiplets in the $(\mathbf{2}, \mathbf{7}+ \mathbf{1})$ of each $\mathfrak{sp}_1 \times \mathfrak{g}_2$ factor (and in the
$(\mathbf{7}+ \mathbf{1}, \mathbf{2})$ of each $\mathfrak{g}_2 \times \mathfrak{sp}_1$ factor), and with flavor group $E_8 \times E_8$.
More precisely we have a \emph{generalized} quiver theory of the form:
\begin{equation}\label{GQE8E8}
\begin{gathered}
\includegraphics[width=0.8\textwidth]{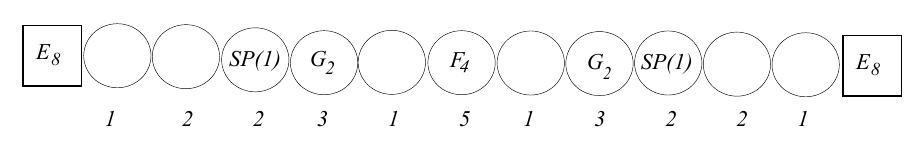}
\end{gathered}
\end{equation}%
where the notation above denotes one tensor multiplet per each circle, the
number below the circle denotes the (negative of) self intersection number
of the corresponding cycle in the F-theory geometry. The two systems
\begin{equation}
\begin{gathered}
\includegraphics[width=0.13\textwidth]{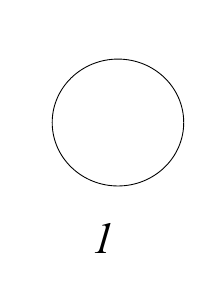}
\end{gathered}\qquad\qquad\begin{gathered}
\includegraphics[width=0.11\textwidth]{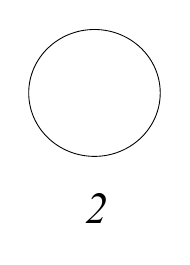}
\end{gathered}
\end{equation}
denote, respectively, the theory of a single small $E_8$ instanton which has a global $E_8$ symmetry (consistent with
the gauge and flavor symmetries attached to it as in line \eqref{GQE8E8}), and the $(2,0)$ theory of $2$ parallel
M5-branes (i.e. the $A_1$ (2,0) system). The configuration of a $-1$ curve next to a $-2$ curve corresponds to the theory of two
small $E_8$ instantons. In the present case, the $-2$ curve touches a curve with $\mathfrak{sp}_1$ gauge symmetry, which is obtained by
gauging a subalgebra of the $\mathfrak{so}_4$ global symmetry.\footnote{For further discussion of the anomaly polynomial for multiple small
$E_8$ instantons, see \cite{Ohmori:2014pca}. The physical interpretation of the $\mathfrak{sp}_1$ gauge symmetry was presented in reference
\cite{Ohmori:2014kda}, after the present work first appeared.}

\begin{figure}
\begin{center}

\includegraphics[width=0.6\textwidth]{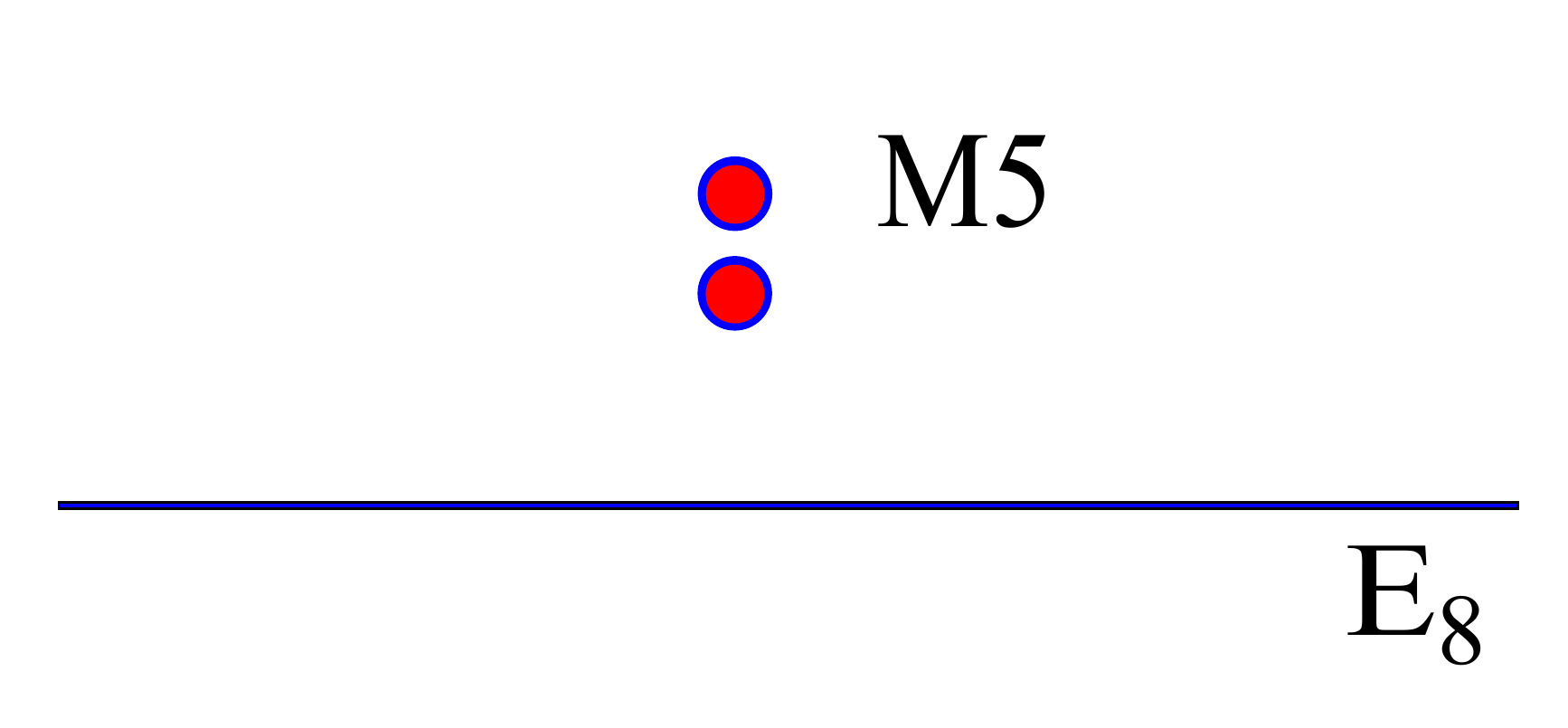}

\includegraphics[width=0.6\textwidth]{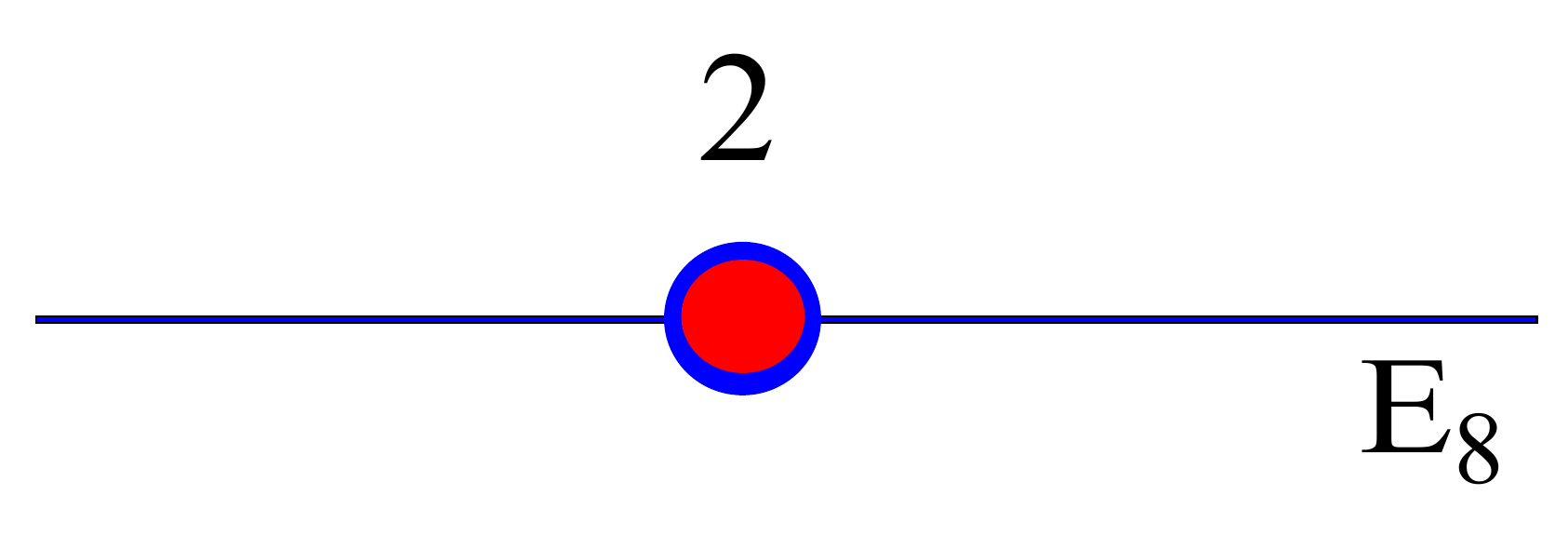}

\includegraphics[width=0.6\textwidth]{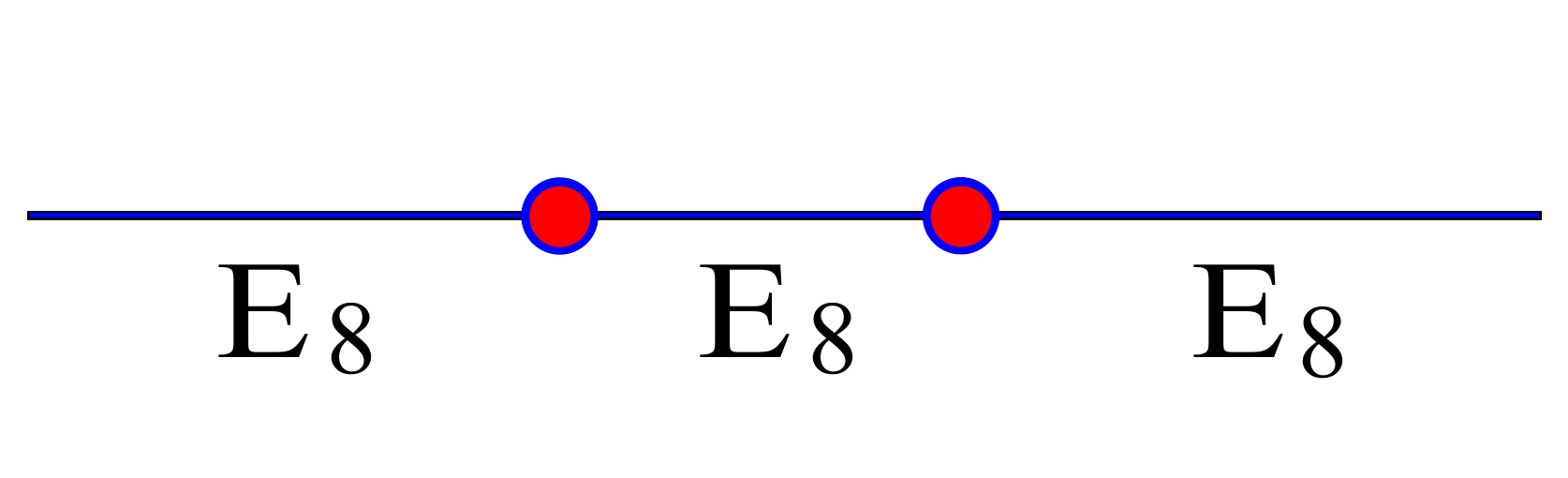}

\includegraphics[width=1.0\textwidth]{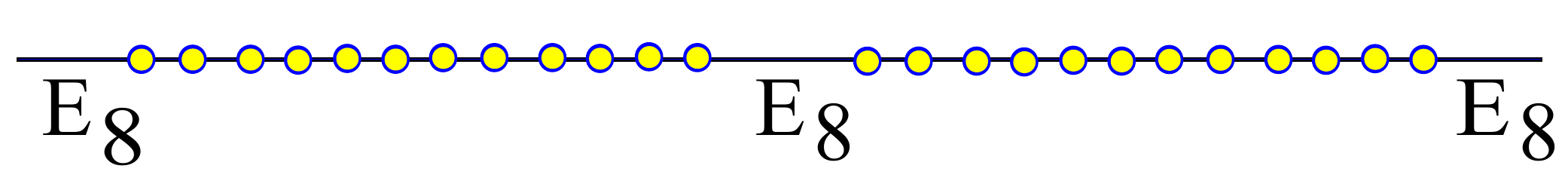}

\includegraphics[width=1.0\textwidth]{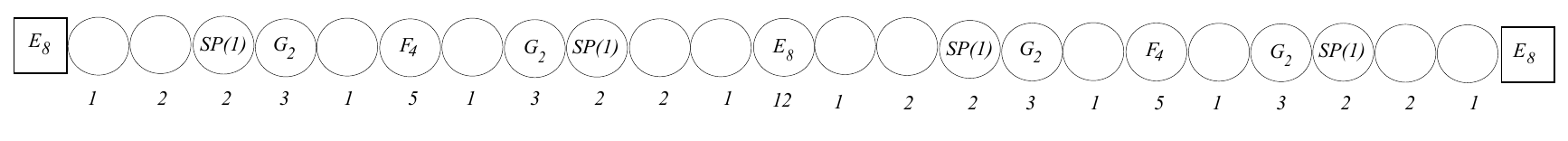}

\caption{Example of two M5-branes probing an $E_8$ singularity. Moving onto the tensor branch gives
rise to $(E_8,E_8)$ conformal matter that are SCFTs themselves with their own tensor branches, as
described by the generalized quiver of line \eqref{GQE8E8}.  This suggests that each M5-brane
on $E_8$ has fractionated to 12 pieces.}\label{fig:M5fractn12}
\end{center}
\end{figure}

\subsubsection{Fractional M5-branes}

This picture suggests that the single M5-brane on the line of $E_8$ singularity has split to 12 points
on it, leading to 11 finite segments whose lengths are controlled by the scalars
of the associated tensor multiplet (see figure \ref{fig:M5fractn12}).    In other words, we have branes with {\it fractional}
M5 brane charge.
Looking at the list of colliding E-singularities, we discover in this way that the fractionalization
of M5-branes for the ADE series are given by
\begin{equation}%
\begin{tabular}
{c|c|c|c|c|c|c|}\cline{2-6}
&  $E_8$ & $E_{7}$ & $E_{6}$ & $D_{p}$ & $A_{k}$ \\\hline
\multicolumn{1}{|l|}{\# of M5 Fractions} &  $12$ & $6$ & $4$ & $2$ & $1$\\\hline
\end{tabular}
\end{equation}
where as we will discuss in the context of D-type singularities, an M5-brane on it can fractionate
to 2, while in the case of M5-branes on an A-type singularity no fractionation occurs.

This raises the question of why the gauge group
factor on each interval is {\it not} the $E_8$ gauge symmetry, but rather the list above.
We propose an answer to this question:\footnote{This proposal was motivated by a question posed by E. Witten at
Strings 2014.} It has already been suggested that by choices of discrete three-form fluxes stuck at seven-dimensional M-theory singularities,
the type of the gauge symmetry can change\cite{de Boer:2001px}.  We thus propose that fractional M5-branes
change the discrete flux from one value to the next, changing the gauge group in the process.  Moreover we propose
that each fractional M5-brane changes the three-form flux fraction by equal amounts.  So for example in the
$E_8$ case, each fractional M5-brane will change the fractional three-form flux by $1/12$.
Our description of fractional M5-branes also matches to the list of groups (up to what we hope is a typo for the $E_6$ entry)
listed in table 14 of reference \cite{deBoer:2001px}.

In fact, the fraction of discrete flux matches \textit{where} we find
the corresponding group in our repeated pattern of exceptional curves!
For example, reading from left to right in the configuration $1,2,2,3,1,5,1,3,2,2,1$
for the conformal $(E_8 , E_8)$ matter, we find that there is an $\mathfrak{sp}_{1}$ gauge symmetry
on the third curve, which would give $ 3/12 =1/4 $, associated with a $\mathbb{Z}_{4}$ flux.
Further, on the fourth curve, we have $4/12=1/3$ flux
giving $\mathfrak{g}_2$ ($\mathbb{Z}_3$ flux), the sixth curve yields $6/12=1/2$,
giving $\mathfrak{f}_4$ ($\mathbb{Z}_2$ flux). Similar considerations hold for $E_7$ and $E_6$ (and $D$ type) conformal matter.
One subtlety, however, is that the labels which correspond to trivial gauge
group are different from \cite{deBoer:2001px}. For example, for $E_8$ there is no $\mathbb{Z}_{5}$. But in addition
there are all the other fractions of $1/12$ which lead to no gauge factors.

\subsection{IIA Realization of $\mathcal{T}(SU(k) , N)$ theories}

\begin{figure}[ptb]
\begin{center}
\includegraphics[width=0.6\textwidth]{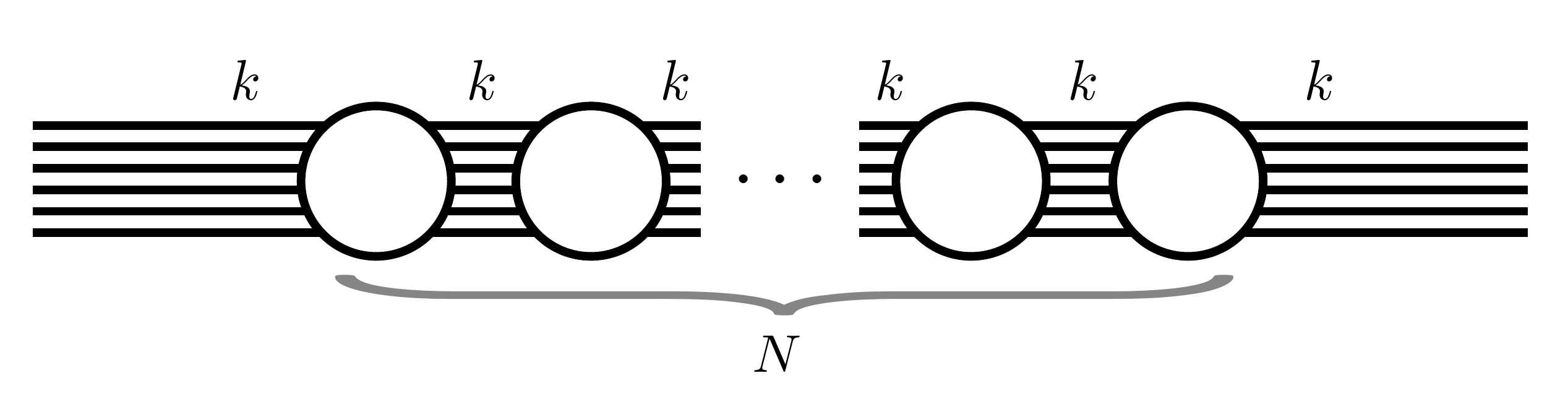}
\end{center}
\caption{Type IIA description of the system of $N$ M5-branes probing the
$\mathbb{C}^{2}/\mathbb{Z}_{k}$ singularity: here we are drawing schematically
the $X^{6}$ direction for the tensor branch. One obtains $N$ NS5-branes
sitting on top of $k$ infinite D6 branes.}%
\label{noD8HZ}%
\end{figure}

Let us now consider more closely the theory of $N$ M5-branes probing an $A_{k-1}$-type singularity
$\mathbb{C}^{2}/\mathbb{Z}_{k}$. A convenient description of the domain wall
discussed above is obtained via a standard duality with Type IIA. The
$A_{k-1}$-type singularity can be thought of as an infinite radius limit of
the charge $k$ Taub-NUT space, or $A_{k-1}$ ALF space. More precisely, the
$TN_{k}$ space has a canonical fibration as a circle of radius $R$, $S_{R}%
^{1}$, over $\mathbb{R}^{3}$: in the limit $R\rightarrow\infty$ one obtains
the $A_{k-1}$ ALE space. M-theory on the $TN_{k}$ geometry describes a system
of $k$ Kaluza-Klein monopoles that dualize to a system of $k$ parallel
infinite D6-branes on the Type IIA side, once one identifies $S_{R}^{1}$ with
the M-theory circle \cite{Sen:1997kz}. In the limit in which the KK
monopoles coincide, one obtains an enhanced $SU(k)$ gauge symmetry; these are
the degrees of freedom of the 7d gauge theory we discussed above. Instead, in
Type IIA the D6-branes fill the $X^{0},X^{1},...,X^{6}$ directions and the
enhanced symmetry comes from the gauge theory living on the worldvolume of the
stack of $k$ coincident D6-branes. Under such a duality the $N$ M5 probes turn
into $N$ NS5 probes of the stack of $k$ D6-branes. On
the worldvolume of the $N$ coincident NS5s lives a well-known but still
quite mysterious 6d $(1,0)$ SCFT with tensionless strings \cite{Hanany:1997gh}.
Let us briefly discuss the field content of this theory. Each NS5
contributes to the worldvolume a $(2,0)$ tensor multiplet. As usual, the
center of mass degrees of freedom decouple and we are left with a system of
$N-1$ tensor multiplets at the superconformal point. Each $(2,0)$ multiplet
decomposes into a $(1,0)$ tensor plus two $(1,0)$ hypers. The scalars of the
multiplets arise from quantizing the motion transverse to the NS5s in the whole
geometry, including the M-theory circle. In particular the vev of the scalars
in the tensor multiplets parameterize the relative distance between the NS5s
along the $X^{6}$ direction. Separating the NS5s we break superconformal
invariance and move onto the tensor branch of the system, eventually
landing on a quiver gauge theory (see figure \ref{noD8HZ})
\begin{equation}
\begin{gathered} \includegraphics[width=0.8\textwidth]{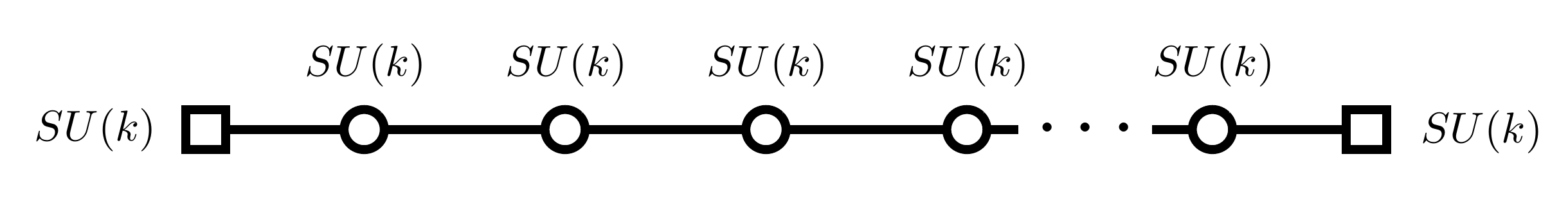} \end{gathered}\label{222quiver}%
\end{equation}
where edges stands for bifundamental hypers, the $N-1$ round nodes for gauge
groups, and the two square nodes for flavor groups as usual. Notice that
naively the brane system in figure \ref{noD8HZ} should correspond to a $U(k)$
quiver gauge theory. In this context, however, the $U(1)$'s are anomalous
because of the nonzero term $F_{U(1)}\wedge\text{Tr}(F^{3}_{\text{SU(k)}})$ in
the anomaly polynomial. To cure this pathology, one couples the compact scalar
corresponding to the M-theory circle to the abelian gauge field, making it
massive \cite{Hanany:1997gh, Douglas:1996sw}. This is the reason why in 6d one
obtains $SU(k)$ gauge groups on the tensor branch.

\subsubsection{IIB / F-theory\ Description}\label{ssub:IIB-AN}

The $\mathcal{T}(SU(k), N)$ theories also have a straightforward realization in
type IIB string theory. To obtain it, we recall that if we T-dualize the
$S^{1}$ at the boundary of an ALE\ space, we reach a collection of coincident
NS5-branes. In such a configuration, we can next consider a stack of D7-branes
which pass through the singular locus of this geometry. Upon T-dualizing this
circle, we see that a D7-brane wrapped over a collapsing $\mathbb{P}^{1}$ of
the geometry $\mathbb{C}^{2}/\mathbb{Z}_{N}$ will become a D6-brane in the dual description. Putting
these elements together, we see that for each such $\mathbb{P}^{1}$, we get a
stack of D6-branes suspended between NS5-branes.

In fact, we can also lift this IIB\ description back to F-theory. The
tensor branch of the 6d $(1,0)$ theory engineered with F-theory is given by a
local system of $-2$ curves in the base of an elliptically fibered Calabi-Yau
threefold. In the case of the system of $N$ M5-branes probing the
$\mathbb{C}^{2}/\mathbb{Z}_{k}$ singularity, each such curve supports a Kodaira
type $I_{k}$ singular fiber, corresponding to enhanced gauge symmetry of type
$SU(k)$. Flavor groups in the F-theory description correspond to non-compact
divisors of the base of the elliptically fibered Calabi-Yau threefold. Since
we know that the system carries an $SU(k)\times SU(k)$ flavor group, we need
two non-compact divisors in the base. Moreover, the fact that the system we
are engineering corresponds to the tensor branch of an SCFT
translates to the requirement that all compact $-2$ curves in the base can be
shrunk simultaneously to zero. Since bifundamental hypers are trapped at the
intersections in between the various $\mathbb{P}^{1}$'s in the base, starting
from the tensor branch description of line \eqref{222quiver}, the
resulting configuration of $-2$ curves for the F-theory description is:
\begin{equation}
\begin{gathered} \includegraphics[width=0.6\textwidth]{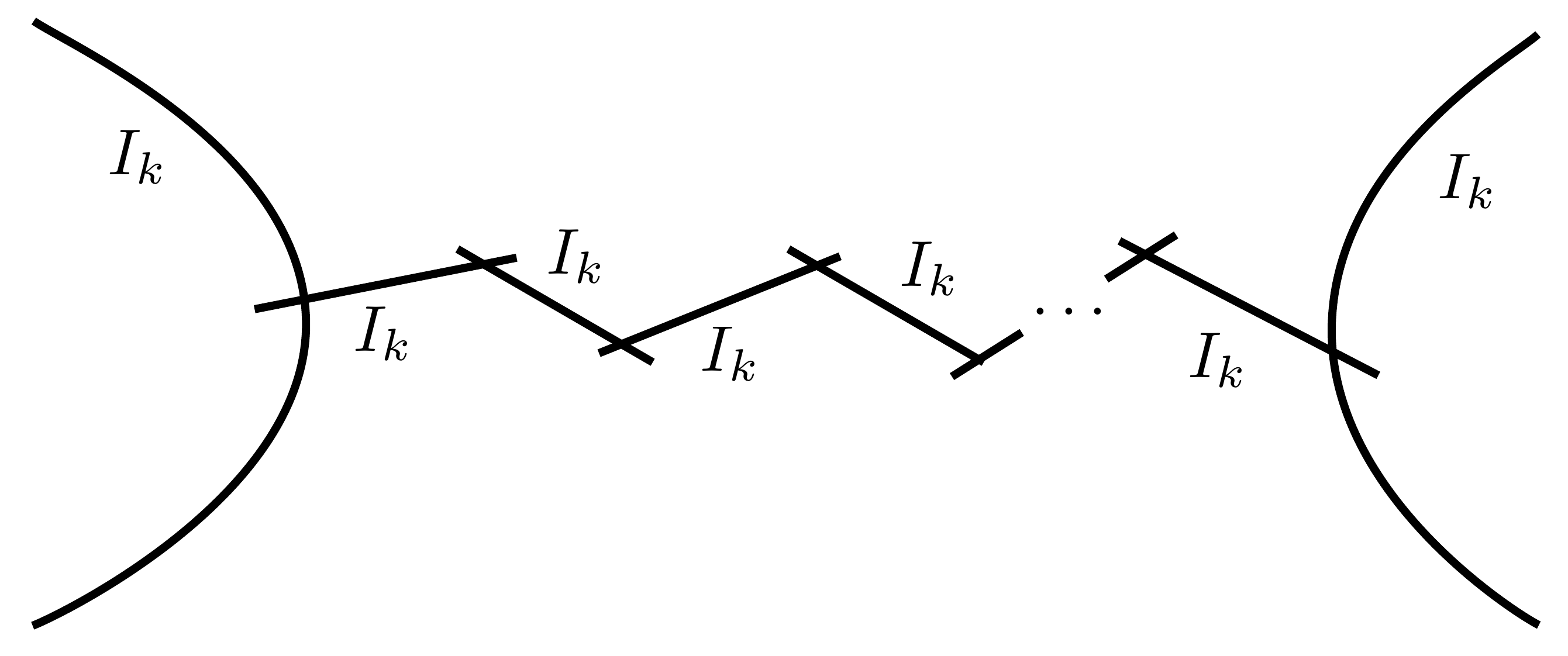} \end{gathered}
\end{equation}
the leftmost and rightmost $-2$ curves supporting $I_{k}$ type Kodaira fibers
are non-compact. Taking the conformal limit of such a system would correspond
to shrinking the compact $-2$ curves to zero size: at the SCFT point the two
non-compact curves corresponding to the flavor symmetries touch at an
$A_{N-1}$ singularity of the base:
\begin{equation}
\begin{gathered} \includegraphics[width=0.5\textwidth]{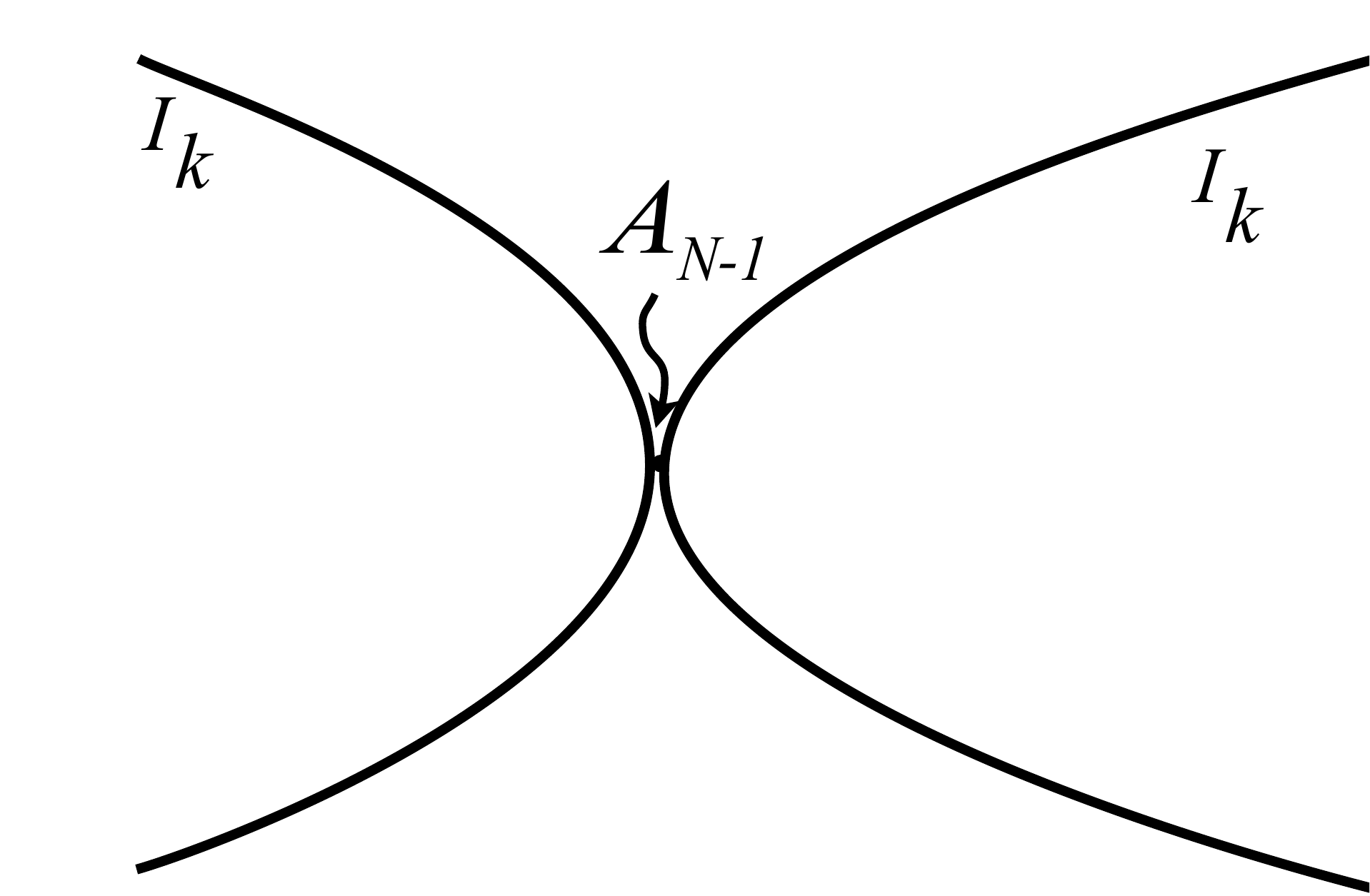} \end{gathered}
\end{equation}
This local configuration of curves corresponds to the SCFT describing the
worldvolume of $N$ M5-branes probing an $A_{k-1}$ singularity in M-theory.

\subsection{IIA Realization of $\mathcal{T}(SO(2p) , N)$ theories}\label{sub:IIA-D}

Let us now turn to the theory of $N$ M5-branes probing a D-type singularity. First
of all, a $D_{p+4}$ singularity gives rise to 7d\ SYM\ theory with gauge group
$SO(2p+8)$ for $p\geq0$. Following the same reasoning of the previous section,
we seek a 6d $(1,0)$ system with $SO(2p+8)\times SO(2p+8)$ flavor group.

Following \cite{Sen:1997kz, Seiberg:1996nz} we can obtain a Type IIA
description of the system by replacing the $D_{p+4}$ singularity with the
corresponding $D_{p+4}$ ALF space. Eventually, one obtains a stack of
$p+4$ parallel D6-branes on top of an O$6_-$ plane, together with $p+4$
mirror images of the D6s below. The resulting 7d theory has $SO(2p+8)$
gauge symmetry. Consider now introducing domain walls. By construction, the
theory living on the wall has $SO(2p+8)\times
SO(2p+8)$ flavor symmetry. However, when an O$6_{\pm}$ plane meets an
NS5-brane, it turns into an O$6_{\mp}$ plane, with a net shift of eight units of
D6 charge. Now, a system of $p+4$ D6-branes parallel to an O$6_{+}$ plane gives rise
to an $\mathfrak{sp}_{p}$ gauge theory. Therefore, we do not just get a set of $N$ NS5-branes
sitting on top of the $p+4$ D6s in the presence of an O$6_{-}$ plane. Indeed, for $N$ odd
such a system would have the wrong flavor symmetry, i.e.
$SO(2p+8)\times Sp(p)$, and this is impossible. The only way out from
this paradox is to conclude that there are $2N$ NS5-branes, and $Sp$ factors
in between our $SO$ factors, for a total of $2N-1$ tensor multiplets.

We have just found that in contrast with M5-brane probes of A-type
singularities, for D-type singularities we find fractional M5-branes. Since
there are only two varieties of these branes for the D-type singularity, we
shall refer to them as $1/2$-M5-branes.

This notion of fractionalization of M5-brane for the D-type singularity further
supports the picture we proposed for fractionalization of M5-branes probing the E-type singularities.
Indeed, as noted in \cite{Witten:1997bs}, the singularity with $\mathbb{Z}_2$ flux
is a simple lift of a D6-O6 system of Type IIA.

\subsubsection{IIB / F-theory\ Description}

Let us now turn to the IIB / F-theory description of this system. As before,
we can consider the effects of T-dualizing our suspended brane configuration
to a related configuration of D7-branes and O7-planes in type IIB string
theory. This leads us to a configuration of seven-branes of $SO$ type wrapping
the $-2$ curves of an A-type singularity. This A-type singularity is, in the
IIB\ description, associated with the IIA NS5-branes used to partition up the
interval in the first place.

Turning to the F-theory lift of this description, we need to consider $-2$
curves intersecting according to the $A_{N-1}$ Dynkin diagram. Each
$\mathbb{P}^{1}$ supports a Kodaira-Tate $I_{p}^{\ast}$ fiber:
\begin{equation}
\begin{gathered} \includegraphics[width=0.6\textwidth]{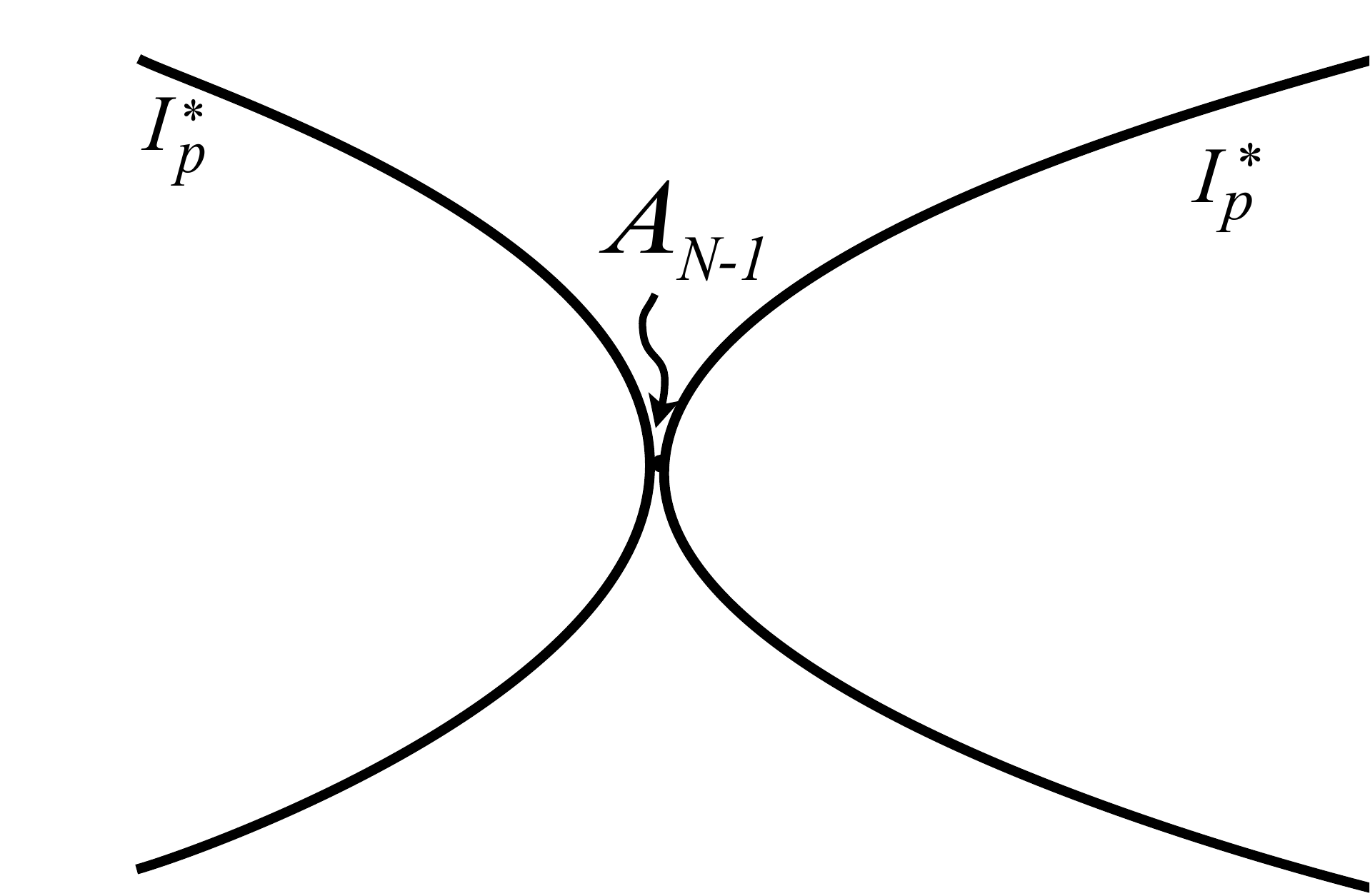} \end{gathered}
\end{equation}
As we already explained, the collision of two such fibers contains conformal
matter, given by an $\mathfrak{sp}_{p}$ gauge theory (coupled to half hypers)
wrapping a collapsing exceptional curve:%
\begin{equation}
\begin{gathered} \includegraphics[width=0.6\textwidth]{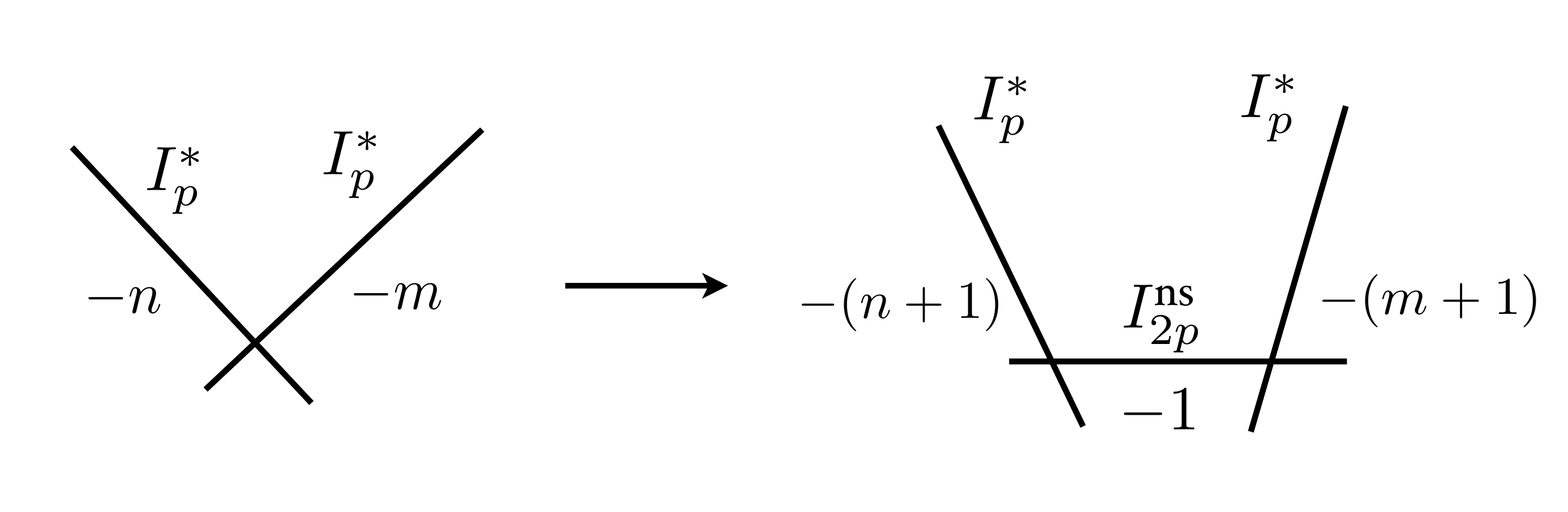} \end{gathered}
\end{equation}
This type of blowup is the only one that occurs for the
configuration of curves we are considering. Proceeding by successive
blowups, eventually we obtain the configuration
\begin{equation}
\begin{gathered} \includegraphics[width=0.6\textwidth]{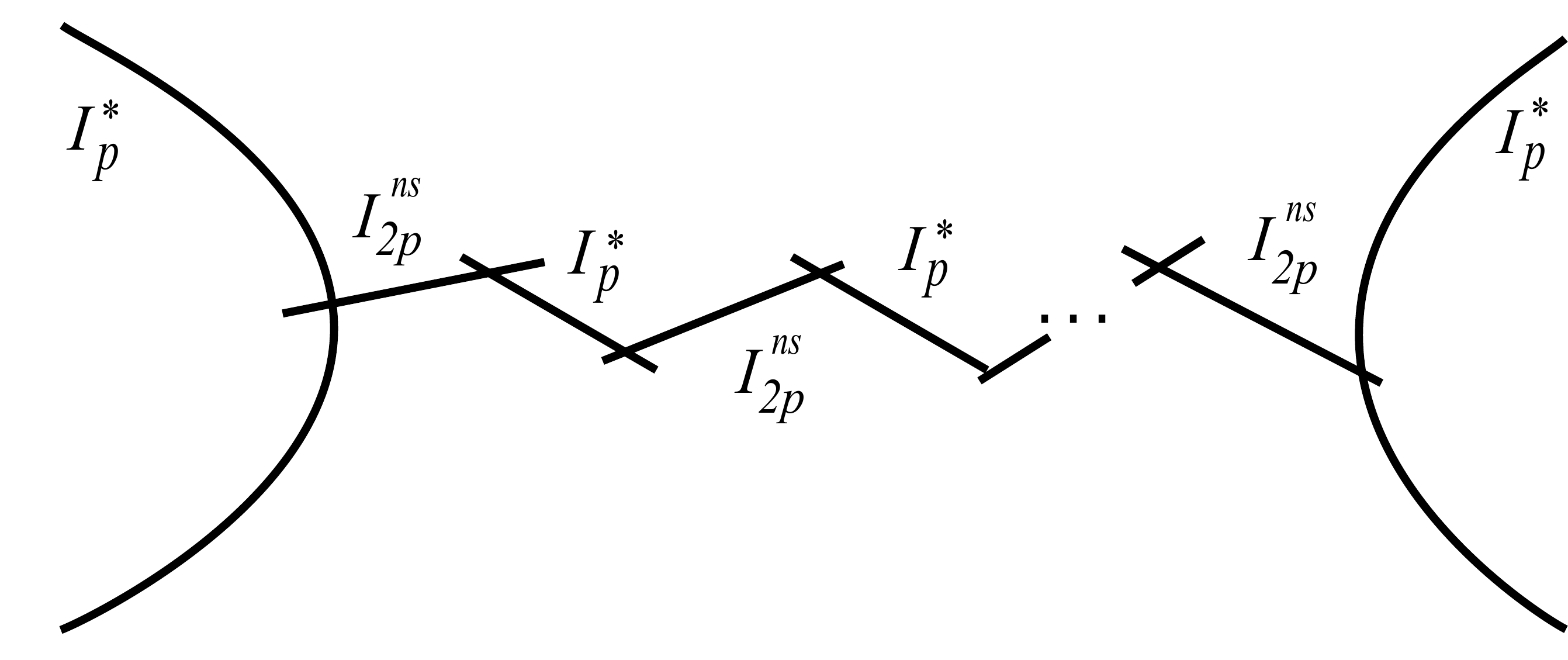} \end{gathered} \label{DkkCoulomb}%
\end{equation}
where each $I^{ns}_{2p}$ (resp$.$ $I_{p}^{\ast}$) fiber is supported on a compact
$-1$ (resp$.$ $-4$) curve in the base. Further, each $I^{ns}_{2p}$ is of non-split type,
so the associated algebra is $\mathfrak{sp}_{p}$. For the leftmost and rightmost non-compact curves
curves we have an ``external'' $I_{p}^{\ast}$ fiber, yielding an
$SO(2p+8)\times SO(2p+8)$ flavor symmetry. The tensor branch of this system can be described by an
$SO$ / $Sp$ quiver theory: the alternating $\mathfrak{sp}_{p}$ and
$\mathfrak{so}_{2p+8}$ factors correspond to the alternation of the $I^{ns}_{2p}$
and the $I_{p}^{\ast}$ singular fibers along the chain of intersecting
$\mathbb{P}^{1}$'s of line \eqref{DkkCoulomb}. At the intersections of the $-4$
with the $-1$ curves we also find localized matter modes in the form of
bifundamental half hypers.

Let us stress here a crucial difference with respect to what we have found
discussing the M5 probes of an $A_{k-1}$ singularity. For the theory of
$N$ M5-branes probing an $A_{k-1}$ singularity, the corresponding F-theory realization
involves dressing the $-2$ curves obtained by resolving the $A_{N-1}$ singularity
by $I_{k}$ fibers, so we find precisely $N-1$ tensor multiplets in the tensor/Coulomb branch of the
system. If, instead, we consider the theory of $N$ M5-branes probing a $D_{p+4}$ singularity, we have seen that
in the F-theory realization we dress the $-2$ curves obtained by resolving the $A_{N-1}$ singularity
with $I_{p}^{\ast}$ fibers. However, the full resolution does not lead to $N-1$ tensor multiplets, but to $2N-1$. This
fact again suggests the existence of a fractional M5 charge: probing the
$D_{p+4}$ singularity we find that a full M5-brane is a compound of two
objects, and this explains why we find almost twice as many tensor multiplets
in both the IIA\ and IIB\ descriptions of these 6d\ systems.

\subsubsection{The Special Case $p=0$}\label{subsub:SpecialK}

For generic values of $p$, we see the M5-branes probing the $D_{p+4}$
singularity lead, in the resolved phase, to a configuration of $-4$ and $-1$
curves, where the $-4$ curves support $SO$ type gauge groups, while the $-1$
curves support $Sp$ type gauge groups. When $p=0$, however, each $-1$ curve
supports no gauge group, since $Sp(0)$ is trivial.

This special case is also closely connected with the \textquotedblleft
rigid\textquotedblright\ theories encountered in \cite{Heckman:2013pva}.
Indeed, let us recall that the alternating pattern of compact curves:%
\begin{equation}
4,1,...,1,4
\end{equation}
supports a collection of $\mathfrak{so}_{8}$ gauge symmetries, each supported
on a $-4$ curve. In the notation of \cite{Heckman:2013pva} (see also Appendix
\ref{tables}), this theory comes from the minimal resolution of the endpoint
$3A_{N}3$. To reach the case of M5-branes probing a D-type singularity, we
decompactify the leftmost and rightmost $-4$ curves, thus converting them to
flavor symmetries.

\section{Novel 5d Dualities} \label{sec:5D}

Though our main focus in this paper is the theory of conformal matter in six dimensions, it is natural to ask what becomes of this system
upon further compactification. Here we study the simplest possibility given by compactifying on an $S^1$. We will argue that this
compactified theory is dual to a more conventional 5d quiver gauge theory, as obtained from D-brane probes of an ADE singularity.

To orient ourselves, let us consider the simplest example of this, namely $N$ M5-branes
probing an $A_k$ singularity.  As we already discussed, this is expected to give a linear quiver
gauge theory with $SU(k+1)^{N-1}$ gauge symmetry, and bifundamental matter.  Compactifying
on a circle will still give rise to the same quiver theory but now in five dimensions.
On the other hand, using the relation between M-theory and type IIA, by viewing the circle
as the `11-th' dimension, this is the same as $N$ D4-branes probing the $A_k$ singularity.
Using the Douglas-Moore construction
\cite{Douglas:1996sw}, we deduce that this should be given by the quiver theory involving
$SU(N)^{k+1}$ arranged along a ring leading to a duality between two different 5d quiver theories.
In fact this duality was explained in \cite{Haghighat:2013gba,Haghighat:2013tka,Hohenegger:2013ala}
using the horizontal/vertical exchange symmetry \cite{Katz:1997eq} of the toric realization of these theories, and used to compute
the partition function of the associated strings of the 6d $(1,0)$  SCFT.

\begin{figure}
\begin{center}

\includegraphics[width=1.0\textwidth]{GQ2M5E8R}

\medskip

\includegraphics[width=0.6\textwidth]{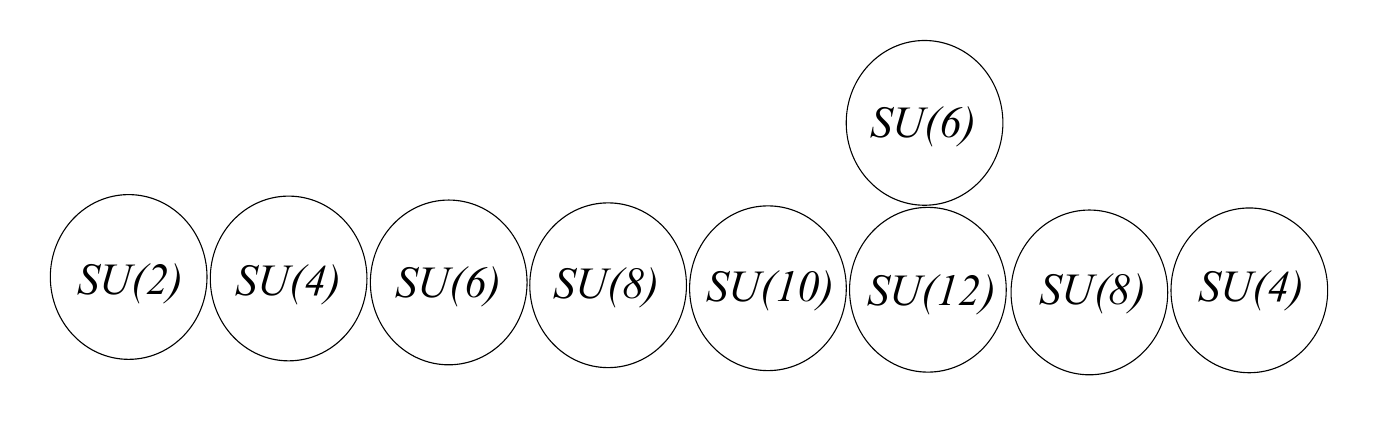}


\caption{An instance of a novel 5d duality. The 6d theory of two M5-branes probing an $E_8$ singularity (\textsc{top}) compactified on $S^1$ is dual to the affine $\widehat{E}_8(2)$ 5d quiver system (\textsc{bottom}).}\label{fig:M5fractnE8}
\end{center}
\end{figure}

For the case of M5-branes probing D-singularities in the same way we get a duality
between the quiver chain $SO\times Sp\times SO....$ and the affine D-quiver arising
from $N$ D4-branes probing D-singularities.  These are already non-trivial dualities.
For the E-series we get novel dualities which were not noted before.  For example,
in the case of 2 M5-branes probing an $E_8$ singularity the resulting duality is shown in figure \ref{fig:M5fractnE8}. Let us note here that the $E_8$ flavor symmetries are only realized in the strong coupling limit: indeed, the $-1$ curves at the very left and very right function as ``unconventional matter'' for the rest of the gauge theory system, which is massive on the tensor branch.

To provide further evidence in support of this duality, we do a basic check:
We shall count the number of vector multiplets on the Coulomb branch in
the five-dimensional theory obtained by dimensional reduction. We shall then
verify that this matches the dimensional reduction of M5-branes, i.e. D4-branes probing our
ADE\ singularity. This will provide a nice check on our proposal that we
have correctly identified the degrees of freedom in the 6d theory.\footnote{For
a similar statement about a different 5d duality see \cite{intriliaddme, PeRaj}.}

In the five-dimensional theory, we can get vector multiplets in one of two ways.
First of all, starting on the tensor branch, we see that each tensor
multiplet, upon dimensional reduction, gives a $U(1)$ vector multiplet. Additionally,
we see that for each 6d\ gauge group factor on the tensor branch, we get some
additional vector multiplets in 5d. So, we predict the number of vector multiplet to be $r+n_{T}$
where $r$ is the total rank of the 6d gauge groups and $n_{T}$ is the number of 6d
tensor multiplets. Moreover, the tensor multiplets either arise from the $(N-1)$ ``intervals''
between the M5-branes and $N$ times the number of tensors for each conformal
matter system.  Similarly the total rank of the gauge group $r$ arises
from two sources: $(N-1)r_G$ from the intervals between M5-branes where $r_G$ is
the rank of the $G_{ADE}$-type singularity, and $N$ times the rank of the gauge group from each conformal matter
sector. In a system with $N$ M5-branes, we have partitioned up our
interval into $N-1$ compact pieces, so the dimension of the Coulomb branch for
the 5d\ theory is:
\begin{equation}\label{testx}
\dim_{5d\text{ Coul}}(G,N)=n_{\text{interval}}+n_{\text{cmatter}},
\end{equation}
The contribution from the intervals is straightforward:
\begin{equation}
n_{\text{interval}}=(N-1)r_{G}+(N-1).
\end{equation}
The first summand comes from the 6d\ vector multiplets, and the additional
summand of $(N-1)$ comes from the tensor multiplet of each interval. The
contribution from the conformal matter sector must be calculated on a case by case
basis:%
\begin{align}
A_{k} &  :n_{\text{cmatter}}=0\\
D_{p} &  :n_{\text{cmatter}}=(p-3)N\\
E_{6} &  :n_{\text{cmatter}}=(2+3)N\\
E_{7} &  :n_{\text{cmatter}}=(5+5)N\\
E_{8} &  :n_{\text{cmatter}}=(10+11)N
\end{align}
in the E-type cases, we have split up the contribution in the parentheses:
The first term is the rank of the gauge group for the
conformal matter sector, and the second summand is the total number of
tensors. Totalling this up, we get a prediction for the dimension of the
Coulomb branch in the lower-dimensional theory:%
\begin{align}
\dim_{5d\text{ Coul}}(A_{k},N) &  =(k+1)N-(k+1)\\
\dim_{5d\text{ Coul}}(D_{p},N) &  =(2p-2)N-(p+1)\\
\dim_{5d\text{ Coul}}(E_{6},N) &  =12N-7\\
\dim_{5d\text{ Coul}}(E_{7},N) &  =18N-8\\
\dim_{5d\text{ Coul}}(E_{8},N) &  =30N-9.
\end{align}
A succinct formula for all of these cases is:%
\begin{equation}
\dim_{5d\text{ Coul}}(G)=\underset{i}{\sum}(Nd_{i}^{(\widehat{G})}%
-1)=Nh_{G}-r_{\widehat{G}},\label{fourDCoul}%
\end{equation}
where the sum on $i$ is over all the nodes of the affine Dynkin diagram, with
$h_{G}$ the dual Coxeter number for $G$, $r_{\widehat{G}}$ the rank of the
affine $\widehat{G}$ Dynkin diagram, and $d_{i}^{(\widehat{G})}$ the affine
Dynkin numbers for each node.

On the other hand, we also know that in the IIA description of the D4-brane
probe system, there is a quiver gauge theory with connectivity of an
affine ADE\ Dynkin diagram \cite{Douglas:1996sw}. Each node of the affine
Dynkin diagram quiver has gauge group $\prod_{i}SU(Nd_{i}^{(\widehat{G})})$.
Thus, the dimension of the Coulomb branch for the D4-brane probe theory is
nothing other than equation (\ref{fourDCoul}). This is a rather non-trivial
check of the 5d duality and on our proposal as a whole.

Finally, it would be interesting to see whether these dualities can also be extended to the case of flavor symmetries $G \times G^{\prime}$,
where $G \neq G^{\prime}$. By a similar token, it would be instructive to extend
these considerations to lower-dimensional systems, perhaps
along the lines of references \cite{Seiberg:1996nz, Ganor:1996pc, Porrati:1996xi, Dey:2013nf}.

\subsection{Hints of a 6d Duality}

The close match in five dimensions between D4-brane probes of ADE singularities, and our more ``exotic'' superconformal matter
sectors naturally suggests a lift all the way back to six dimensions.\footnote{ For a similar but ultimately different 6d physical system with a completion to a little string theory see \cite{intriliaddme}. The main distinction with our case is that there the configuration of curves involves $1,2,2,\dots,2,2,1$, while here the configuration of curves is of $2,2,\dots,2,2$ type. The configuration of curves $1,2,2,\dots,2,2,1$ cannot all
be simultaneously contracted at finite distance in moduli space, and so does not correspond to the tensor branch of an SCFT. This is the reason why the two dualities are ultimately very different in nature. However, we find interesting that the main test passed by the dualities proposed in \cite{intriliaddme} is formally the same as the one in our \eqref{testx}.} On the one hand, we have a rather conventional 6d quiver
gauge theory, of the type studied in \cite{Blum:1997fw}. On the other hand, we have a strongly interacting non-Lagrangian conformal fixed point.
Indeed, if we compare the superconformal indices on $S^5 \times S^1$ for these two theories, we are guaranteed to get the same answer. This is because the 5d theories on $S^5$ (after reduction on the $S^1$) are already dual. While this does not constitute a proof of a 6d duality, it is
at least suggestive and would be interesting to understand better.

\section{Partial Higgs Branches of the $\mathcal{T}(G,N)$ Theories \label{sec:HIGGS}}

So far, our discussion has focussed on SCFTs which have a geometric avatar in
both M- and F-theory. Starting from these \textquotedblleft master
theories\textquotedblright\ we can also move down to a number of lower
theories by partial Higgsing, that is, by activating operator vevs in the 6d SCFT which break part of the
flavor symmetry. Our plan in this section
will be to characterize how to pass from our $(G,G)$ theories down to
theories with a broken flavor symmetry.

Our interest in this section will be on partial Higgsing of the flavor symmetry. A full Higgsing to the diagonal $G_{diag} \subset G_L \times G_R$
symmetry group would correspond to moving the M5-branes off the singularity. We are interested instead in keeping all the M5-branes on the singularity, so
we exclude this possibility in what follows. This means in particular that the complex geometry of the F-theory compactification will stay put.
The resulting class of deformations in F-theory specify ``T-brane data'' (see e.g. \cite{TBRANES, Anderson:2013rka}).

The main result from our analysis is that we can characterize the resulting SCFTs
as $\mathcal{T}(G,\mu_{L},\mu_{R},N)$, where $N$ is the number of domain
walls, and $\mu_{L}$ and $\mu_{R}$ specify two independent nilpotent elements $\mu_{L} \in \mathfrak{g}_{L}$
and $\mu_{R} \in \mathfrak{g}_{R}$ (and their orbits), yielding a breaking pattern of the
\textquotedblleft ambient\textquotedblright\ flavor symmetry $G_{L}\times
G_{R}$. There are no anomaly cancellation constraints since
the choice of nilpotent element dictates a breaking pattern
into the interior of the theory on the tensor branch.

To understand the possible ways to break the flavor symmetry, it is
instructive to see the effects of these contributions on the worldvolume
theory of the flavor branes. For specificity, we work in terms of the F-theory
picture with flavor symmetry $G$. We seek to understand how background
values for fields in the seven-brane show up in the 6d\ SCFT. What we will
show is that these background values induce vevs for operators in the
theory. Conversely, a choice of operator vev leads to a specific choice of
boundary condition for the flavor branes.

Consider, therefore, the worldvolume theory for our leftmost flavor brane.
This is a seven-brane with gauge symmetry $G$ wrapping the curve $\Sigma
\simeq\mathbb{C}^{\ast}$, that is, a cylinder. To study possible boundary
conditions for this system, it is convenient to view this curve as a compact
$\mathbb{P}^{1}$ with two marked points, which we label as $p_{0}$ and
$p_{\infty}$. The intersection of the flavor brane with another seven-brane
occurs at $p_{0}$, while $p_{\infty}$ is far away from this intersection point.

Allowing for singular behavior for our fields at the marked points,
the BPS equations for the seven-brane are governed by the Hitchin system
coupled to defects:
\begin{equation}
\overline{\partial}_{A} \Phi=\underset{p}{\sum}\mu_{\mathbb{C}}^{(p)}%
\delta_{(p)}\ ,\qquad F+[\Phi,\Phi^{\dagger}]=\underset{p}{\sum}%
\mu_{\mathbb{R}}^{(p)}\delta_{(p)}\label{eq:hitchin}%
\end{equation}
for an adjoint-valued $(1,0)$ form $\Phi$ and the worldvolume connection $A$,
with $F$ its field-strength. Here, we allow for possibly singular behavior at
a marked point $p$. For each point $p$, $\mu_{\mathbb{C}}^{(p)}$ and $\mu_{\mathbb{R}}^{(p)}$ specify
a triplet of moment maps. When matter is localized at an intersection point,
the associated sources can be interpreted as vevs for matter fields
\cite{Anderson:2013rka, BHVI}.

Letting $z$ denote a local coordinate on $\Sigma$ such
that $z = 0$ is a marked point, the local behavior of a solution to this system
of equations can be determined by solving the holomorphic constraint (i.e. the F-term)
modulo complexified gauge transformations. Doing so, we get:
\begin{equation}
\Phi \sim \mu_{\mathbb{C}} \frac{dz}{z}
\end{equation}
where we have presented the solution in a holomorphic gauge so that $A_{(0,1)}$
is trivial (see e.g. \cite{TBRANES, GukovWittenDefect}). To
pass to a unitary gauge in which F- and D-terms modulo unitary gauge
transformations are satisfied, we need to conjugate by an appropriate
position dependent element $h(z , \overline{z})$ of
the complexified group, $\Phi \rightarrow h^{-1} \cdot \Phi \cdot h$ and
$A_{(0,1)} \rightarrow h^{-1} \cdot \overline{\partial} h$.
This amounts to replacing $\mu_{\mathbb{C}}$ by a position
dependent profile $\mu_{\mathbb{C}}(z,\overline{z})$. This position dependence is
related to the fact that the flux profile of the seven-brane yields a funnel solution which opens up near the
point $p_{\infty}$. When we turn to the IIA realization of this system, we will encounter this behavior again
as an ``ordering constraint'' on the partitioning of semi-infinite D6-branes (see also \cite{Gaiotto:2008sa}).
Solutions to Hitchin system with a simple pole were first considered in \cite{Simpson}.
For further discussion, we refer the interested reader to section 3.3 of reference \cite{GukovWittenDefect}. Higher order poles
(i.e. irregular singularities) can also be studied in the same fashion, and correspond to activating vevs for multiple operators \cite{BHVI}.

To characterize these solutions more globally, it is convenient to introduce
the complexified connection:%
\begin{equation}
\mathcal{A}=A+\Phi+\Phi^{\dag}.
\end{equation}
Solutions to Hitchin's system are given by flat complexified connections \cite{HitchinSelf}.
In the case at hand where $\Sigma$ is a cylinder, we see that there
is precisely one closed one-cycle to integrate around, so there is
one holonomy we get to specify:
\begin{equation}
\mathcal{H}= P \exp\left(  -
{\displaystyle\oint}
\mathcal{A}\right)  ,
\end{equation}
valued in $G_{%
\mathbb{C}
}$. The conjugacy class for the holonomy $\mathfrak{C}_{\mathcal{H}}$ in $G_{\mathbb{C}}$ is gauge invariant data,
and fixes a choice of vacuum. Observe that encircling the point $p_{\infty}$ and specifying the
holonomy there amounts to also specifying the holonomy around $p_{0}$ since $\mathcal{H}_{\infty} \cdot \mathcal{H}_{0} = \mathbf{id}$.

Now, in the 6d SCFT, the boundary data associated with $\mu_{\mathbb{C }}$ and $\mu_{\mathbb{R}}$ is
interpreted as giving vevs to operators of the theory. Said differently, we see that the background values of the flavor
seven-branes generate a non-zero Higgsing in the resulting theory. Conversely,
if we ask what the lift of operator vevs in the 6d\ SCFT translates to in the
brane construction, we need to alter the boundary conditions for the Hitchin system fields
out at $p_{\infty}$. Finally, we note that although this fixes the breaking pattern for the flavor symmetry,
it is important to note that in unitary gauge, the relative size of
the Hitchin system fields at $p_{\infty}$ and $p_{0}$ will in general be different, being controlled
by the Hermitian pairing for the Higgs bundle.

Thus, our characterization of vacua boils down to possible choices for boundary conditions at the marked points, which are in
turn controlled by $\mu_{\mathbb{C}}$ and $\mu_{\mathbb{R}}$. Now, recall that we are also restricting attention
to deformations of the 6d SCFT which cannot be understood as unfolding the singularity associated with the
seven-brane. In the decomposition of $\mu_{\mathbb{C}} = \mu_s + \mu_n$ into a semi-simple (i.e. in the Cartan) and nilpotent piece,
it is well-known that the semi-simple part shows up as just such an unfolding (see e.g. \cite{BHVI}). For our present purposes,
it is therefore enough to focus attention on the case $\mu_s = 0$, so that $\mu_{\mathbb{C}}$ nilpotent. In
other words, $\mu_\mathbb{C}$ acts as a raising operator, and $\mu_{\mathbb{R}} \propto [\mu_{\mathbb{C}} , \mu^{\dag}_{\mathbb{C}}]$ acts
as a Cartan generator for an $\mathfrak{su}(2)$ subalgebra of $\mathfrak{g}$, specifying
a homomorphism (by abuse of notation) $\mu: \mathfrak{su}(2) \rightarrow \mathfrak{g}$.

Summarizing, the basic data associated with a
partial Higgs branch of our 6d SCFT is a choice of nilpotent element $\mu$, or even more precisely, its
orbit $O_{\mu} \subset \mathfrak{g}_{\mathbb{C}}$. Conversely, if we specify in $G_{\mathbb{C}}$ a conjugacy
class $\mathfrak{C}_{\mathcal{H}}$ for the holonomy, then we have implicitly also fixed a
choice of Higgs branch.

Now, since we can independently choose the boundary data for our two flavor branes
$G_{L}$ and $G_{R}$, we see that the Higgs branches of our theories are
labeled by a pair of homomorphisms $(\mu_{L},\mu_{R})$. We shall refer to these theories
as $\mathcal{T}(G,\mu_{L},\mu_{R},N).$ The residual flavor symmetry
for $\mathcal{T}(G , \mu_L, \mu_R, N)$ is the commutant of the image
$\mu_{L}(SU(2)) \times \mu_{R}(SU(2)) \subset G_L \times G_R $.\footnote{ This residual symmetry can
be extracted from known results in the literature (see e.g. \cite{carterfinite}), and was actually already considered in the
physics literature in the related context of class S theories in four
dimensions; see for example \cite{Chacaltana:2012zy,Chacaltana:2014jba}.}

Our characterization of vacua by nilpotent elements of the complexified flavor
symmetry is basically a special case of the broader notion of
\textquotedblleft T-brane data\textquotedblright\ in an F-theory
compactification \cite{TBRANES, Chiou:2011js, glueI, glueII, Anderson:2013rka}%
. These are non-abelian intersecting brane configurations in which the
adjoint-valued Higgs field of the seven-brane is nilpotent (i.e. upper
triangular for $\mathfrak{sl}(n,\mathbb{C})$). As they are nilpotent, such
vacua do not appear in holomorphic Casimir invariants, and so are not visible
in the complex geometry of the Calabi-Yau threefold. Rather, they can be seen
in the limiting behavior of its intermediate Jacobian \cite{Anderson:2013rka}.
What we have just seen is that T-brane data
leads to a rich class of 6d\ SCFTs.\footnote{ Such T-brane configurations also
provide a way to construct four-dimensional superconformal field
theories. These are realized by D3-branes probing a T-brane background. For
additional discussion of these worldvolume theories, see \cite{Funparticles,
FCFT, HVW, D3gen, Heckman:2012jm}.}

Having illustrated how boundary data of the Hitchin system realization in F-theory
feeds into the 6d SCFT, let us now return to the
M-theory realization in terms of 6d\ domain walls in 7d\ super
Yang-Mills theory with gauge group $G$. Here, our flavor symmetry is supported
on a pair of semi-infinite intervals, so it is convenient to label this coordinate as
$x_{6}$. The two marked points $p_{\infty}$
and $p_{0}$ are now replaced by a choice of boundary conditions for this
7d\ Yang-Mills sector: One which is far away from all of the compact intervals
at $x^{6}=x_{\infty}^{6}$, and one which touches the various compact
intervals. In the worldvolume of the 7d theory, there is a pole for the Nahm
equations:%
\begin{equation}
\partial_{A}\Phi^{i}=\epsilon^{ijk}[\Phi^{j},\Phi^{k}]\ ,\label{NAHMA}%
\end{equation}
where now $\Phi^{i}$ for $i=x,y,z$ are the triplet of real scalars in 7d\ SYM,
and $\partial_{A}=\frac{\partial}{\partial x^{6}}-A_{6}$ is the worldvolume
covariant derivative in the direction $x^{6}$ along the semi-infinite
interval. Near $x^{6}=x_{\infty}^{6}$, the fields have asymptotic behavior:
\begin{equation}
\Phi^{i} \sim \frac{t^{i}}{x^{6}-x_{\infty}^{6}}\ ,\label{eq:Nahmpole}%
\end{equation}
where the $t^{i}$ are Hermitian generators of an $\mathfrak{su}(2)$ subalgebra
of the gauge algebra $\mathfrak{g}$. Of course, this is the same data we already encountered in the F-theory description.
Further, we see that the pole for the Higgs field in the Hitchin system out at
$p_{\infty}$ is now reflected in a pole for the $\Phi^{i}$ at $x^{6} = x_{\infty}^{6}$.
With this boundary condition fixed, we also see that the profile of
the field configuration in 7d\ SYM will now interpolate from
$x^{6}=x_{\infty}^{6}$ inwards to the first M5-brane, inducing a vev for
operators in the 6d\ SCFT.

\subsection{IIA Realizations \label{sub:IIASUk}}

\begin{figure}[th]
\centering
\includegraphics[scale=.5]{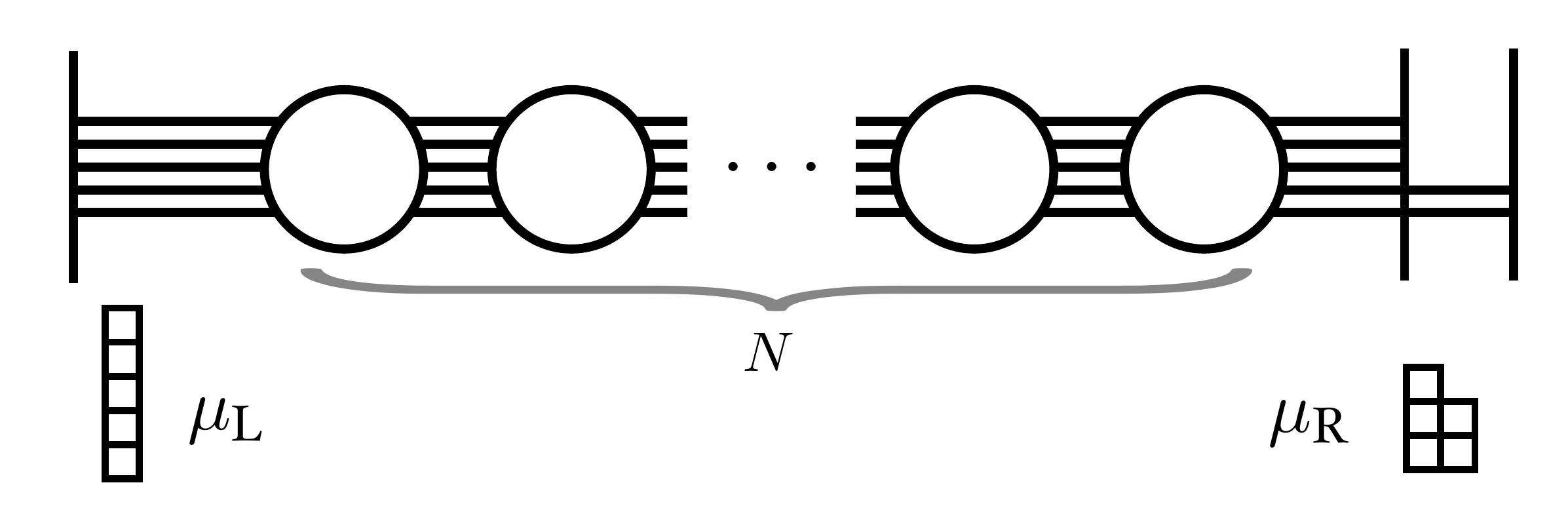}  \caption{A brane configuration similar to
figure \ref{noD8HZ}, where now the D6s end on D8s in two different ways on the
two sides, corresponding to two different Young diagrams. This configuration
is T-dual to the one in figure \ref{fig:D7Hpoles}.}
\label{fig:D6D8}
\end{figure}

In the special case where we have a IIA realization, we can be even more
explicit. For specificity, we focus on the case $G=SU(k)$ studied in
\cite{Gaiotto:2014lca}. In that context, the relevant brane systems are the
ones considered in \cite{BlumINT,Hanany:1997gh, Gaiotto:2014lca}. The relevant
brane configurations are similar to the ones of our ``master theory'', except
that now we take the D6 to end on several D8s; let $\mu_{\mathrm{L}}^{a}$
($\mu_{\mathrm{R}}^{a}$) be the number of D6s ending on the $a$-th D8 on the
left (right). When each D6 ends on a separate D8, so that all the
$\mu_{\mathrm{L,R}}^{a}=1$, the D8s impose Dirichlet boundary conditions for
the gauge field on the D6s, and Neumann for the scalars.

An example of a more general situation, with not all the $\mu_{\mathrm{L,R}%
}^{a}=1$, is depicted for example in figure \ref{fig:D6D8}. Actually, however,
such a picture is too naive: one now expects that the D6s and the D8s fuse
into a single D8/D6 bound state. This is described on the worldvolume of the
D6s by a pole for the Nahm equations, just as in equation \eqref{NAHMA}.

The presence of the Nahm pole can be interpreted as the fact that the D6s open
up into the D8s. To see this, consider first for simplicity the
\textquotedblleft full\textquotedblright\ pole where all $k$ D6s end on a
single D8, as on the left of figure \ref{fig:D6D8}. Notice that for
$x^{6}\rightarrow x_{\infty}^{6}$ we have $\Phi_{i}\Phi_{i}\rightarrow
\frac{k(k-1)}{(x^{6}-x_{\infty}^{6})^{2}}\mathbf{id}$; a slice at constant
$x^{6}$ is well approximated by a fuzzy sphere of radius $\sim\frac{k}%
{x^{6}-x_{\infty}^{6}}$. In the more general case where not all the D6s end on
the same D8 (as on the right side of figure \ref{fig:D6D8}), each D8
represents a Jordan block, whose size is the number $\mu^{i}$ of D6s ending
on it; we then have several fuzzy spheres, of radii $\sim\frac{\mu_{i}}%
{x^{6}-x_{\infty}^{6}}$. An exception is the case of a Jordan block of size 1,
which is of course simply a zero; in that case, the fuzzy sphere actually has
radius 0, and the D6 actually ends on a D8: the two do not fuse together.
This block then behaves as in the case we mentioned at the beginning of this
subsection: the D8s in this block impose Dirichlet boundary conditions for the
gauge field on the D6s, and Neumann for the scalars.

Thus in general we should imagine fuzzy funnels coming out of the NS5 system;
for a cartoon (in the case where the NS5s coincide) see figure 5 of reference
\cite{Gaiotto:2014lca}. Implicit in that cartoon is also the fact that the D8s in
pictures such as figure \ref{fig:D6D8} should be ordered so that the $\mu
_{\mathrm{L}}^{a}$ and $\mu_{\mathrm{R}}^{a}$ should decrease as one goes
outside (this was found in \cite{Gaiotto:2008sa} by comparing the moduli
spaces of solutions to Nahm equations to the moduli spaces of the
corresponding field theories).

A second constraint is that there should be enough NS5s \cite{Gaiotto:2014lca}:
when one moves the branes around to reach a quiver description, one should
not remain with some extra decoupled sectors. This reads
\begin{equation}
\label{eq:Nbound}N \ge\mu^{1}_{\mathrm{L}} + \mu^{1}_{\mathrm{R}}\ ,
\end{equation}
where $\mu^{1}_{\mathrm{L}}$ and $\mu^{1}_{\mathrm{R}}$ are the number of D6s
ending on the two innermost D8s or, in other words, the tallest columns of
the two Young diagrams (this would amount to $N \ge5+3=8$ in the example of
figure \ref{fig:D6D8}).

In order to see the connection of the IIA discussion in this subsection to the
previous discussion in F-theory it is enough to T-dualize along a direction
transverse to the D6s, say $x^{7}$, generalizing our discussion in section
\ref{ssub:IIB-AN}. The NS5s turn into geometry: each pair of them give rise to
a nontrivial two-cycle. The D6s suspended in between the NS5s now become a D7
stack (or in other words D7s with nonabelian gauge group $SU(k)$) wrapping the
two-cycle. This is the tensor branch of the model; the SCFT point
is reached by putting the NS5s on top of each other, and in the T-dual
picture this amounts to collapsing the corresponding nontrivial two-cycles to
zero size. At each of the ends of the diagram, however, things are a little
different. Let us first look at the case where each of the D6s ends on a
separate D8. In this case, we get an $SU(k)$ flavor symmetry, because no Nahm pole is possible. Further,
if we T-dualize, we get a single stack of D7-branes wrapping a non-compact two-cycle.

\begin{figure}[th]
\centering
\includegraphics[scale=.5]{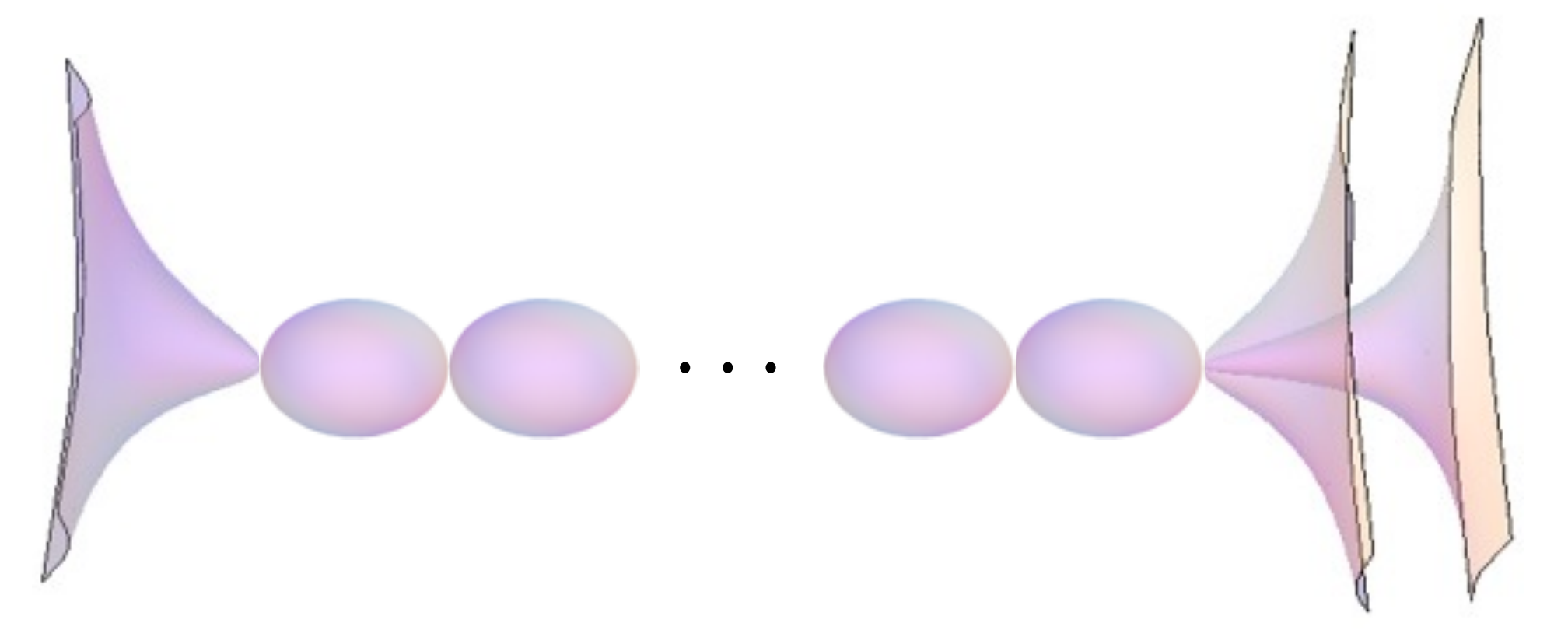}  \caption{The T-dual in type IIB / F-theory of figure
\ref{fig:D6D8}. The original description in terms of partitions has been smoothed out to a flux profile over the
Hitchin system curve, resulting in a funnel-like solution in the geometry.}%
\label{fig:D7Hpoles}%
\end{figure}

In more general cases, however, we have seen in IIA that the D6s and the D8s
fuse together into a single object, as reflected in non-trivial Nahm pole data.
We expect this to have its own counterpart
on the IIB side. This can be described on the worldvolume of the former D6 as
follows: one of the transverse scalars (say $\Phi^{3}$) becomes in IIB a gauge
field, and we can complexify the two remaining scalars: $\Phi\equiv\Phi
_{1}+i\Phi_{2}$. The Nahm equations (\ref{NAHMA}) now turn into the (gauge fixed) Hitchin
equations (\ref{eq:hitchin}); the presence of a Nahm pole (\ref{eq:Nahmpole})
turns into the presence of a Hitchin pole. The D7 with a Hitchin pole can now
be thought of as the fusion of a system of D7s. The structure of the D8/D6 system with a Nahm
pole is therefore encoded in the structure of the block decomposition of the
Hitchin field, that, in turn, encode the T-brane data on the F-theory side.
The situation is now depicted schematically in figure \ref{fig:D7Hpoles}.

In IIA\ string theory, we can also realize theories with $SO$ and $Sp$ gauge
symmetry. Compared with the A-type case, the Type IIA description of the
system is almost analogous, the main difference being the presence of the
O$6_{-}$ plane. Again introducing D8s on the left and on the right of the NS5s
that impose Dirichlet boundary conditions for the gauge field on the D6s, and
Neumann for the scalars we recover the system with $SO_{L}\times SO_{R}$
flavor symmetry and the T-dual configuration of $I_{p}^{\ast}$ fibers we
discussed in the previous section. Consider now introducing Nahm poles for the
D6/D8 systems on the left and on the right. These are specified by a pair of
homomorphisms $\mu_L$ and $\mu_R$ from $\mathfrak{su}(2)$ to $\mathfrak{so}_{2p+8}$.
By T-duality these very same homomorphisms characterize the T-brane data on the F-theory side.

\subsection{Alternative realizations for some $SU(k)$ cases
\label{sub:altIIASUk}}

Notice that so far we have kept all the NS5s in the region where $F_{0}=0$. In
figure \ref{fig:D6D8}: there are no D6s ending on any NS5. In general, the net
number of D6s ending on a NS5 equals $k_{\mathrm{D6,\,L}}-k_{\mathrm{D6,\,L}%
}=n_{0}=2\pi F_{0}$ (which is an integer by flux quantization). This can be
seen either by anomaly cancellation in the 6d field theory, or by a Bianchi
identity in the brane picture. Because of this, given any valid D8-D6-NS5
configuration, we can decide to move all the NS5s in the region $F_{0}=0$ as
we have done so far. It is also possible to move them all to a different
region, say where $F_{0}=1$. This is not especially natural in general, but it
is a useful alternative for some particular theories, namely for those that
saturate the bound (\ref{eq:Nbound}).

\begin{figure}[t!]
\centering
\includegraphics[scale=.5]{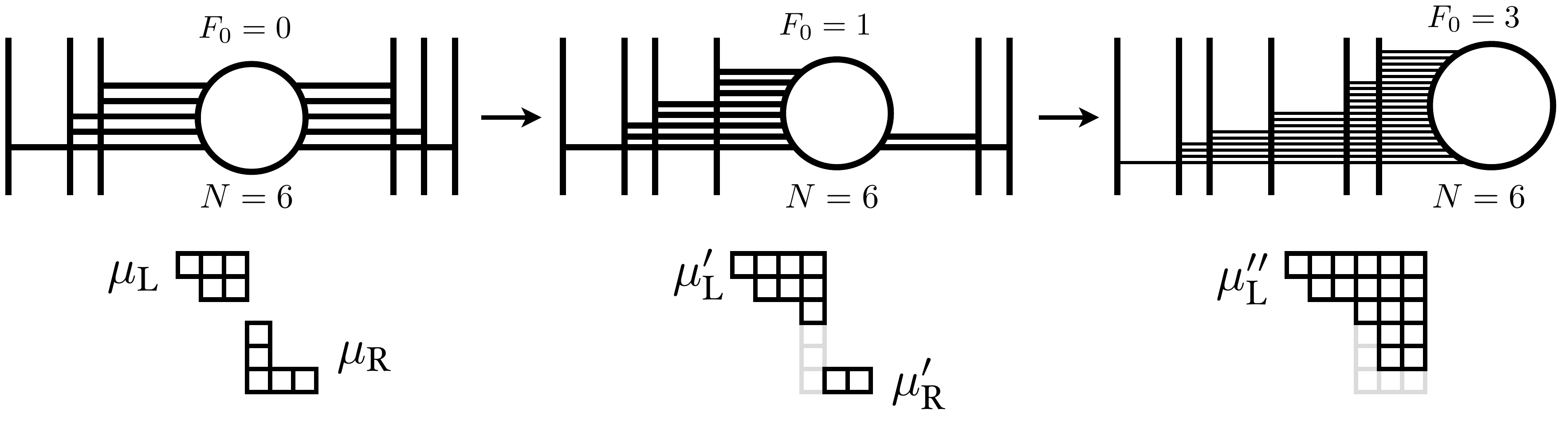}  \caption{When one moves the NS5s to a region
where $F_{0}\neq0$, an intuitive algorithm relates the new Young diagrams
$\mu^{\prime}_{\mathrm{L}}$ and $\mu^{\prime}_{\mathrm{R}}$ to the old ones
$\mu_{\mathrm{L}}$ and $\mu_{\mathrm{R}}$. Continuing the process, one can reach a situation
where all the NS5s are on one side.}%
\label{fig:F01}%
\end{figure}

This will actually become clearer when we discuss gravity duals in section
\ref{sec:HOLO}. Meanwhile, we notice here that there is a nice visual device
(applicable whether the bound (\ref{eq:Nbound}) is saturated or not) to see
how the T-brane data change when one moves the NS5s to a region where $F_{0}
\neq0$; it is depicted in figure \ref{fig:F01}. Basically, we add a column
$(N-\mu^{1}_{\mathrm{R}})$-box tall to the $\mu_{\mathrm{L}}$ Young diagram,
and erase the first column to $\mu_{\mathrm{R}}$, obtaining two new Young
diagrams $\mu^{\prime}_{\mathrm{L}}$ and $\mu^{\prime}_{\mathrm{R}}$. Notice
that actually $\mu^{\prime}_{\mathrm{L}}$ is still a Young diagram only thanks
to the bound (\ref{eq:Nbound}). Moreover, a new bound $N \ge\mu^{\prime
}_{\mathrm{L}}{}^{1} + \mu^{\prime}_{\mathrm{R}}{}^{1}$ is now also satisfied.
So in a sense the bound (\ref{eq:Nbound}) is true more generally than in the
way we originally formulated it.

In the previous subsection our field theories $\mathcal{T}(SU(k),\mu_{\mathrm{L}},\mu_{\mathrm{R}},N)$
were presented as produced by Higgsing
from the theory $\mathcal{T}(SU(k),N)$ describing $N$ NS5s on top of $k$ D6s.
Now we see that we can actually also think of them as arising by Higgsing from
a more general theory $\mathcal{T}( SU(k_L) , SU(k_R) , N )$
describing $N$ NS5s in a region where $F_{0}\neq0$, with $k_{\mathrm{L}}$
semi-infinite D6s on their left and $k_{\mathrm{R}}$ semi-infinite D6s on
their right (where $k_{\mathrm{R}}-k_{\mathrm{L}}=n_{0}N$). As we mentioned,
this might actually be more natural for the theories which saturate
(\ref{eq:Nbound}), as we will see in section \ref{sec:HOLO}. It would be
interesting to develop further such alternative characterizations for the
$SO/Sp$ type examples which also have IIA\ realizations.

\section{SCFTs from the Ho\v{r}ava-Witten Wall \label{sec:SMALL}}

So far we have discussed  $(1,0)$ 6d SCFTs which are realized by M5-branes probing an ADE
singularity.  Another well known example of $(1,0)$ theories involve M5-branes approaching the Ho\v{r}ava-Witten
wall, namely the theory of small $E_8$ instantons (a.k.a E-string theories) in
heterotic string theory \cite{WittenSmall, Ganor:1996mu, Seiberg:1996vs}.

The F-theory realization of E-strings has been studied
in \cite{MorrisonVafaII, Witten:1996qb}. In F-theory, the single E-string
theory is given by working on the base $\mathcal{O}(-1) \rightarrow \mathbb{P}^1$,
that is, the base is the local geometry of a single $\mathbb{P}^1$
with self-intersection $-1$. This arises from blowing up the
intersection point $u=v=0$ in the base of the geometry:
\begin{equation}
y^{2}=x^{3}+uv^{5},
\end{equation}
where the $E_{8}$ flavor symmetry is localized at $v=0$. Multiple M5-branes
probing the $E_{8}$ wall leads to a similar class of theories. In heterotic
string theory, this is the theory of $N$ coincident small $E_8$ instantons. In
F-theory, this is given by a configuration of curves:%
\begin{equation}
\lbrack E_{8}]\underset{N}{\underbrace{%
\begin{array}
[c]{cccc}%
1 & 2 & ... & 2
\end{array}
}}\label{smallinst}%
\end{equation}
where the $-1~$curve again enjoys an $E_{8}$ flavor symmetry.

The Higgs branch of these theories corresponds (in the heterotic
description)\ to dissolving some of the instantons of this background back
into flux. In the F-theory description this is captured by a T-brane
configuration for the seven-brane with $E_{8}$ symmetry. The moduli space for
this system is that of $N$ instantons for $E_{8}$ gauge theory on the
four-manifold wrapped by the nine-brane.


We can also combine the ingredients of these two classes of $(1,0)$ theories,
by considering an ADE singularity intersecting the $E_8$ wall and bringing in M5-branes to probe
this intersection.  The aim of this section is to study the resulting $(1,0)$ theory and in particular
elucidate its tensor branch and partial Higgs branches.

\subsection{Orbifolds}

Just as we considered the case of M5-branes probing a singularity, we can also
consider the case where the $E_{8}$ nine-brane intersects an ADE\ singularity
$\mathbb{C}^{2}/\Gamma_{G}$. Such configurations still preserve $(1,0)$
supersymmetry in six dimensions, and are therefore excellent candidates for
realizing additional SCFTs. The F-theory realization of these configurations
has been studied in \cite{Aspinwall:1997ye}, and is given by decorating the
configuration of line (\ref{smallinst}) by a non-minimal Kodaira-Tate fiber:%
\begin{equation}
\lbrack E_{8}]\underset{N}{\underbrace{%
\begin{array}
[c]{cccc}%
\mathfrak{g} & \mathfrak{g} & ... & \mathfrak{g}\\
1 & 2 & ... & 2
\end{array}
}}[G]\label{smallinstorb}%
\end{equation}
that is, it consists of $N$ compact curves (corresponding to $N$ M5-branes) and one curve which has been
decompactified, as denoted by $[G]$. The matter content of these models has been worked out in \cite{intriliaddme} by requiring anomaly cancellation. We shall refer to this class of theories as $\mathcal{T}(E_8 , G_{R} , N)$,
in the obvious notation.

One way to see that this is indeed the correct characterization is to use the
standard rules for heterotic / F-theory duality. Starting from heterotic on an
ADE\ singularity, we can instead consider a non-compact elliptic $K3$ surface
$T^{2}\rightarrow K3_{het}\rightarrow\mathbb{C}$ with prescribed
Kodaira-Tate fiber over a \ marked point of the base. This elliptic $K3$ is then
the gluing region for the stable degeneration limit of a Calabi-Yau threefold
in the dual F-theory description:%
\begin{equation}
X=X_{L}\cup_{K3_{het}}X_{R},
\end{equation}
so in other words, there is a collision between the $\mathfrak{g}$-type
Kodaira-Tate fiber from the right, and the $II^{\ast}$ fiber type supported
over $[E_{8}]$.

As presented, (\ref{smallinstorb}) is not a completely resolved phase of
the F-theory geometry. This follows because at the collision of the $-1$ curve
with the $[E_{8}]$ component, the singularity type passes beyond the allowed
order of vanishing for a Kodaira-Tate fiber. To pass to a resolved geometry,
we would need to perform further blowups at this point. Alternatively, we can
move from the conformal fixed point onto the Higgs branch, passing to a lower theory.
Let us now turn to a characterization of each such branch.

\subsubsection{Tensor Branch}
\label{subsub:tbranch}

To get started with thinking about the tensor branch, it turns out to be useful
to use the M-theory frame.  We consider the geometry
 $\mathbb{R}/\mathbb{Z}_{2}\times\mathbb{C}^{2}%
/\Gamma_{ADE}$.  Note that the locus of the ADE singularity
transverse to the 6d spacetime is a half-line  $\mathbb{R}/\mathbb{Z}_{2}$, where
the boundary of the half-line is where the singularity intersects the $E_8$ wall.
We can in addition introduce $N$ M5-branes on the singularity half-line.  The SCFT is
obtained by bringing the M5-branes to the $E_8$ wall, i.e., the boundary
of the half-line.  To go to the tensor branch we need to first separate the M5-branes
along the half-line.  We end up with $N$ segments, each carrying the corresponding
ADE gauge symmetry.  Between every adjacent interval we get the conformal
matter of the ADE type we have already discussed.  The only new ingredient
is the extra degree of freedom corresponding to the matter localized where the
ADE singularity meets the Ho\v{r}ava-Witten wall.  This matter
will have $E_8\times G$ global symmetry. This is the main new ingredient
we need to understand in the context of this new class of SCFTs.  To figure out
what this new conformal matter is we will use the F-theory setup.

In the F-theory setup, the theory on the tensor branch is given by performing all possible blowups so
that the elliptic fibers are all in Kodaira-Tate form. This has been worked
out in \cite{Aspinwall:1997ye}, though the particular points we emphasize here
are somewhat different. The analysis  To begin, suppose that we have a $\mathfrak{g}%
=\mathfrak{su}_{k}$ type gauge symmetry on each curve, corresponding to an
$I_{k}$ type singularity. Performing a further blowup on the intersection of
the $-1$ curve with the $[E_{8}]$ locus, we get a new $-1$ curve, but which
now supports an $I_{k-1}$ singularity. Proceeding in this way, we get $k$
additional blowups until we finally reach the configuration:%
\begin{equation}\label{eq:rampplateau}
\lbrack E_{8}]\underset{k}{\underbrace{%
\begin{array}
[c]{ccccc}
& \mathfrak{su}_{1} & \mathfrak{su}_{2} &  & \mathfrak{su}_{k-1}\\
1 & 2 & 2 & ... & 2
\end{array}
}}\underset{N}{\underbrace{%
\begin{array}
[c]{cccc}%
\mathfrak{su}_{k} \oplus \mathbf{k} & \mathfrak{su}_{k}  &  & \mathfrak{su}_{k}\\
2 & 2 & ... & 2
\end{array}
}}[SU(k)].
\end{equation}
with the usual bifundamental matter between the adjacent quivers nodes. Additionally,
there is one extra hypermultiplet attached to the leftmost $\mathfrak{su}_{k}$
factor, i.e. at the ``plateau'' of $\mathfrak{su}_k$ factors.
In F-theory language, this comes from the collision of the zero section
with the leftmost $\mathfrak{su}_{k}$ factor, so that on each such $\mathfrak{su}_{k}$ there are a total of
$2k$ hypermultiplets, as required for 6d gauge anomaly cancelation \cite{intriliaddme}.

Focussing on the region with gauge groups of increasing rank, the new conformal
matter system with $E_8\times SU(k)$ flavor symmetry is the quiver
system given by the ramp:
$$
\begin{gathered}\includegraphics[scale=.8]{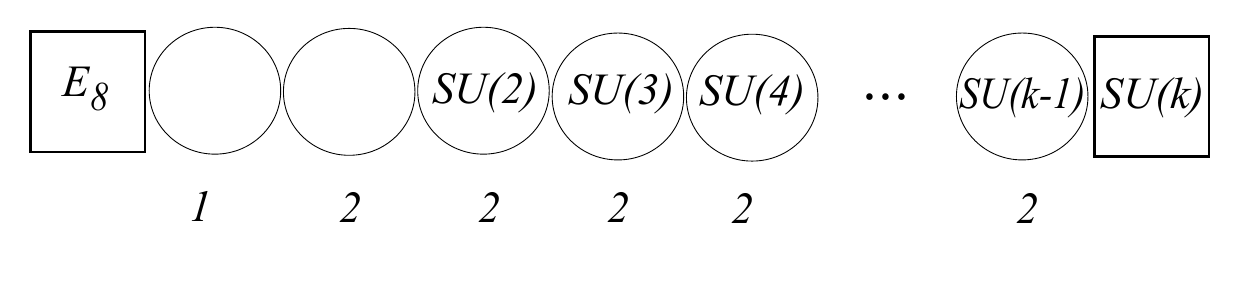}\end{gathered}
$$
This ramp was also reproduced in IIA language in \cite{Hanany:1997gh}. There,
the end of the ramp corresponds to the presence of an additional D8-brane just
before the plateau region to the right. We can view this quiver as arising
from the splitting of the intersection point between $SU(k)$ singularity and the wall to $k+1$ points.

\begin{figure}
\begin{center}
\includegraphics[scale=.55]{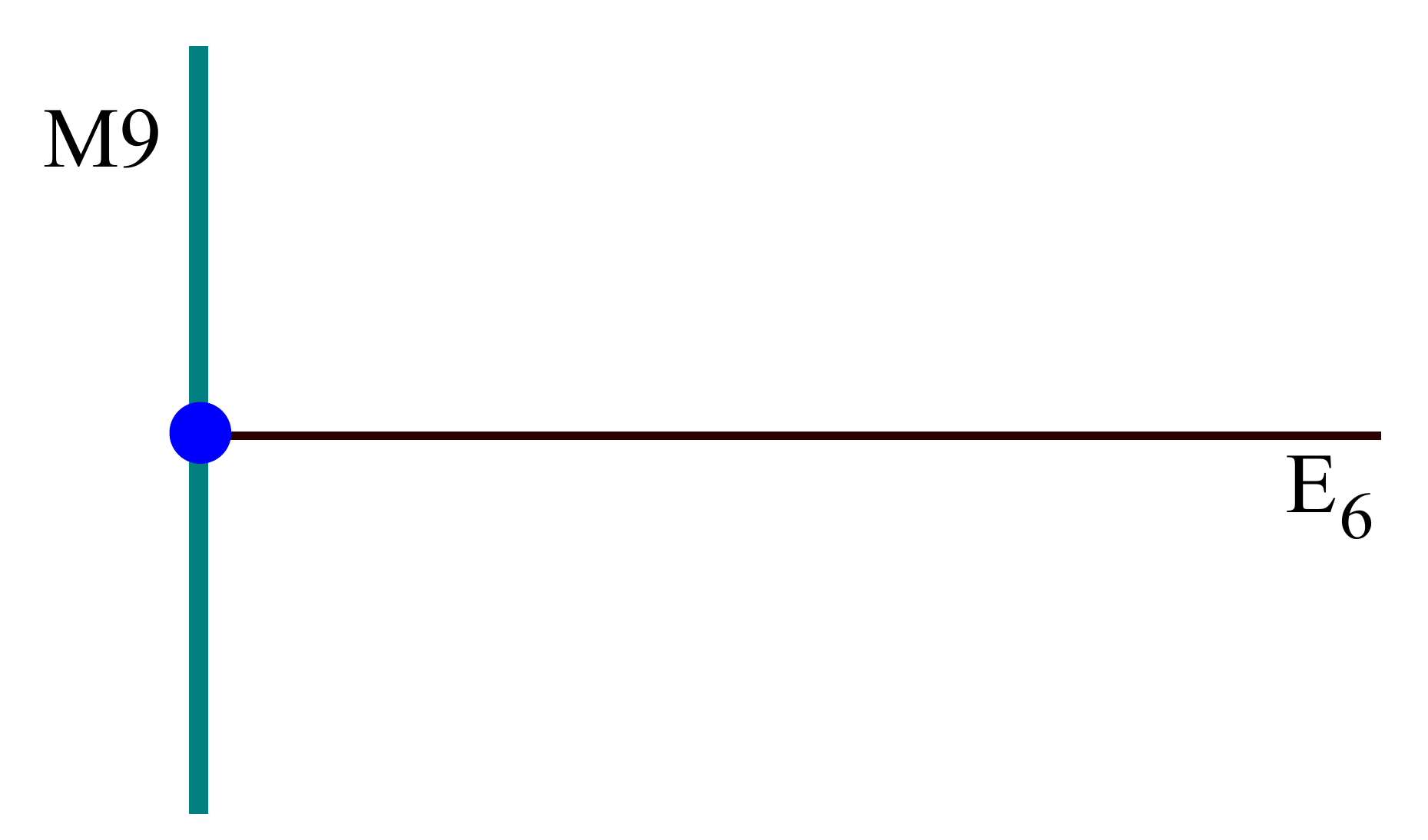}

\medskip

\includegraphics[scale=.55]{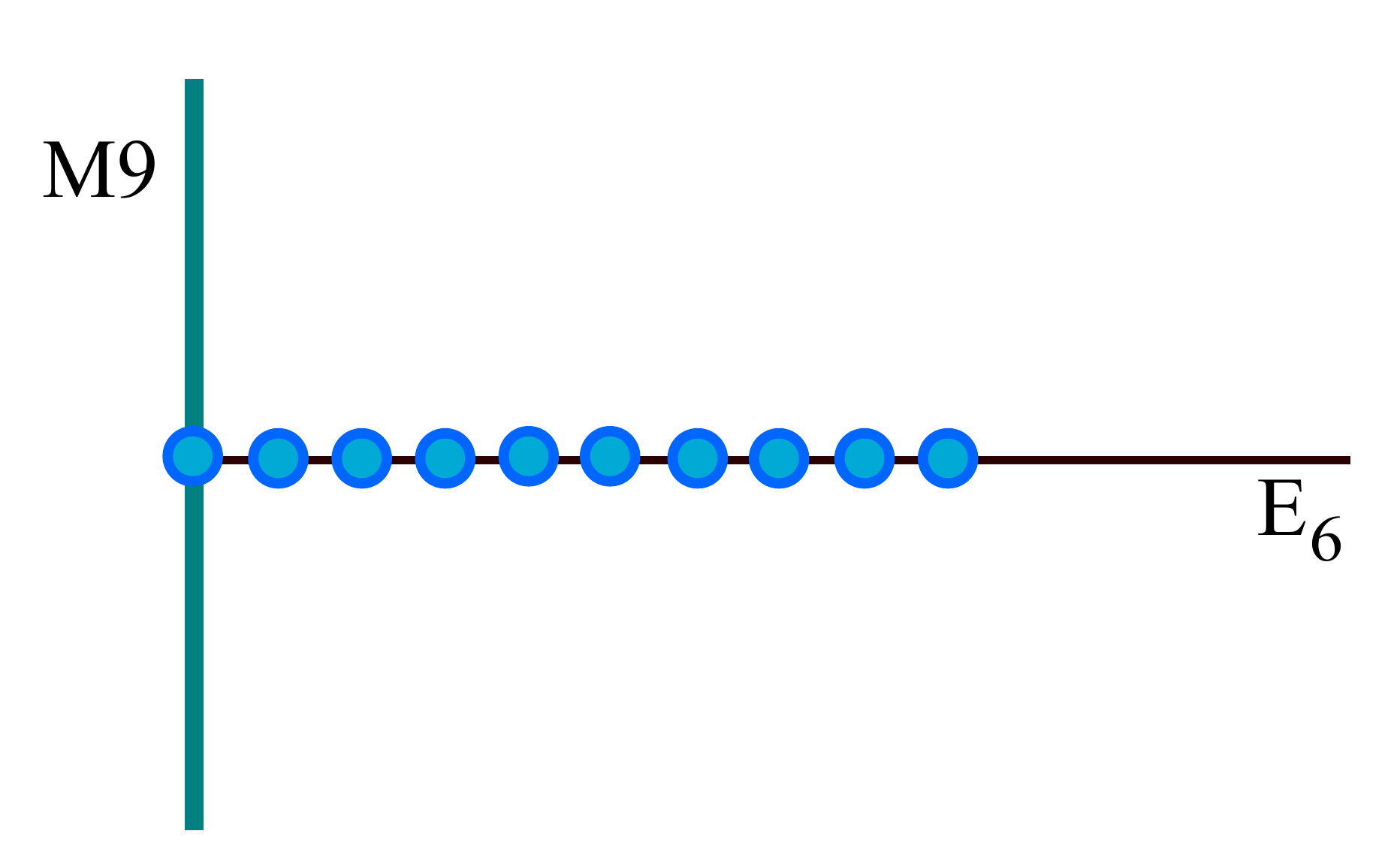}

\medskip

\includegraphics[scale=1.2]{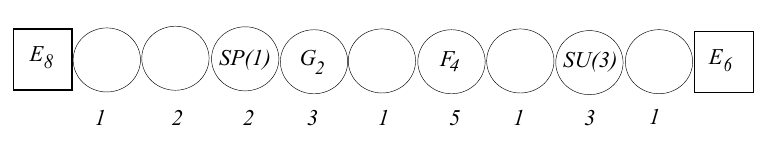}

\caption{\textsc{up}: M9-brane at the orbifold  $E_6$ singularity. \textsc{down}: SCFT matter trapped at the intersection in between the M9 and the singularity that gives the $(E_8,E_6)$ ramp.  The boundary point has fractionated to 10 points.}\label{fig:CoulombM9}
\end{center}
\end{figure}

In the case of a $D_{p+4}$-type orbifold, the initial configuration of curves
and singular fibers is:%
\begin{equation}
\lbrack E_{8}]\underset{N}{\underbrace{%
\begin{array}
[c]{cccc}%
\mathfrak{so}_{2p+8} & \mathfrak{so}_{2p+8} &  & \mathfrak{so}_{2p+8}\\
1 & 2 & ... & 2
\end{array}
}}[SO(2p+8)]
\end{equation}
As we already saw in the case of M5-branes probing a D-type singularity, a
blowup is required at each collision of $I_{p}^{\ast}$ fibers. Performing the
requisite sequence of blowups to reach the\ resolved phase, we have:%
{\footnotesize \begin{align*}
&  [E_{8}]\underset{N}{\underbrace{%
\begin{array}
[c]{cccc}%
\mathfrak{so}_{2p+8} & \mathfrak{so}_{2p+8} &  & \mathfrak{so}_{2p+8}\\
1 & 2 & ... & 2
\end{array}
}}[SO(2p+8)]\\
&  \rightarrow\lbrack E_{8}]\underset{p+4}{\underbrace{%
\begin{array}
[c]{ccccccc}
&  & \mathfrak{sp}_{1} & \mathfrak{g}_{2} & \mathfrak{so}_{9} &  &
\mathfrak{so}_{2p+7}\\
1 & 2 & 2 & 2 & 2 & ... & 2
\end{array}
}}\underset{N}{\underbrace{%
\begin{array}
[c]{cccc}%
\mathfrak{so}_{2p+8} & \mathfrak{so}_{2p+8} &  & \mathfrak{so}_{2p+8}\\
2 & 2 & ... & 2
\end{array}
}}[SO(2p+8)]\\
&  \rightarrow\lbrack E_{8}]\underset{2p+5}{\underbrace{%
\begin{array}
[c]{ccccccccc}
&  & \mathfrak{sp}_{1} & \mathfrak{g}_{2} & \mathfrak{sp}_{0} & \mathfrak{so}%
_{9} &  & \mathfrak{so}_{2p+7} & \mathfrak{sp}_{p} \oplus \frac{1}{2} \mathbf{2p} \\
1 & 2 & 2 & 3 & 1 & 4 & ... & 4 & 1
\end{array}
}}\underset{2N}{\underbrace{%
\begin{array}
[c]{cccccc}%
\mathfrak{so}_{2p+8} & \mathfrak{sp}_{p} & \mathfrak{so}_{2p+8} &  &
\mathfrak{so}_{2p+8} & \mathfrak{sp}_{p}\\
4 & 1 & 4 & ... & 4 & 1
\end{array}
}}[SO(2p+8)]\ ;
\end{align*}}
that is, there is again a ramp in the rank of the gauge groups reaching the plateau involving the D-quivers with $D\times D$
conformal matter we have already studied. In this case, there is one additional half hypermultiplet attached to the rightmost
$\mathfrak{sp}_{p}$ factor of the ramp. All the rest of the $\mathfrak{sp}$ factors to the left come from the collision of $I^{\ast , ns}_{n-1}$ and $I^{\ast , ns}_{n}$
fibers \cite{Aspinwall:1997ye}, and so lead to a non-split $I_{2n - 1}$ fiber. As explained in \cite{Bershadsky:1996nu}, this leads to no additional
matter multiplets (beyond the half hypers trapped at the $SO / Sp$ intersections), and the gauge symmetry is $\mathfrak{sp}_{n-1}$. Thus, the ramp
contains $p + 1$ $\mathfrak{sp}_{n}$ factors which start at the left with $\mathfrak{sp}_{0}$, and increase one rank at a time until the right of the
ramp, with the final $\mathfrak{sp}_{p}$ factor. Note that in the plateau region, all $\mathfrak{so}_{2p+8}$ factors have $4p$ half hypermultiplets
in the fundamental, as required by 6d gauge anomaly cancelation \cite{intriliaddme}. Again, these ramps can also be reproduced using IIA methods \cite{Hanany:1999sj}.
So the new conformal matter with $E_8\times SO(2p+8)$ symmetry is the ramping up quiver with $(2p+5)$ tensor multiplets.
In particular we can view this as fractionating of the boundary point to $2p+6$ points. In the special case, $p = 0$, the
ramp truncates to just the $1,2,2,3,1,$ factor, that is, there is no alternating $SO / Sp$ quiver.

Finally, consider the case of M5-branes probing an E-type singularity (see figure \ref{fig:CoulombM9}). In this
case, the minimal resolution for $N$ instantons probing the $E_{6}$ orbifold
is:%
{\footnotesize
\begin{equation}
\lbrack E_{8}]\underset{N}{\underbrace{%
\begin{array}
[c]{cccc}%
\mathfrak{e}_{6} & \mathfrak{e}_{6} &  & \mathfrak{e}_{6}\\
1 & 2 & ... & 2
\end{array}
}}[E_{6}]\rightarrow\lbrack E_{8}]\underset{9}{\underbrace{%
\begin{array}
[c]{ccccccccc}
&  & \mathfrak{sp}_{1} & \mathfrak{g}_{2} &  & \mathfrak{f}_{4} &  &
\mathfrak{su}_{3} & \\
1 & 2 & 2 & 3 & 1 & 5 & 1 & 3 & 1
\end{array}
}}\underset{4N}{\underbrace{%
\begin{array}
[c]{cccccc}%
\mathfrak{e}_{6} \oplus \mathbf{27} &  & \mathfrak{e}_{6} &  & \mathfrak{e}_{6} & \\
5 & 131 & 6 & ... & 6 & 131
\end{array}
}}[E_{6}]
\end{equation}}
So the new conformal matter system with $E_8\times E_6$ symmetry is the first 9 factors.  In particular
the intersection point between the $E_6$ singularity and the wall has fractionated to 10 points. Observe also that
in the plateau region, the leftmost $E_6$ factor also couples to one additional hypermultiplet in the $\mathbf{27}$ of
$E_6$. This is just an E-type generalization of the phenomenon already encountered for A- and D-type orbifolds
in joining the ``ramp'' to the ``plateau'' regions.

For the minimal resolution for $N$ instantons probing the $E_{7}$ orbifold, we have:
{\footnotesize
\begin{equation}
\lbrack E_{8}]\underset{N}{\underbrace{%
\begin{array}
[c]{cccc}%
\mathfrak{e}_{7} & \mathfrak{e}_{7} &  & \mathfrak{e}_{7}\\
1 & 2 & ... & 2
\end{array}
}}[E_{7}] \rightarrow \lbrack E_{8}]\underset{10}{\underbrace{%
\begin{array}
[c]{cccccccccc}
&  & \mathfrak{sp}_{1} & \mathfrak{g}_{2} &  & \mathfrak{f}_{4} &  &
\mathfrak{g}_{2} & \mathfrak{su}_{2} & \\
1 & 2 & 2 & 3 & 1 & 5 & 1 & 3 & 2 & 1
\end{array}
}}\underset{6N}{\underbrace{%
\begin{array}
[c]{ccccc}%
\mathfrak{e}_{7} \oplus \frac{1}{2} \mathbf{56} &  &  & \mathfrak{e}_{7} & \\
7 & 12321 & ... & 8 & 12321
\end{array}
}}[E_{7}]
\end{equation}}
So the new conformal matter system with $E_8\times E_7$ symmetry is the first 10 factors,
and the boundary point has fractionated to 11 points. Additionally, there is a half hypermultiplet
in the $\mathbf{56}$ of $E_7$, precisely on the leftmost node of the plateau, i.e. in
the region where we join the ramp to the plateau.

For the minimal resolution for $N$ instantons probing the $E_{8}$ orbifold, we have:
{\footnotesize
\begin{equation}
\lbrack E_{8}]\underset{N}{\underbrace{%
\begin{array}
[c]{cccc}%
\mathfrak{e}_{8} & \mathfrak{e}_{8} &  & \mathfrak{e}_{8}\\
1 & 2 & ... & 2
\end{array}
}}[E_{8}] \rightarrow \lbrack E_{8}]\underset{11}{\underbrace{%
\begin{array}
[c]{ccccccccccc}
&  & \mathfrak{sp}_{1} & \mathfrak{g}_{2} &  & \mathfrak{f}_{4} &  &
\mathfrak{g}_{2} & \mathfrak{sp}_{1} &  & \\
1 & 2 & 2 & 3 & 1 & 5 & 1 & 3 & 2 & 2 & 1
\end{array}
}}\underset{12N}{\underbrace{%
\begin{array}
[c]{ccccc}%
\mathfrak{e}_{8} \oplus \mathrm{small} &  &  & \mathfrak{e}_{8} & \\
(11) & B_{11} & ... & (12) & B_{11}%
\end{array}
}}[E_{8}]
\end{equation}}
where $B_{11}$ denotes the configuration of curves for $(E_8 , E_8)$ conformal matter.
The new conformal matter system at the intersection of the $E_8$ singularity and the $E_8$ wall
is the system involving the first 11 factors.  In other words the boundary point has fractionated
to 12 points. Finally, in this case, in the region where we join the ramp to the plateau region, we see that there
is the $E_8$ version of a half hyper, namely a single small instanton on the leftmost node of the plateau region.

\subsubsection{Partial Higgs Branches}

Instead of passing to the resolved phase, we can instead consider moving to a
lower theory by passing to partial Higgs branches, as we discussed in the
case of M5-branes probing ADE singularities. In this case, it is simplest to
treat all of the theories%
\begin{equation}
\lbrack E_{8}]\underset{N}{\underbrace{%
\begin{array}
[c]{cccc}%
\mathfrak{g} & \mathfrak{g} & ... & \mathfrak{g}\\
1 & 2 & ... & 2
\end{array}
}}[G],
\end{equation}
in a uniform fashion.

Again the partial Higgsing is associated with a choice of breaking pattern for the left
and right flavor groups. The right flavor group has exactly the same structure
as we have already discussed, and its vacua are characterized by the orbit of a
nilpotent element $\mu_{R}$ in the algebra $\mathfrak{g}_{\mathbb{C}}$.

For the left flavor symmetry, that is, the global $E_8$ factor, the corresponding
breaking patterns are characterized somewhat differently. This is because they originate
(in heterotic language) from a 10d rather than a 7d Yang-Mills sector.
It is simplest to work in terms of the dual heterotic description.\footnote{ In
the F-theory realization, one must include some additional
data beyond just the Hitchin system with $E_8$ gauge group. The reason is that
the Hitchin system on a $\mathbb{P}^1$ with marked points
is an appropriate local dual for a smooth $K3_{het}$, since the local geometry of the base in the
fibration $T^2 \rightarrow K3_{het} \rightarrow \mathbb{P}^{1}$ is $\mathcal{O}(-2) \rightarrow \mathbb{P}^1$.
Once we allow singularities, however, the reduction to the $\mathbb{P}^1$ will also include extra data associated with the singular
fiber of $K3_{het}$. This must be included to fully characterize the moduli space of the flavor seven-brane.}
There, we are considering heterotic strings on the orbifold $\mathbb{C}^{2}/\Gamma_{G}$. $E_8$ instanton configurations in this case comes
with the standard moduli of an instanton, for example the size and position,
but also requires specifying boundary data \textquotedblleft off at
infinity\textquotedblright.
 The boundary of $\mathbb{C}^{2}/\Gamma_{G}$ is
$S^{3}/\Gamma_{G}$, and the behavior of the instanton density at infinity is
captured by a flat connection at infinity, which translates to a flat $E_8$ connection
on $S^3/\Gamma_{G}$.  Such flat bundles are in one to one correspondence with
an element $\gamma \in$
Hom$(\pi_{1}(S^{3}/\Gamma_{G}),E_{8})\simeq$  Hom$(\Gamma_{G},E_{8})$.

Summarizing, we see that all of these theories are characterized
as $\mathcal{T}(E_{8} , G_{R} , \gamma_L , \mu_R , N)$.

\section{Holographic Duals and Scaling Limits \label{sec:HOLO}}

In this section we study scaling limits for the conformal field theories just
constructed in which the number of probe M5-branes becomes large. The
most straightforward case to consider is that of $N$ M5-branes in flat space.
The near horizon limit for this geometry is well-known, and is given by the
11D\ supergravity background $AdS_{7}\times S^{4}$ with $N$ units of four-form
flux threading the $S^{4}$. The probe theory for an ADE\ singularity is then
given by
\begin{equation}
	AdS_{7}\times S^{4}/\Gamma_{G}\ ,
\end{equation}
where the ADE\ subgroup of $SU(2)$ specified by $\Gamma_{G}$ has
fixed points at the north and south pole of the $S^{4}$ \cite{Zaffaddme, Tarta}.

Long ago, the holographic dual of the (1,0) theory of $N$ small instantons was found
to be $AdS_7\times S^4/\mathbb{Z}_2$ (see \cite{Berkooz:1998bx}). In the
case of $N$ small instantons probing an ADE singularity, we
instead have the gravity dual
\begin{equation}
	AdS_{7}\times S^{4}/\mathbb{Z}_2\times \Gamma_{G} \ .
\end{equation}
The $%
\mathbb{Z}
_{2}$ fixed point locus is the equator of the $S^{4}$, while the two fixed
points of $S^{4}/\Gamma_{G}$ are now identified. Along the fixed point locus,
we also see that there is an $E_{8}$ nine-brane wrapped on $AdS_{7}\times
S^{3}/\Gamma_{G}$.

\begin{figure}[ht]
	$$
	\begin{gathered}
		\includegraphics[scale=0.6]{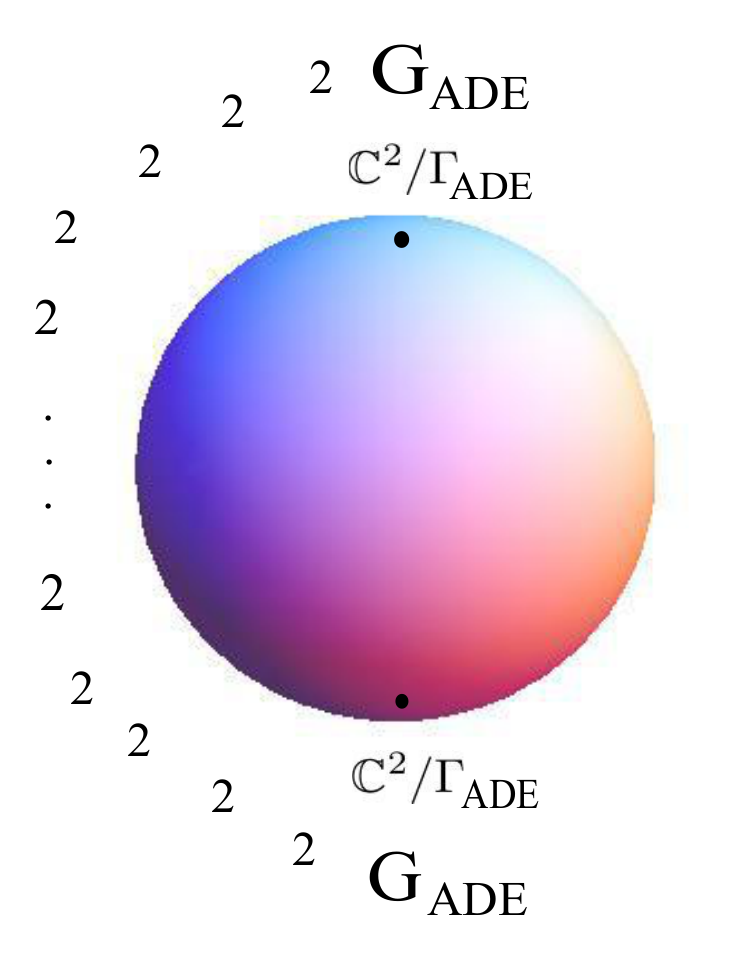}
		\end{gathered}\qquad\qquad
		\begin{gathered}
		\includegraphics[scale=0.6]{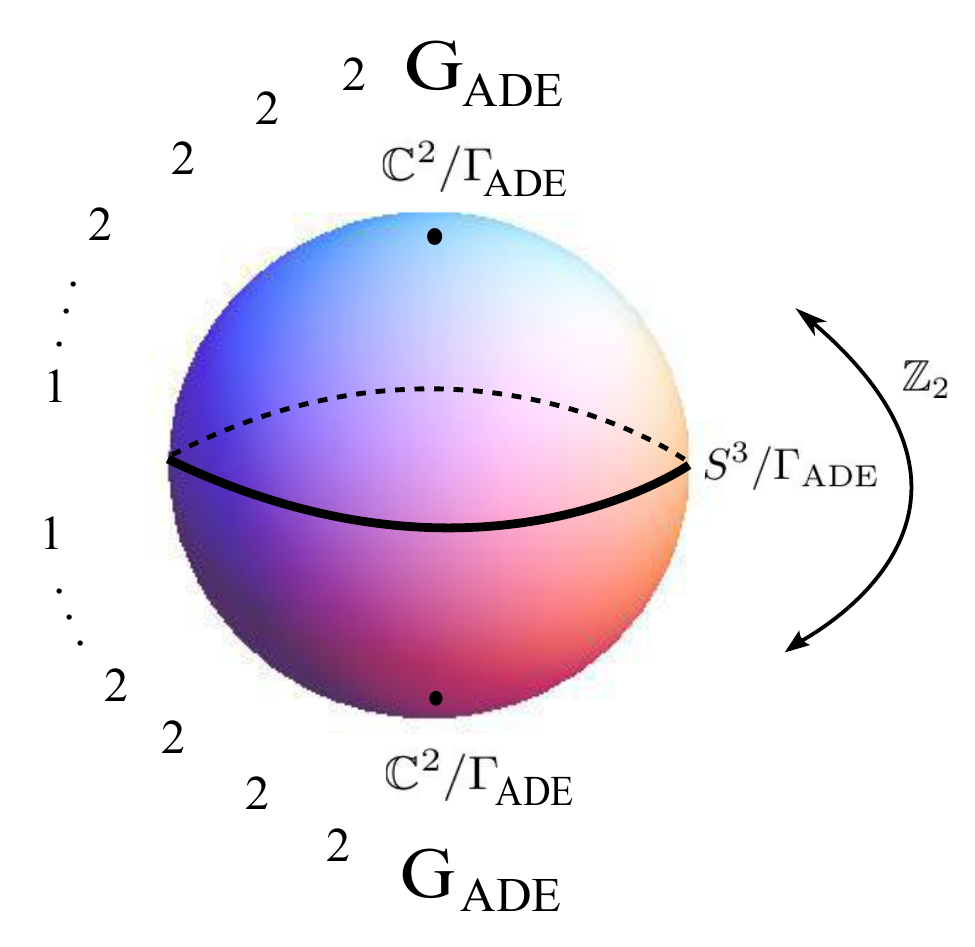}
		\end{gathered}
	$$
	\caption{\textsc{left}: A schematic view of the deconstruction of the great arc of $S^4/\Gamma_{ADE}$. \textsc{right}: A schematic view of the deconstruction of the great semi-arc of $S^4/\mathbb{Z}_2 \times \Gamma_{ADE}$.}
	\label{fig:ArcDeconstruction}
\end{figure}

An interesting feature of the F-theory geometry is that the $-2$ curves in the
sequence $2,...,2$ are literally a deconstruction of a great arc of the
associated $S^{4}$. More precisely, this is the interval obtained from taking
$S^{4}$ as an $S^{3}$ fibration over a finite interval. Further, in the
orbifold $S^{4}/\Gamma_{G}$, we see that just as the F-theory geometry
predicts, there is a flavor symmetry at the north pole, and another at the
south pole. These are simply the 7d Super Yang-Mills theories. Starting from
such a holographic dual, we can also see that the process of decompactifying
leads in the IR to a new dual with a different number of flux units. For
example, in the configuration:%
\begin{equation}
\underset{N}{\underbrace{2,...,2}},\underset{M}{\underbrace{2,...,2}},
\end{equation}
we see that decompactifying curves at the interface of the $N$ and $M$
partitions will decouple the two CFTs. In the gravity dual, this corresponds
to introducing a stack of M5-branes located at a specific AdS radius (i.e. the
energy scale in the CFT dual), and at a particular point on the great circle
of the $S^{4}$ (and/or its orbifolds). From this perspective, we can also see
that the large $N$ limit of one of our configurations need not yield a
semi-classical gravity dual. Indeed, if we attempt to gauge the flavor
symmetries, and then push them to strong coupling, we end up collapsing the
radius of the AdS space. So in other words, not every large $N$ $(1,0)$ theory
will have a semi-classical gravity dual.

We have also seen that there are a large number of additional SCFTs which can
be generated by starting from one of our \textquotedblleft
master\textquotedblright\ theories. This is accomplished by moving on to the
Higgs branch, i.e. by activating vevs for some operators of these theories. As
we have already seen, this is captured by a choice of nilpotent element of an
algebra, and in the case of the small instanton theories also involves a
choice of flat $E_{8}$ connection in $S^{3}/\Gamma_{G}$. It is therefore
natural to ask whether these theories with a Higgs branch also have a gravity dual.

First of all, we can see that in most cases, this data will remain hidden from
the strict $N=\infty$ description of the gravity dual. This is because the
data of the Higgs branches is localized near the north and south poles in the
case of the $S^{4}/\Gamma_{G}$ duals, and in the case of the $S^{4}/\mathbb{Z}_2 \times \Gamma_{G}$
duals is localized near the equator as well. In
particular, since the Higgsing only involves a small number of $-2$ curves (in
F-theory language), the actual portion of the great arc sensitive to these
effects is of order $1/N$, in units where the great arc between the north and
south poles is order one.

To see the effects of the Higgs branches in this limit, we therefore need to
take a scaling limit in which
\begin{equation}
\nu=\frac{\left\vert \Gamma_{G}\right\vert }{N}%
\end{equation}
is held fixed. This can be done for both the A- and D-type orbifold
singularities, but not for the E-type singularities. Taking such a limit, we see
that we have essentially \textquotedblleft stretched out\textquotedblright%
\ the contribution from the 7d\ Super Yang-Mills sector on both the north pole
and on the south pole, so we can expect to see the different Higgs branches,
i.e. T-brane data directly in the form of the gravity dual solutions.

\subsection{The $\mathbb{Z}_k$ Case}

We can be a little more specific in the case $G=SU(k)$, $\Gamma_G = \mathbb{Z}_k$. This case was analyzed in \cite{Zaffaddme, Tarta, Apruzzi:2013yva, Gaiotto:2014lca}, and we will review it here. The solutions corresponding to the presence of the T-branes can be described in IIA, by switching on the so-called Romans mass parameter $F_0$.

First of all, when all T-brane data is switched off,
we can reduce $AdS_{7}\times S^{4}/\mathbb{Z}_k$ to IIA. This
can be done using the characterization of $S^{4}$ as an $S^{3}$ fibered over an interval $I$, and observing that the
$S^{3}$ admits an $S^{1}$ fibration $S^{1}\rightarrow S^{3}\rightarrow S^{2}$,
along which the $\mathbb{Z}_{k}$ acts. Reducing along this $S^1$, we get a IIA geometry $AdS_{7}\times Y$, where $Y$ is an $S^{2}$ fibration over the interval $I$. The fibration does not shrink smoothly at the two endpoints of the interval; rather, there are two singularities, which correspond to the presence of two stacks of $k$ D6s and of $k\ \overline{\rm D6}$s.\footnote{ This is to be expected, since the $S^1$ we are reducing along has two zeros on $S^4$. Note also that the D6s and $\overline{\rm D6}$s
are mutually supersymmetric since they sit at opposite poles of the $S^4$.}. Thus $Y$ is not
a smooth space; we can think of it as a ``football''. For more details on this
solution, see section 5.1 of reference \cite{Apruzzi:2013yva}.

\begin{figure}[ht]
	\centering
		\includegraphics[scale=0.5]{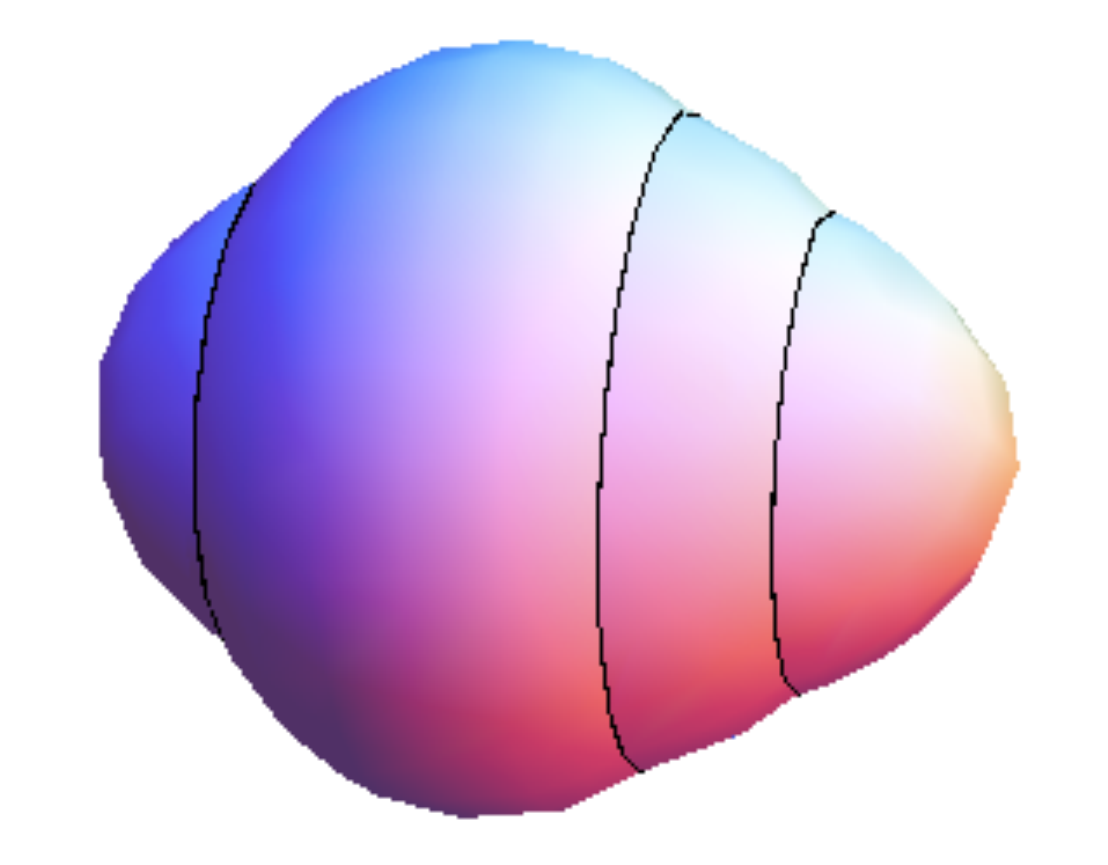}
	\caption{A cartoon of the internal space (topologically an $S^3$) for one of the solutions in \cite{Gaiotto:2014lca, Apruzzi:2013yva}; it represents the near-horizon geometry
of the brane configuration in figure \ref{fig:D6D8}.}
	\label{fig:nhD8D6}
\end{figure}

Recall that this solution is dual in F-theory to a chain of ordinary D7s, leading to the CFT given in line \eqref{222quiver}. Now let us see what happens if we add T-brane data on the left- and right-most D7s. As we discussed in section \ref{sec:HIGGS}, these correspond in IIA to brane configurations involving D6s ending on D8s, in a way intuitively summarized by two Young diagrams $\mu_{\rm L}$, $\mu_{\rm R}$ which are the same as the T-brane data. It was argued in \cite{Gaiotto:2014lca} that the near-horizon limit of those brane configurations is given by the $F_0\neq 0$ solutions in \cite{Apruzzi:2013yva}. The internal manifold $Y$ is now no longer a football: its two ``spikes'' at the North and South Pole now expand into several ``creases'' due to the presence of D8/D6 bound states; we show a cartoon in figure \ref{fig:nhD8D6}. Each of these bound states represents a D8 in a brane picture such as the one in figure \ref{fig:D6D8}; the D6 charge of the bound state represents the number of D6s ending on the D8, and a Jordan block in the T-brane Hitchin pole in the IIB picture.

The physical radius $r$ of a D8/D6 with charge $\mu$ is determined by a relationship of the type
\begin{equation}\label{eq:myers}
	e^{-\phi} r \sim \mu\ .
\end{equation}
This is reminiscent of the Myers effect relationship; indeed we argued in section \ref{sub:IIASUk} that the D8s and D6s fuse into ``fuzzy funnels''. Actually equation (\ref{eq:myers}) might cause some confusion in the case where $\mu=1$, since we saw in section \ref{sub:IIASUk} that this particular case does not correspond to a fuzzy funnel, while in the gravity dual we just described it would appear to have a certain finite radius. However, in the range of applicability of the gravity solution, the radius of a D8/D6 with $\mu=1$ is always smaller than $l_s$, thus making it indistinguishable from a D6.

Another case that deserves a special mention is when the bound (\ref{eq:Nbound}) is saturated in the brane picture. In the $AdS_7$ solution, what happens in this case is that the leftmost and rightmost D8/D6s join, and the region $F_0=0$ disappears. It is in this sense that we anticipated in section \ref{sub:altIIASUk} that in these cases it might be more natural to move the NS5s to a region where $F_0\neq0$. In that section, we also called ${\cal T}(SU(k_L) , SU(k_R) , N)$ the theory engineered by the configuration with unequal numbers of semi-infinite D6s coming out of a stack of NS5s in a region where $F_0\neq 0$. This case was only briefly mentioned in \cite{Gaiotto:2014lca}; the gravity dual in this case looks like an ``asymmetric football'', with two unequal D6 singularities.

\subsection{Adding Orientifolds}

A similar analysis can also be performed in the presence of orientifolds, for example, the quotient $S^4 / \Gamma_{D}$ for D-type $SU(2)$ subgroups
also admits a scaling limit.  Such solutions arise by orientifolding the solutions
for the $\mathbb{Z}_k$ case with an orientifold whose space action is the antipodal map on the $S^2$.
We defer a full analysis of the corresponding gravity duals along the
lines presented in \cite{Gaiotto:2014lca, Apruzzi:2013yva} to future work.

Perhaps more interesting is that some features of an orientifold construction also persist for the theories with an $E_8$ wall, that is, on the duals of the form $AdS_7 \times S^4 / \mathbb{Z}_2 \times \Gamma_G$. We now explain what becomes of the $\mathbb{Z}_2$ invariant wall upon reducing to a IIA configuration. To remain at weak coupling in the reduction,
we need to take $G$ to be in the A- or D-series. For illustrative purposes, we focus on the A-series.

In this case, we will now only see half of the football. In eleven dimensions, the ``end of the world'' boundary supports an $E_{8}$ gauge symmetry. When reduced to IIA, this becomes an O8 with a stack of eight D8-branes on top. Only an $SO(16)$ gauge symmetry is visible in the brane construction; the further enhancement to $E_{8}$ is due to a strong coupling effect, and is not directly visible in terms of perturbative IIA\ ingredients. This is very analogous to what is found in the duality between type\ I\ and heterotic strings \cite{Polchinski:1995df}. The presence of the O8-D8 system now changes the theory (\ref{222quiver}) on one of the two sides: it adds a ``ramp'' of $SU$ groups with linearly decreasing gauge groups, ending with an E-string theory. The result is the theory (\ref{eq:rampplateau}). This ``ramp'' was first found in \cite{Aspinwall:1997ye} in F-theory with essentially the computation we presented in section \ref{subsub:tbranch}, the matter content was determined in \cite{intriliaddme}. This computation was then reproduced in \cite{Hanany:1997gh} directly in IIA.

We can now introduce D8/D6 bound states in this case as well, replacing the D6 singularity of our half-football with several creases corresponding to a single Hitchin pole. This possibility was only briefly mentioned in \cite{Gaiotto:2014lca, Apruzzi:2013yva}, but  presenting such gravity solutions would not be particularly challenging. Starting from the previous case without an O8, one would have to consider configurations where the two partitions $\mu_{\rm L}$, $\mu_{\rm R}$ are equal, and mod out by the reflection that exchanges them.


\section{Conclusions \label{sec:CONC}}

In this paper we have studied aspects of $(1,0)$ superconformal theories
which arise in M-theory when M5-branes probe an ADE singularity, possibly
near an $E_8$ wall. The central theme of this work has been the idea that
such theories have tensor branches which involve a generalization
of quiver gauge theories in which the matter sector defines a strongly coupled SCFT. This has led
us to an identification of specific theories for domain walls in 7d SYM theory, as well
as matter systems arising from the collision of an ADE singularity with the $E_8$ wall of heterotic M-theory.
We have also discovered the phenomenon of fractionating of M5-branes on an ADE singularity
as well the fractionating of the intersection point of an ADE singularity with the $E_8$ wall.
We have also shown that the partial Higgs branches of these theories are characterized by discrete
algebraic data in each flavor symmetry factor which in F-theory is
encoded in the choice of a T-brane configuration. Taking scaling limits of these
theories, we have also shown when to expect a supergravity dual for these
configurations. In the remainder of this section we discuss some further
avenues of investigation.

As we have argued, many 6d SCFTs can be understood in terms of a generalized notion of a
quiver, in which the matter sector is itself a strongly coupled SCFT. However, a direct derivation
of the matter sector from some putative generalization of brane probes of
orbifolds is still to be understood.  We have also presented evidence that fractional
M5-branes can be understood as activating a fractional flux on the singularity, and
changing the gauge symmetry.  It would be interesting to further understand this.

In the context of F-theory, there are also some immediate generalizations of
such theories which involve non-simply laced flavor symmetries. One expects
that in the M-theory description, circle reduction with a
twist will lead to conformal matter for non-simply laced symmetry factors.

Though our primary focus has been on 6d SCFTs, we have also seen that
compactifying some of these theories to lower dimensions leads to novel dual theories for affine quiver
gauge theories.  It would be interesting to explore this further.

Finally, we have also seen that many of these generalized quiver theories have
a holographic dual description. Nevertheless, some aspects, especially the partial Higgs branches
for the E-type probe theories are washed away at $N = \infty$. It would be
very interesting to see how semi-classical $1/N$ corrections to these
duals recover these more detailed structures.

\section*{Acknowledgements}

We thank B. Haghighat, A. Hanany, D. R. Morrison, T. Rudelius, E. Witten and A. Zaffaroni for helpful discussions.
JJH and AT thank the Perimeter Institute 2014 workshop on Supersymmetric
Quantum Field Theories in Five and Six Dimensions for hospitality during which
some of this work was completed. AT\ also thanks the high energy theory group
at Harvard for hospitality during which some of this work was completed. The
work of MDZ, JJH and CV is supported by NSF grant PHY-1067976. AT\ is supported in part by INFN, by the MIUR-FIRB grant RBFR10QS5J \textquotedblleft String Theory and Fundamental Interactions\textquotedblright, and by the European Research Council under the European Union's Seventh Framework Program (FP/2007-2013) - ERC Grant Agreement n. 307286 (XD-STRING).


\newpage

\appendix

\section{Non-Higgsable Clusters}

In this Appendix we briefly review some elements of non-Higgsable clusters in
F-theory. For additional discussion, see \cite{Heckman:2013pva, Morrison:2012np}.

The central idea of \cite{Morrison:2012np} is to study the minimal singularity
type along a given $\mathbb{P}^{1}$ in the base, which is dictated by the
order of vanishing for $f$ and $g$ along the curve. Interestingly, this is
fully specified by the self-intersction of such a curve inside the base $B$. A
\textquotedblleft non-Higgsable cluster\textquotedblright\ consists of all
such configurations where the singularity type cannot be deformed by a
smoothing (i.e. by a Higgsing operation in the field theory). These NHCs have
been determined in \cite{Morrison:2012np}, and consist of a configuration of
up to three curves. The NHCs consist of the ADE\ Dynkin diagrams for $-2$ curves, as well as
some additional cases. For a single curve we can have a self-intersection $-n$
for $2\leq n\leq12$, and for two curves, we have a single intersection, with
one curve of self-intersection $-3$ and one with self-intersection $-2$.
For three curves, we have a three node linear graph, with two curves having
self-intersection $-2$, and one curve with self-intersection $-3$. For each
such cluster there is a corresponding gauge group, and possibly some
additional matter fields. The six-dimensional theory associated with each type
of cluster is:%
\begin{align}
&
\begin{array}
[c]{lllllllll}%
\begin{array}
[c]{c}%
\text{Theory:}\\
\text{Curve:}%
\end{array}
&
\begin{array}
[c]{c}%
\\
2
\end{array}
&
\begin{array}
[c]{c}%
\mathfrak{su}_{3}\\
3
\end{array}
&
\begin{array}
[c]{c}%
\mathfrak{so}_{8}\\
4
\end{array}
&
\begin{array}
[c]{c}%
\mathfrak{f}_{4}\\
5
\end{array}
&
\begin{array}
[c]{c}%
\mathfrak{e}_{6}\\
6
\end{array}
&
\begin{array}
[c]{c}%
\mathfrak{e}_{7}\oplus\frac{1}{2}\mathbf{56}\\
7
\end{array}
&
\begin{array}
[c]{c}%
\mathfrak{e}_{7}\\
8
\end{array}
&
\begin{array}
[c]{c}%
\mathfrak{e}_{8}\\
12
\end{array}
\end{array}
\\
&  \text{Theory:\ }\mathfrak{g}_{2}\times\mathfrak{su}_{2}\oplus\frac{1}%
{2}(\mathbf{7}+\mathbf{1},\mathbf{2})\text{ \ \ / \ \ Curves:\ }3,2\\
&  \text{Theory:\ }\mathfrak{g}_{2}\times\mathfrak{sp}_{1}\oplus\frac{1}%
{2}(\mathbf{7}+\mathbf{1},\mathbf{2})\text{ \ \ / \ \ Curves:\ }3,2,2\\
&  \text{Theory:\ }\mathfrak{su}_{2}\times\mathfrak{so}_{7}\times
\mathfrak{su}_{2}\oplus\frac{1}{2}(\mathbf{2,8,1})\oplus\frac{1}%
{2}(\mathbf{1,8,2})\text{ \ \ / \ \ Curves: }2,3,2
\end{align}
To form bigger configurations of curves, one then joins these clusters by curves of self-intersection $-1$. If
the $-1$ curve intersects two curves $\Sigma_L$ and $\Sigma_R$ with respective gauge symmetries $\mathfrak{g}_L$ and
$\mathfrak{g}_{R}$, existence of an elliptic fibration with fibers in Kodaira-Tate form requires
$\mathfrak{g}_L \times \mathfrak{g}_R \subset \mathfrak{e}_8$. See reference \cite{Heckman:2013pva} for further
discussion of this gluing condition.

\section{$(G_{L},G_{R})$ Conformal Matter}

In this Appendix we calculate the conformal matter sector associated with a
general pairing of flavor symmetries $(G_{L},G_{R})$, for $G_{L}$ and $G_{R}$
a flavor symmetry. We start in an F-theory compactification in which there is
a component of the discriminant locus supporting Lie algebras $\mathfrak{g}%
_{L}$ and $\mathfrak{g}_{R}$, respectively. If we cannot reach this
configuration from Higgsing of an adjoint-valued field in a higher rank gauge
symmetry such that $\mathfrak{g}_{L}\times\mathfrak{g}_{R}\subset
\mathfrak{g}_{\text{parent}}$, we must blow up the intersection point.
Continuing in this fashion, we compute the minimal conformal matter between
two such symmetry factors. We focus on the case of ADE\ flavor symmetries
which generate conformal matter, i.e. we exclude the cases of $A\times A$ and
$A\times D$ type collisions as they lead to weakly coupled hypermultiplets.
Finally, it would be interesting to extend this analysis to non-simply laced algebras.

\subsection{$E\times E$ Conformal Matter}

Recall the algorithm for resolving a collision of two E-type loci. We start
with two non-compact divisors, and start blowing up the intersection point.
The first blowup produces a single $-1$ curve. Then, if we still do not
satisfy the gauging rule outlined in \cite{Heckman:2013pva}, we continue to blow up further.
The procedure is completely algorithmic, and we collect the theories on the
tensor branch:%
\begin{align}
(E_{8},E_{8}) &  :%
\begin{tabular}
[c]{|c|ccccccccccccc|}\hline
Gauge Symm: &  &  & $\mathfrak{sp}_{1}$ &  & $\mathfrak{g}_{2}$ &  &
$\mathfrak{f}_{4}$ &  & $\mathfrak{g}_{2}$ &  & $\mathfrak{sp}_{1}$ &  &
\\
Curve: & $1$ & $2$ & $2$ &  & $3$ & $1$ & $5$ & $1$ & $3$ &
 & $2$ & $2$ & $1$\\
Hyper: &  &  &  & $\frac{1}{2}(\mathbf{2},\mathbf{7}+\mathbf{1})$
&  &  &  &  &  & $\frac{1}{2}(\mathbf{7}+\mathbf{1,2})$ &  &  & \\\hline
\end{tabular}
\\
(E_{8},E_{7}) &  :%
\begin{tabular}
[c]{|c|cccccccccccc|}\hline
Gauge Symm: &  &  & $\mathfrak{sp}_{1}$ &  & $\mathfrak{g}_{2}$ &  &
$\mathfrak{f}_{4}$ &  & $\mathfrak{g}_{2}$ &  & $\mathfrak{su}_{2}$ & \\
Curve: & $1$ & $2$ & $2$ &  & $3$ & $1$ & $5$ & $1$ & $3$ &
 & $2$ & $1$\\
Hyper: &  &  &  & $\frac{1}{2}(\mathbf{2},\mathbf{7}+\mathbf{1})$
&  &  &  &  &  & $\frac{1}{2}(\mathbf{7}+\mathbf{1,2})$ &  & \\\hline
\end{tabular}
\\
(E_{8},E_{6}) &  :%
\begin{tabular}
[c]{|c|cccccccccc|}\hline
Gauge Symm: &  &  & $\mathfrak{sp}_{1}$ &  & $\mathfrak{g}_{2}$ &  &
$\mathfrak{f}_{4}$ &  & $\mathfrak{su}_{3}$ & \\
Curve: & $1$ & $2$ & $2$ & & $3$ & $1$ & $5$ & $1$ & $3$ &
$1$\\
Hyper &  &  &  & $\frac{1}{2}(\mathbf{2},\mathbf{7}+\mathbf{1})$
&  &  &  &  &  & \\\hline
\end{tabular}
\\
(E_{7},E_{7}) &  :%
\begin{tabular}
[c]{|c|cccccccc|}\hline
Gauge Symm &  & $\mathfrak{su}_{2}$ &  & $\mathfrak{so}_{7}$ &  &
$\mathfrak{su}_{2}$ &  & \\
Curve & $1$ & $2$ &  & $3$ &  & $2$ & $1$ &
\\
Hyper &  &  & $\frac{1}{2}(\mathbf{2,8})$ &  & $\frac{1}%
{2}(\mathbf{8,2})$ &  &  & \\\hline
\end{tabular}
\\
(E_{7},E_{6}) &  :%
\begin{tabular}
[c]{|c|cccccccc|}\hline
Gauge Symm: &  & $\mathfrak{su}_{2}$ &  & $\mathfrak{so}_{7}$ &  &
$\mathfrak{su}_{2}$ &  & \\
Curve: & $1$ & $2$ & & $3$ &  & $2$ & $1$ &
\\
Hyper: &  &  & $\frac{1}{2}(\mathbf{2,8})$ &  & $\frac{1}%
{2}(\mathbf{8,2})$ &  &  & \\\hline
\end{tabular}
\\
(E_{6},E_{6}) &  :%
\begin{tabular}
[c]{|c|ccc|}\hline
Gauge Symm: &  & $\mathfrak{su}_{3}$ & \\
Curve: & $1$ & $3$ & $1$\\\hline
\end{tabular}
\ .
\end{align}

\subsection{$E\times A$ Conformal Matter}

Next, consider the case of a collision of an E-type locus with an A-type
algebra. This can actually occur in two-different ways in F-theory, so to
distinguish them, we shall refer to $E\times A$ matter, and $E\times H$
matter. A-type symmetries arise from a stack of parallel D7-branes in weakly coupled IIB string theory.
H-type symmetries arise from a non-perturbative bound state of seven-branes of different $(p,q)$ type. Whereas $A_k$
symmetries exist for arbitrary $k$, $H_{k}$ only exists for $1 \leq k \leq 3$.

Consider first the case of an A-type singularity, which is realized by an
$I_{n}$ locus for the discriminant. This has already been worked out in
\cite{Aspinwall:1997ye}, see also \cite{intriliaddme}. Basically, we consider a collision of two components
of the discriminant locus, one supporting an $II^{\ast}$ fiber, and the other supporting
an $I_{n}$ singular fiber. Each such collision can be blown up in the
base, thereby leading to an exceptional curve with lower singularity type. In
the case of the A- series, we have an $I_{n-1}$ fiber after one such blowup.
This leads to the minimal resolution on the tensor branch. For a collision
with an $E_{8}$ seven-brane, this yields:%
\begin{equation}
(E_{8},A_{k-1}):%
\begin{tabular}
[c]{|c|cccccccc|}\hline
Gauge Symm: &  & $\mathfrak{su}_{1}$ &  & $\mathfrak{su}_{2}$ &  &  &
$\mathfrak{su}_{k-1}$ & \\
Curve: & $1$ & $2$ &  & $2$ &  & ... & $2$ & \\
Hyper: &  &  & $(\overline{\mathbf{1}},\mathbf{2})$ &  & $(\overline
{\mathbf{2}},\mathbf{3})$ &  &  & $(\overline{\mathbf{k-1}}\mathbf{,k}%
)$\\\hline
\end{tabular}
\end{equation}
where in this case, there is a full hypermultiplet in the bifundamental
trapped at the intersection of each A-type gauge group.

Similar considerations hold for collisions with $E_{7}$ and $E_{6}$. However,
in these cases there is no need to fully resolve the collision of an $I_{n}$
fiber with the singularity, since it can also just lead to ordinary matter.
Taking this into account, we get the following collisions:%
\begin{align}
(E_{7},A_{k-1})  & :%
\begin{tabular}
[c]{|c|cccccccc|}\hline
Gauge Symm: &  & $\mathfrak{su}_{1}$ &  & $\mathfrak{su}_{2}$ &  &  &
$\mathfrak{su}_{k-1}$ & \\
Curve: &  & $1$ &  & $2$ &  & ... & $2$ & \\
Hyper: & $\frac{1}{2}(\mathbf{56,1)}$ &  & $(\overline{\mathbf{1}},\mathbf{2})$
&  & $(\overline{\mathbf{2}},\mathbf{3})$ &  &  & $(\overline{\mathbf{k-1}%
}\mathbf{,k})$\\\hline
\end{tabular}
\\
(E_{6},A_{k-1})  & :%
\begin{tabular}
[c]{|c|cccccc|}\hline
Gauge Symm: &  & $\mathfrak{su}_{2}$ &  &  & $\mathfrak{su}_{k-1}$ & \\
Curve: &  & $1$ &  & ... & $2$ & \\
Hyper: & $(\overline{\mathbf{27}},\mathbf{2})$ &  & $(\overline{\mathbf{2}%
},\mathbf{3})$ &  &  & $(\overline{\mathbf{k-1}}\mathbf{,k})$\\\hline
\end{tabular}
\end{align}

Consider next the case of an H-type singularity, that is, by having an E-type
locus collide with a fiber of type $II$, $III$ or $IV$. These respectively
generate $\mathfrak{su}_{1}$, $\mathfrak{su}_{2}$ and $\mathfrak{su}_{3}$, and
so we refer to all of them as $H_{n-1}$ for $\mathfrak{su}_{n}$. In these
special cases, there is initially no minimal singularity type on the
exceptional curve after blowing up. Following the minimal resolution
algorithm, we have, for $n=1,2,3$ and $k=1,2$ the $E\times H$ theories:%
\begin{align}
(E_{8},H_{n}) &  :%
\begin{tabular}
[c]{|c|cccccc|}\hline
Gauge Symm: &  &  & $\mathfrak{sp}_{1}$ &  & $\mathfrak{g}_{2}$ & \\
Curve: & $1$ & $2$ & $2$ &  & $3$ & $1$\\
Hyper: &  &  &  & $\frac{1}{2}(\mathbf{2},\mathbf{7}+\mathbf{1})$
&  & \\\hline
\end{tabular}
\\
(E_{7},H_{2}) &  :%
\begin{tabular}
[c]{|c|ccccc|}\hline
Gauge Symm: &  & $\mathfrak{su}_{2}$ &  & $\mathfrak{g}_{2}$ & \\
Curve: & $1$ & $2$ &  & $3$ & $1$\\
Hyper: &  &  & $\frac{1}{2}(\mathbf{2},\mathbf{7}+\mathbf{1})$ &
& \\\hline
\end{tabular}
\\
(E_{7},H_{k}) &  :%
\begin{tabular}
[c]{|c|c|}\hline
Gauge Symm: & \\
Curve: & $1$\\\hline
\end{tabular}
\\
(E_{6},H_{n}) &  :%
\begin{tabular}
[c]{|c|c|}\hline
Gauge Symm: & \\
Curve: & $1$\\\hline
\end{tabular}
\end{align}
Observe that in the case of the $E_{7}\times H$ theories, a different number
of blowups are required for some of the cases.

\subsection{$E\times D$ Conformal Matter}

Consider next the collision of an E-type locus with a D-type locus. In
F-theory, a D-type singularity comes from an $I_{k}^{\ast}$ Kodaira-Tate
fiber. For $E_8$, and $k\geq1$, this yields:%
\begin{align}
(E_{8},D_{k+4}) &  :%
\begin{tabular}
[c]{|c|cccccc|}\hline
Gauge Symm: &  &  & $\mathfrak{sp}_{1}$ &  & $\mathfrak{g}_{2}$ & \\
Curve: & $1$ & $2$ & $2$ & & $3$ & $1$\\
Hyper: &  &  &  & $\frac{1}{2}(\mathbf{2},\mathbf{7+1})$ &  &
\\\hline
\end{tabular}
\ \times\\
&  \times%
\begin{tabular}
[c]{|ccccc|}\hline
$\mathfrak{so}_{9}$ &  & $\mathfrak{sp}_{1}$ &  & $\mathfrak{so}_{11}$\\
$4$ & & $1$ & & \\
& $\frac{1}{2}(\mathbf{9},\mathbf{2})$ &  & $\frac{1}{2}(\mathbf{2,11)}$ &
\\\hline
\end{tabular}
\ \times...\times%
\begin{tabular}
[c]{|cccc|}\hline
$\mathfrak{so}_{2k+7}$ &  & $\mathfrak{sp}_{k} $ & \\
&  & $1$ & \\
& $\frac{1}{2}(\mathbf{2k+7,2k})$ & $\frac{1}{2} \mathbf{2k}$ & $\frac{1}{2}(\mathbf{2k},\mathbf{2k+8}%
)$\\\hline
\end{tabular}
\end{align}
where we have introduced a line break for typographical purposes. The minimal
resolutions for the cases $E_7$ and $E_6$ correspond
to replacing the top line by $1,2,3,1$ and $1,3,1$, respectively.\footnote{We thank T. Rudelius for
alerting us to a typo in a previous version of this statement.} The case
$k=0$ follows by omitting the second line. Here, there is a half
hypermultiplet trapped at each $\mathfrak{so}/\mathfrak{sp}$ intersection.
Furthermore, the $\mathfrak{sp}$ factors increase from $\mathfrak{sp}_{1}$ up
to $\mathfrak{sp}_{k}$. Finally, for the rightmost $\mathfrak{sp}_{k}$ factor, there is
one additional half hypermultiplet in the fundamental.

\subsection{$D\times D$ Conformal Matter}

Finally, we have the collisions of A- or D-type singularities with each other.
In all cases other than the collision of two D-type singularities, we get a
weakly coupled hypermultiplet. We therefore focus on the D-type collisions. When $k+l$ is even, we
have:
\begin{equation}
(D_{k+4},D_{l+4}):%
\begin{tabular}
[c]{|c|ccc|}\hline
Gauge Symm: &  & $\mathfrak{sp}_{r}$ & \\
Curve: &  & $1$ & \\
Hyper: & $\frac{1}{2}(\mathbf{2l+8} , \mathbf{2r})$ &  & $\frac{1}{2}(\mathbf{2r} , \mathbf{2l+8})$\\\hline
\end{tabular}
\end{equation}
where $ r = (k+l) / 2$.

When $k+l$ is odd, the analysis is more subtle, because after blowing up
the collision of the $I^{\ast}_{k}$ and $I^{\ast}_{l}$ fibers, we get a non-split $I_{k + l - 1}$
fiber. Letting $r = (k + l  + 1) / 2$, only an $\mathfrak{sp_{r-1}}$ can be identified in purely geometric
terms \cite{Aspinwall:1997ye}. Nevertheless, as explained in \cite{Bershadsky:1996nu, intriliaddme}, 6d
anomaly cancelation and consistent Higgsing dictates the structure
of the resulting conformal matter sector to be:
\begin{equation}
(D_{k+4},D_{l+4}):%
\begin{tabular}
[c]{|c|ccc|}\hline
Gauge Symm: &  & $\mathfrak{sp}_{r}$ & \\
Curve: &  & $1$ & \\
Hyper: & $\frac{1}{2}(\mathbf{2l+8} , \mathbf{2r})$ & $\mathbf{2r}$  & $\frac{1}{2}(\mathbf{2r} , \mathbf{2l+8})$\\\hline
\end{tabular}
\end{equation}
that is, there is an extra hypermultiplet in the fundamental of $\mathfrak{sp}_{r}$. This is rather analogous to the
fact that in F-theory on Calabi-Yau fourfolds, the structure of a Yukawa point \cite{BHVI} which is transparently realized
in gauge theory terms can sometimes be obscure just from the resolution of singular fibers \cite{Esole:2011sm}.

\section{ADE\ Subgroups of $SU(2)$}

In this Appendix we summarize some of the relevant properties of discrete
subgroups of $SU(2)$, and the corresponding orbifold singularities which they
generate. For additional discussion, see \cite{Slodowy}. We first start by
considering the set of generators of the exceptional binary polyhedral groups
as subgroups of $SU(2)$. For each generator, the subscript indicates the order
of the element. Also, we let $\xi_{(k)}=\exp(2\pi i/k)$ denote a primitive
$k$th root of unity.

\begin{itemize}
\item $\mathbb{A}_{k}$, the cyclic group of order $k+1$, with generator:%
\begin{equation}
\omega_{(k+1)}\equiv\left(
\begin{matrix}
\xi_{(k+1)} & 0\\
0 & \xi_{(k+1)}^{-1}%
\end{matrix}
\right)  ,
\end{equation}

\item $\mathbb{D}_{p}$, for $p\geq4$, the binary dihedral group of order
$4p-8$, with generators:%
\begin{equation}
\omega_{(2p)}\equiv\left(
\begin{matrix}
\xi_{(2p)} & 0\\
0 & \xi_{(2p)}^{-1}%
\end{matrix}
\right)  \text{ \ \ and \ \ }\tau_{(4)}\equiv\left(
\begin{matrix}
0 & 1\\
-1 & 0
\end{matrix}
\right)  ,
\end{equation}

\item $\mathbb{T}$, the binary tetrahedral group has order $24$, with
generators:%
\begin{equation}
\omega_{(4)}=\left(
\begin{matrix}
\xi_{(4)} & 0\\
0 & \xi_{(4)}^{-1}%
\end{matrix}
\right)  \text{ \ \ and \ \ }\kappa_{(6)}=\frac{1}{\sqrt{2}}\left(
\begin{matrix}
\xi_{(8)}^{7} & \xi_{(8)}^{7}\\
\xi_{(8)}^{5} & \xi_{(8)}%
\end{matrix}
\right)  ,
\end{equation}

\item $\mathbb{O}$, the binary octahedral group has order $48$, with the same
generators as $\mathbb{T}$, as well as an additional generator:
\begin{equation}
\omega_{(4)}=\left(
\begin{matrix}
\xi_{(4)} & 0\\
0 & \xi_{(4)}^{-1}%
\end{matrix}
\right)  \text{, \ \ }\kappa_{(6)}=\frac{1}{\sqrt{2}}\left(
\begin{matrix}
\xi_{(8)}^{7} & \xi_{(8)}^{7}\\
\xi_{(8)}^{5} & \xi_{(8)}%
\end{matrix}
\right)  \text{, \ \ }\omega_{(8)}=\left(
\begin{matrix}
\xi_{(8)} & \\
0 & \xi_{(8)}^{-1}%
\end{matrix}
\right)
\end{equation}

\item $\mathbb{I}$, the binary icosahedral group has order $120$, with
generators:
\begin{equation}
\omega_{(10)}=-\left(
\begin{matrix}
\xi_{(5)}^{3} & 0\\
0 & \xi_{(5)}^{2}%
\end{matrix}
\right)  \text{ \ \ and \ \ }\kappa_{(4)}=\frac{1}{\xi_{(5)}^{2}-\xi
_{(5)}^{3}}\left(
\begin{matrix}
\xi_{(5)}+\xi_{(5)}^{4} & 1\\
1 & -\xi_{(5)}-\xi_{(5)}^{4}%
\end{matrix}
\right)  .
\end{equation}

\end{itemize}

For each of these discrete subgroups of $SU(2)$, we get a corresponding
orbifold singularity $\mathbb{C}^{2}/\Gamma$. We summarize the corresponding
hypersurface, discrete subgroup, and order of the subgroup in the following
list (see e.g. \cite{Slodowy}):
\begin{equation}%
\begin{tabular}
[c]{|c|l|c|c|}\hline
& singularity & $\Gamma$ & $|\Gamma|$\\\hline
$A_{k}$ & $y^{2}=x^{2}+z^{k+1}$ & $\mathbb{Z}_{k+1}$ & $k+1$\\\hline
$D_{p}$ & $y^{2}=x^{2}z+z^{p-1}$ & $\mathbb{D}_{p-2}$ & $4p-8$\\\hline
$E_{6}$ & $y^{2}=x^{3}+z^{4}$ & $\mathbb{T}$ & $24$\\\hline
$E_{7}$ & $y^{2}=x^{3}+xz^{3}$ & $\mathbb{O}$ & $48$\\\hline
$E_{8}$ & $y^{2}=x^{3}+z^{5}$ & $\mathbb{I}$ & $120$\\\hline
\end{tabular}
\ \ \label{ADEsing}%
\end{equation}

\section{6d $(1,0)$ Minimal Models of Type $\alpha A_{N} \beta$}
\label{tables}

In this Appendix we study the emergence of conformal matter sectors as building blocks of the
6d $(1,0)$ models classified in \cite{Heckman:2013pva}. We consider in detail the case of the
$A_N$ models. The models of $D_N$ type and exceptional outliers can be treated in a similar way.

Let us start by introducing some notations. Let us denote $C^{*}$ the mirror of the configuration $C$, e.g. for
$C=2,2,3,1,5$, $C^{*} = 5,1,3,2,2$. Notice that a given configuration is
palindromic iff $C^{*}= C$. Moreover, le us write $C^{N}$ for a given
configuration of curves that is repeated $N$ times, e.g. $(1,4)^{3} =
1,4,1,4,1,4$. In the following tables we describe the minimal resolutions of
the minimal models of type $\alpha\, A_{N} \, \beta^{*}$. For all pairs
$(\alpha, \beta)$ one can form out of
\[
\mathcal{I} \equiv\big\{ 7,\, 3\,3,\, 2\,4,\, 2\,2\,3,\, 2\,2\,2\,3 ,\,
2\,2\,2\,2\,3 \big\}\quad\text{and}\quad\mathcal{J} \equiv
\big\{ 6,\,5,\,4,\,3,\, 2,\, 2\,3,\, 2\,2,\, 2\,2\,2,\, 2\,2\,2\,2 \big\}
\]
The models of type $\alpha\, A_{N} \, \beta^{*}$ behave literally as \emph{generalized linear quivers}. Each model has a central core or \emph{plateaux} built of generalized exceptional bifundamentals and two external tails on the left and on the right that complete these systems without (non-abelian) flavor symmetries in a superconformal fashion. Recall from \cite{Heckman:2013pva} that these models are non-Higgsable, therefore the only allowed bifundamentals are those without a Higgs branch. As we discussed in the main body of the text these are the exceptional bifundamentals
\begin{equation}
\begin{aligned}
& B(E_8) \equiv 1,2,2,3,1,5,1,3,2,2,1\\
& B(E_7) \equiv 1,2,3,2,1\\
& B(E_6) \equiv 1,3,1\\
\end{aligned}
\end{equation}
and the bifundamental with $SO(8)\times SO(8)$ flavor symmetry we have discussed in \S.\ref{subsub:SpecialK}. The infinite series that we find can be organized in four types, according to the type of bifundamental that occurs in the core of the
generalized linear quiver. For pairs in $\mathcal{I} \times\mathcal{I}$ and $\mathcal{I} \times
\mathcal{J}$ only bifundamentals $B(E_8)$ and $B(E_7)$ shows up; the other two types
arises only when one considers pairs in $\mathcal{J} \times\mathcal{J}$. The
descendents 2, 22, 222, and 2222 can be grouped in the same $A_{N}$ family of
type $\alpha A_{N}$, however, as we will see below, typically, $\alpha2$,
$\alpha22$, and $\alpha222$, behaves differently from $\alpha2222$, and only
in this last case the systems develops a specific type. We refer
to these models as ``isolated'' below.




In the tables below the we list the types, blow ups, number of curves in the
minimal blow up of the base, $N_{T}$, algebras for all $A_{N}$
minimal models. In addition, in the column marked by a $W$, we give also the ``wannabe'' flavor symmetry
obtained by making the leftmost and rightmost cycles non-compact.

Typically the models of type $E_8$ have many $\mathfrak{sp}_{1}$ factors in
their Lie algebras. To avoid lenghty and redundant tables, in this case we
have marked with a $\star$ the models in which there are genuine
$\mathfrak{su}_{2}$ factors (e.g. nHc's of type $3,2$ or $2,3$) and used the
isomorphism of the two rank one Lie algebras. This however happens rarely, as
one can see explicitly from the blow ups, typically when $\alpha$ or $\beta$
equals $2,2,2,3$.

Moreover, we have noticed that all models of type $E_7$ occur when
$\alpha$ or $\beta$ equals $5$ or $2,2,3$. Similarly, models of type
\emph{III} or \emph{IV} arise only if $\alpha$ or $\beta$ are $3$, $4$, or
$2,3$. We have marked with a $\spadesuit$ models with $\tfrac{1}{2}$ hypers in
the $\mathbf{56}$ of $E_{7}$.



\begin{landscape}
\begin{table}
\begin{center}
\caption{ Minimal resolutions: $(\, \cI , \cI \,)$ infinite families i.e. rigid theories}
{\footnotesize
\begin{tabular}{cccccccc}
Series  \phantom{$\Bigg|$}   & Type & Blow up & $N_T$ &Algebra & $W$ &\\
\hline
$7 \, A_N \, 7 $ \phantom{$\Bigg|$} & $E_8$  & {\scriptsize$12,( B(E_8),12)^{N+1}$} &{\scriptsize$12(N+1)+1$} & $(\mathfrak{e}_8)^{N+2} \oplus (\mathfrak{sp}_1 \oplus \mathfrak{g_2})^{2N+2} \oplus (\mathfrak{f}_4)^{N+1}$ & $\mathfrak{e_8} \oplus \mathfrak{e}_8$\\
$2 \, 4 \, A_N \, 7  $  \phantom{$\Bigg|$} & $E_8$  & {\scriptsize $3,2,2,1,12, ( B(E_8),12)^{N+1}$}  &{\scriptsize$12(N+1)+5$}&$(\mathfrak{e}_8)^{N+2} \oplus (\mathfrak{sp}_1 \oplus \mathfrak{g_2})^{2N+3} \oplus (\mathfrak{f}_4)^{N+1}$ & $\mathfrak{g_2} \oplus \mathfrak{e}_8$\\
$3 \, 3 \, A_N \, 7  $  \phantom{$\Bigg|$} & $E_8$  & {\scriptsize $5,1,3,2,2,1,12, ( B(E_8),12)^{N+1}$}  &{\scriptsize$12(N+1)+7$}&$(\mathfrak{e}_8)^{N+2} \oplus (\mathfrak{sp}_1 \oplus \mathfrak{g_2})^{2N+3} \oplus (\mathfrak{f}_4)^{N+2}$ & $\mathfrak{f_4} \oplus \mathfrak{e}_8$\\
$2 \, 2 \, 3 \, A_N \, 7  $  \phantom{$\Bigg|$} & $E_8$  & {\scriptsize $3,1,5,1,3,2,2,1,12, ( B(E_8),12)^{N+1}$}  &{\scriptsize$12(N+1)+9$}&$\mathfrak{su}_3 \oplus (\mathfrak{e}_8)^{N+2} \oplus (\mathfrak{sp}_1 \oplus \mathfrak{g_2})^{2N+3} \oplus (\mathfrak{f}_4)^{N+2}$ & $\mathfrak{su}_3 \oplus \mathfrak{e}_8$\\
$2 \, 2 \,2\, 3 \, A_N \, 7  $  \phantom{$\Bigg|$} & $E_8$  & {\scriptsize $2,3,1,5,1,3,2,2,1,12, ( B(E_8),12)^{N+1}$}  &{\scriptsize$12(N+1)+10$}&$ (\mathfrak{e}_8)^{N+2} \oplus (\mathfrak{sp}_1 \oplus \mathfrak{g_2})^{2N+4} \oplus (\mathfrak{f}_4)^{N+2}$ & $\mathfrak{su}_2 \oplus \mathfrak{e}_8$&$\star$\\
$2 \, 2 \,2\, 2 \,3 \, A_N \, 7  $  \phantom{$\Bigg|$} & $E_8$  & {\scriptsize $2,2,3,1,5,1,3,2,2,1,12, ( B(E_8),12)^{N+1}$}  &{\scriptsize$12(N+1) + 11$}&$(\mathfrak{e}_8)^{N+2} \oplus (\mathfrak{sp}_1 \oplus \mathfrak{g_2})^{2N+4} \oplus (\mathfrak{f}_4)^{N+2}$ & $\mathfrak{e}_8$\\
\hline
$2\,4\,A_N  \, 4 \, 2$ \phantom{$\Bigg|$} & $E_8$  & {\scriptsize $ 3,2,2,1,12,( B(E_8),12)^{N+1},1,2,2,3$} &{\scriptsize$12(N+1) + 9$}& $(\mathfrak{e}_8)^{N+2} \oplus (\mathfrak{sp}_1 \oplus \mathfrak{g_2})^{2N+4} \oplus (\mathfrak{f}_4)^{N+1}$ & $\mathfrak{g_2} \oplus \mathfrak{g}_2$\\
$3\,3\,A_N  \, 4 \, 2$ \phantom{$\Bigg|$} & $E_8$  & {\scriptsize $ 5,1,3,2,2,1,12,( B(E_8),12)^{N+1},1,2,2,3$} &{\scriptsize$12(N+1)+11$}& $(\mathfrak{e}_8)^{N+2} \oplus (\mathfrak{sp}_1 \oplus \mathfrak{g_2})^{2N+4} \oplus (\mathfrak{f}_4)^{N+2}$ & $\mathfrak{f_4} \oplus \mathfrak{g}_2$\\
$2\,2\,3\,A_N  \, 4 \, 2$ \phantom{$\Bigg|$} & $E_8$  & {\scriptsize $ 3,1,5,1,3,2,2,1,12,( B(E_8),12)^{N+1},1,2,2,3$}&{\scriptsize$12(N+2)+1$} & $\mathfrak{su}_3 \oplus (\mathfrak{e}_8)^{N+2} \oplus (\mathfrak{sp}_1 \oplus \mathfrak{g_2})^{2N+4} \oplus (\mathfrak{f}_4)^{N+2}$ & $\mathfrak{su}_3 \oplus \mathfrak{g}_2$\\
$2\,2\,2\,3\,A_N  \, 4 \, 2$ \phantom{$\Bigg|$} & $E_8$ & {\scriptsize $ 2,3,1,5,1,3,2,2,1,12,( B(E_8),12)^{N+1},1,2,2,3$} &{\scriptsize$12(N+2)+2$}& $(\mathfrak{e}_8)^{N+2} \oplus (\mathfrak{sp}_1 \oplus \mathfrak{g_2})^{2N+5} \oplus (\mathfrak{f}_4)^{N+2}$ & $\mathfrak{su_2} \oplus \mathfrak{g}_2$&$\star$\\
$2\,2\,2\,2\,3\,A_N  \, 4 \, 2$ \phantom{$\Bigg|$} & $E_8$  & {\scriptsize $ 2,2,3,1,5,1,3,2,2,1,12,( B(E_8),12)^{N+1},1,2,2,3$} &{\scriptsize$12(N+2)+3$}& $ (\mathfrak{e}_8)^{N+2} \oplus (\mathfrak{sp}_1 \oplus \mathfrak{g_2})^{2N+5} \oplus (\mathfrak{f}_4)^{N+2}$ & $\mathfrak{g}_2$\\
\hline
\end{tabular}
}
\end{center}
\end{table}
\end{landscape}

\newpage

\begin{landscape}
\begin{center}
\phantom{ Wozzeck }
\vspace{1.5cm}
Table 1 -- continued
\vspace{0.3cm}
{\footnotesize
\begin{tabular}{ccccccc}
Series  \phantom{$\Bigg|$}  & Type  & Blow up &  $N_T$ &Algebra & $W$ &\\
\hline
$3 \,3 \, A_N \, 3 \,3  $ \phantom{$\Bigg|$} & $E_8$  & {\scriptsize $ \begin{aligned}5,1,3,2,2,1,12,( B(E_8),12)^{N+1},&\\ 1,2,2,3,1,5&\end{aligned}$} & {\scriptsize$12(N+2)+1$} &$(\mathfrak{e}_8)^{N+2} \oplus (\mathfrak{sp}_1 \oplus \mathfrak{g_2})^{2N+4} \oplus (\mathfrak{f}_4)^{N+3}$ & $\mathfrak{f_4} \oplus \mathfrak{f}_4$\\
$2\,2\, 3 \, A_N \, 3 \,3  $ \phantom{$\Bigg|$} & $E_8$  & {\scriptsize $ \begin{aligned} 3,1,5,1,3,2,2,1,12,( B(E_8),12)^{N+1},&\\1,2,2,3,1,5&\end{aligned}$} & {\scriptsize$12(N+2)+3$}& $\mathfrak{su}_3\oplus (\mathfrak{e}_8)^{N+2} \oplus (\mathfrak{sp}_1 \oplus \mathfrak{g_2})^{2N+4} \oplus (\mathfrak{f}_4)^{N+3}$ & $\mathfrak{su_3} \oplus \mathfrak{f}_4$\\
$2\, 2\,2\, 3 \, A_N \, 3 \,3  $ \phantom{$\Bigg|$} & $E_8$  & {\scriptsize $ \begin{aligned} 2,3,1,5,1,3,2,2,1,12,( B(E_8),12)^{N+1},&\\1,2,2,3,1,5&\end{aligned}$} & {\scriptsize$12(N+2)+4$}& $ (\mathfrak{e}_8)^{N+2} \oplus (\mathfrak{sp}_1 \oplus \mathfrak{g_2})^{2N+5} \oplus (\mathfrak{f}_4)^{N+3}$ & $\mathfrak{su_2} \oplus \mathfrak{f}_4$&$\star$\\
$2\, 2\, 2\,2\, 3 \, A_N \, 3 \,3  $ \phantom{$\Bigg|$} & $E_8$  & {\scriptsize $ \begin{aligned} 2,2,3,1,5,1,3,2,2,1,12,( B(E_8),12)^{N+1},&\\1,2,2,3,1,5&\end{aligned}$} & {\scriptsize$12(N+2)+5$}& $ (\mathfrak{e}_8)^{N+2} \oplus (\mathfrak{sp}_1 \oplus \mathfrak{g_2})^{2N+5} \oplus (\mathfrak{f}_4)^{N+3}$ & $\mathfrak{f}_4$\\
\hline
$2\,2\,3 \, A_N \, 3 \, 2 \, 2 $ \phantom{$\Bigg|$}  & $E_7$  & {\scriptsize$2,3,2,1,8,(B(E_7),8)^{N+1},1,2,3,2$} &{\scriptsize$6(N+1)+ 9$}& $(\mathfrak{e}_7)^{N+2} \oplus (\mathfrak{so_7})^{N+3} \oplus (\mathfrak{su}_2)^{2N+6}$ & $\mathfrak{su}_2 \oplus \mathfrak{su}_2$\\
$2\, 2\,2\, 3 \, A_N \, 3 \,2 \,2  $ \phantom{$\Bigg|$} & $E_8$  & {\scriptsize $\begin{aligned} 2,3,1,5,1,3,2,2,1,12,( B(E_8),12)^{N+1},&\\ 1,2,2,3,1,5,1,3& \end{aligned}$} & {\scriptsize$12(N+2)+6$}& $\mathfrak{su}_3 \oplus (\mathfrak{e}_8)^{N+2} \oplus (\mathfrak{sp}_1 \oplus \mathfrak{g_2})^{2N+5} \oplus (\mathfrak{f}_4)^{N+3}$ & $\mathfrak{su_2} \oplus \mathfrak{su}_3$&$\star$\\
$2\, 2\,2\,2\, 3 \, A_N \, 3 \,2 \,2  $ \phantom{$\Bigg|$} & $E_8$  & {\scriptsize $\begin{aligned} 2,2,3,1,5,1,3,2,2,1,12,( B(E_8),12)^{N+1},&\\ 1,2,2,3,1,5,1,3&\end{aligned}$} & {\scriptsize$12(N+2)+7$}& $\mathfrak{su}_3 \oplus (\mathfrak{e}_8)^{N+2} \oplus (\mathfrak{sp}_1 \oplus \mathfrak{g_2})^{2N+5} \oplus (\mathfrak{f}_4)^{N+3}$ & $\mathfrak{su}_3$\\
\hline
$2\, 2\,2\, 3 \, A_N \, 3 \,2 \,2\,2  $ \phantom{$\Bigg|$} & $E_8$  & {\scriptsize $\begin{aligned} 2,3,1,5,1,3,2,2,1,12,( B(E_8),12)^{N+1},&\\ 1,2,2,3,1,5,1,3,2&\end{aligned}$} & {\scriptsize$12(N+2)+7$}&$ (\mathfrak{e}_8)^{N+2} \oplus (\mathfrak{sp}_1 \oplus \mathfrak{g_2})^{2N+6} \oplus (\mathfrak{f}_4)^{N+3}$ & $\mathfrak{su_2} \oplus \mathfrak{su}_2$&$\star$\\
$2\,2\, 2\,2\, 3 \, A_N \, 3 \,2 \,2\,2  $ \phantom{$\Bigg|$} & $E_8$  & {\scriptsize $\begin{aligned}2, 2,3,1,5,1,3,2,2,1,12,( B(E_8),12)^{N+1},&\\ 1,2,2,3,1,5,1,3,2&\end{aligned}$} & {\scriptsize$12(N+2)+8$}& $ (\mathfrak{e}_8)^{N+2} \oplus (\mathfrak{sp}_1 \oplus \mathfrak{g_2})^{2N+6} \oplus (\mathfrak{f}_4)^{N+3}$ & $\mathfrak{su_2}$&$\star$\\
\hline
$2\,2\, 2\,2\, 3 \, A_N \, 3 \,2 \,2\,2\,2  $ \phantom{$\Bigg|$} & $E_8$  & {\scriptsize $\begin{aligned}2, 2,3,1,5,1,3,2,2,1,12,( B(E_8),12)^{N+1},&\\1,2,2,3,1,5,1,3,2,2&\end{aligned}$} & {\scriptsize$12(N+2)+9$}& $ (\mathfrak{e}_8)^{N+2} \oplus (\mathfrak{sp}_1 \oplus \mathfrak{g_2})^{2N+6} \oplus (\mathfrak{f}_4)^{N+3}$ & \\
\hline
\end{tabular}
}
\end{center}
\end{landscape}

\newpage

\begin{landscape}
\begin{table}
\begin{center}
\caption{ Minimal resolutions: $(\cI,\cJ)$ isolated theories }
{\footnotesize
\begin{tabular}{ccccccc}
Theory \phantom{$\Bigg|$}   & Blow up & $N_T$&Algebra & $W$  \\
\hline
$7 \, 2$ \phantom{$\Bigg|$}  & {\scriptsize 10,1,2,2,3} & 5 & $  \mathfrak{e}_8 \oplus \mathfrak{sp}_1 \oplus \mathfrak{g}_2 $  & $\mathfrak{e}_8 \oplus \mathfrak{g}_2 $  \\
$7 \, 2 \, 2 $ \phantom{$\Bigg|$}  & {\scriptsize 11,1,2,2,3,1,5,1,3} & 9 & $  \mathfrak{e}_8 \oplus \mathfrak{sp}_1 \oplus \mathfrak{g}_2 \oplus \mathfrak{f}_4 \oplus \mathfrak{su}_3$  & $\mathfrak{e}_8 \oplus \mathfrak{su}_3 $ \\
$7 \, 2 \, 2 \, 2 $ \phantom{$\Bigg|$}  & {\scriptsize 11,1,2,2,3,1,5,1,3,2} & 10 &$  \mathfrak{e}_8 \oplus \mathfrak{sp}_1 \oplus \mathfrak{g}_2  \oplus \mathfrak{f}_4 \oplus \mathfrak{g}_2 \oplus \mathfrak{su}_2$  & $\mathfrak{e}_8 \oplus \mathfrak{su}_2 $  \\
\hline
$2 \, 4 \, 2$ \phantom{$\Bigg|$}  & {\scriptsize 3,1,6,1,3} & 5 & $  \mathfrak{e}_6 \oplus (\mathfrak{su}_3)^2$ & $\mathfrak{su}_3\oplus\mathfrak{su}_3$  \\
$2\, 4 \, 2 \, 2 $ \phantom{$\Bigg|$}  & {\scriptsize 3,2,1,8,1,2,3,2} & 8 & $  \mathfrak{e}_7 \oplus (\mathfrak{su}_2)^3 \oplus \mathfrak{g}_2 \oplus \mathfrak{so}_7$  & $\mathfrak{g}_2 \oplus \mathfrak{su}_2 $  \\
$2\, 4 \, 2 \, 2 \, 2 $ \phantom{$\Bigg|$}  & {\scriptsize 3,2,2,1,11,1,2,2,3,1,5,1,3,2} & 14 & $  \mathfrak{e}_8 \oplus (\mathfrak{sp}_1 \oplus \mathfrak{g}_2)^2  \oplus \mathfrak{f}_4 \oplus \mathfrak{g}_2 \oplus \mathfrak{su}_2$  & $\mathfrak{g}_2 \oplus \mathfrak{su}_2 $  \\
\hline
$3 \, 3 \, 2$ \phantom{$\Bigg|$}  & {\scriptsize 5,1, 3,1,6,1,3} & 7 &$  \mathfrak{e}_6 \oplus (\mathfrak{su}_3)^2\oplus \mathfrak{f}_4$ & $\mathfrak{su}_3\oplus\mathfrak{su}_3$ \\
$3\, 3 \, 2 \, 2 $ \phantom{$\Bigg|$}  & {\scriptsize 5,1, 3,2,1,8,1,2,3,2} & 10 & $  \mathfrak{e}_7 \oplus (\mathfrak{su}_2)^3 \oplus \mathfrak{g}_2 \oplus \mathfrak{so}_7\oplus \mathfrak{f}_4$  & $\mathfrak{f}_4 \oplus \mathfrak{su}_2 $  \\
$3\, 3 \, 2 \, 2 \, 2 $ \phantom{$\Bigg|$}  & {\scriptsize 5,1, 3,2,2,1,11,1,2,2,3,1,5,1,3,2} & 16 & $  \mathfrak{e}_8 \oplus (\mathfrak{sp}_1 \oplus \mathfrak{g}_2)^2  \oplus (\mathfrak{f}_4)^2 \oplus \mathfrak{g}_2 \oplus \mathfrak{su}_2$  & $\mathfrak{f}_4 \oplus \mathfrak{su}_2 $  \\
\hline
$2\,2\, \, 3 \, 2 $ \phantom{$\Bigg|$}  & {\scriptsize 2,3,1, 5,1,3} & 6 &$\mathfrak{su}_2 \oplus \mathfrak{g}_2 \oplus \mathfrak{f}_4 \oplus \, \mathfrak{su}_3$ & $\mathfrak{su}_2\oplus\mathfrak{su}_3$ \\
$2\,2\, \, 3 \, 2 \,2 $ \phantom{$\Bigg|$}  & {\scriptsize 2,3,1, 5,1,3,2} & 7 &$(\mathfrak{su}_2 \oplus \mathfrak{g}_2)^2 \oplus \mathfrak{f}_4$ & $\mathfrak{su}_2\oplus\mathfrak{su}_2$\\
$2\,2\, \, 3 \, 2 \,2 \,2 $ \phantom{$\Bigg|$}  & {\scriptsize 2,3,1, 5,1,3,2,2} & 8 & $\mathfrak{su}_2 \oplus \, \mathfrak{sp}_1\oplus (\mathfrak{g}_2)^2 \oplus \mathfrak{f}_4$ & $\mathfrak{su}_2$ \\
\hline
$2\,2\,2\, \, 3 \, 2 $ \phantom{$\Bigg|$}  & {\scriptsize 2,2,3,1, 5,1,3} & 7 & $\mathfrak{sp}_1 \oplus \mathfrak{g}_2 \oplus \mathfrak{f}_4 \oplus \, \mathfrak{su}_3$ & $\mathfrak{su}_3$ \\
$2\,2\,2 \, 3 \, 2 \,2 \,2 $ \phantom{$\Bigg|$}  & {\scriptsize 2,2,3,1, 5,1,3,2,2} & 9 & $( \mathfrak{sp}_1\oplus \mathfrak{g}_2)^2 \oplus \mathfrak{f}_4$ &   \\
\hline
\end{tabular}
}
\end{center}
\end{table}
\end{landscape}

\newpage

\begin{landscape}
\begin{table}
\begin{center}
\caption{ Minimal resolutions: $(\, \cI , \cJ \,)$ infinite families. }
{\footnotesize
\begin{tabular}{cccccccc}
Series  \phantom{$\Bigg|$}   & Type  & Blow up & $N_T$ & Algebra & $W$ \\
\hline
$7 \, A_N \, 6 $ \phantom{$\Bigg|$} & $E_8$  & {\scriptsize$12,( B(E_8),12)^{N},B(E_8),11$} & {\scriptsize$12(N+1)+1$} & $(\mathfrak{e}_8)^{N+2} \oplus (\mathfrak{sp}_1 \oplus \mathfrak{g_2})^{2N+2} \oplus (\mathfrak{f}_4)^{N+1}$ & $\mathfrak{e_8} \oplus \mathfrak{e}_8$\\
$2\,4 \, A_N \, 6 $ \phantom{$\Bigg|$} & $E_8$  & {\scriptsize$3,2,2,1,12,( B(E_8),12)^{N}, B(E_8),11$} & {\scriptsize$12(N+1)+5$}& $(\mathfrak{e}_8)^{N+2} \oplus (\mathfrak{sp}_1 \oplus \mathfrak{g_2})^{2N+3} \oplus (\mathfrak{f}_4)^{N+1}$ & $\mathfrak{e_8} \oplus \mathfrak{g}_2$\\
$3\,3 \, A_N \, 6 $ \phantom{$\Bigg|$} & $E_8$ & {\scriptsize$ 5,1,3,2,2,1,12,( B(E_8),12)^{N}, B(E_8),11$} & {\scriptsize$12(N+1)+7$}& $(\mathfrak{e}_8)^{N+2} \oplus (\mathfrak{sp}_1 \oplus \mathfrak{g_2})^{2N+3} \oplus (\mathfrak{f}_4)^{N+2}$ & $\mathfrak{e_8} \oplus \mathfrak{f}_4$\\
$2\,2 \,3 \, A_N \, 6 $ \phantom{$\Bigg|$} & $E_8$ & {\scriptsize$ 3,1,5,1,3,2,2,1,12,( B(E_8),12)^{N},B(E_8),11$} & {\scriptsize$12(N+1)+9$}& $\mathfrak{su}_3\oplus(\mathfrak{e}_8)^{N+2} \oplus (\mathfrak{sp}_1 \oplus \mathfrak{g_2})^{2N+3} \oplus (\mathfrak{f}_4)^{N+2}$ & $\mathfrak{e_8} \oplus \mathfrak{su}_3$\\
$2\,2\,2 \,3 \, A_N \, 6 $ \phantom{$\Bigg|$} & $E_8$  & {\scriptsize$2,3,1,5,1,3,2,2,1,12,( B(E_8),12)^{N},B(E_8),11$} & {\scriptsize$12(N+1)+10$}& $(\mathfrak{e}_8)^{N+2} \oplus (\mathfrak{sp}_1 \oplus \mathfrak{g_2})^{2N+4} \oplus (\mathfrak{f}_4)^{N+2}$ & $\mathfrak{e_8} \oplus \mathfrak{su}_2$&$\star$\\
$2\,2\,2\,2 \,3 \, A_N \, 6 $ \phantom{$\Bigg|$} & $E_8$  & {\scriptsize$ 2,2,3,1,5,1,3,2,2,1,12,( B(E_8),12)^{N}, B(E_8),11$}& {\scriptsize$12(N+1)+11$} & $(\mathfrak{e}_8)^{N+2} \oplus (\mathfrak{sp}_1 \oplus \mathfrak{g_2})^{2N+4} \oplus (\mathfrak{f}_4)^{N+2}$ & $\mathfrak{e_8}$\\
\hline
$7 \, A_N \, 5 $ \phantom{$\Bigg|$} & $E_8$ & {\scriptsize$12,( B(E_8),12)^{N},B(E_8),10$} & {\scriptsize$12(N+1)+1$}& $(\mathfrak{e}_8)^{N+2} \oplus (\mathfrak{sp}_1 \oplus \mathfrak{g_2})^{2N+2} \oplus (\mathfrak{f}_4)^{N+1}$ & $\mathfrak{e_8} \oplus \mathfrak{e}_8$\\
$2\, 4 \, A_N \, 5 $ \phantom{$\Bigg|$} & $E_8$  & {\scriptsize$3,2,2,1,12,( B(E_8),12)^{N}, B(E_8),10$} & {\scriptsize$12(N+1)+5$}& $(\mathfrak{e}_8)^{N+2} \oplus (\mathfrak{sp}_1 \oplus \mathfrak{g_2})^{2N+3} \oplus (\mathfrak{f}_4)^{N+1}$ & $\mathfrak{e_8} \oplus \mathfrak{g}_2$\\
$3\, 3 \, A_N \, 5 $ \phantom{$\Bigg|$} & $E_8$  & {\scriptsize$5,1,3,2,2,1,12,( B(E_8),12)^{N}, B(E_8),10$}& {\scriptsize$12(N+1)+7$} & $(\mathfrak{e}_8)^{N+2} \oplus (\mathfrak{sp}_1 \oplus \mathfrak{g_2})^{2N+3} \oplus (\mathfrak{f}_4)^{N+2}$ & $\mathfrak{e_8} \oplus \mathfrak{f}_4$\\
$2\,2\,3\,A_N\,5$  \phantom{$\Bigg|$} & $E_7$  & {\scriptsize$ 2,3,2,1,8,(B(E_7),8)^{N+1}$} & {\scriptsize$6(N+1)+5$}& $(\mathfrak{e}_7)^{N+2} \oplus (so_7)^{N+2} \oplus (\mathfrak{su}_2)^{2N+4}$ & $\mathfrak{e_7} \oplus \mathfrak{su}_2$\\
$2\,2\,2\, 3 \, A_N \, 5 $ \phantom{$\Bigg|$} & $E_8$ & {\scriptsize$2,3,1,5,1,3,2,2,1,12,( B(E_8),12)^{N}, B(E_8),10$}& {\scriptsize$12(N+1)+10$} & $(\mathfrak{e}_8)^{N+2} \oplus (\mathfrak{sp}_1 \oplus \mathfrak{g_2})^{2N+4} \oplus (\mathfrak{f}_4)^{N+2}$ & $\mathfrak{e_8} \oplus \mathfrak{su}_2$&$\star$\\
$2\,2\,2\,2\, 3 \, A_N \, 5 $ \phantom{$\Bigg|$} & $E_8$  & {\scriptsize$2,2,3,1,5,1,3,2,2,1,12,( B(E_8),12)^{N}, B(E_8),10$}& {\scriptsize$12(N+1)+11$} & $(\mathfrak{e}_8)^{N+2} \oplus (\mathfrak{sp}_1 \oplus \mathfrak{g_2})^{2N+4} \oplus (\mathfrak{f}_4)^{N+2}$ & $\mathfrak{e_8}$\\
\hline
\end{tabular}
}
\end{center}
\end{table}
\end{landscape}

\newpage

\begin{landscape}
\begin{center}
\phantom{ Wozzeck }
Table 3: continued
\vspace{0.3cm}
{\footnotesize
\begin{tabular}{cccccccc}
Series  \phantom{$\Bigg|$}   & Type &  Blow up & $N_T$&Algebra & $W$ & \\
\hline
$7 \, A_N \, 4 $ \phantom{$\Bigg|$} & $E_8$  & {\scriptsize$12,( B(E_8),12)^{N},1,2,2,3,1,5,1,3,2,1,8$} & {\scriptsize$12(N+1)$} & $\mathfrak{e}_7\oplus (\mathfrak{e}_8)^{N+1} \oplus (\mathfrak{sp}_1 \oplus \mathfrak{g_2})^{2N+2} \oplus (\mathfrak{f}_4)^{N+1}$ & $\mathfrak{e_8} \oplus \mathfrak{e}_7$&$\star$\\
$2 \,4 \, A_N \, 4 $ \phantom{$\Bigg|$} & $E_8$  & {\scriptsize$\begin{aligned} 3,2,2,1,12,( B(E_8),12)^{N},&\\ 1,2,2,3,1,5,1,3,2,1,8&\end{aligned}$} & {\scriptsize$12(N+1)+4$} &  $\mathfrak{e}_7\oplus (\mathfrak{e}_8)^{N+1} \oplus (\mathfrak{sp}_1 \oplus \mathfrak{g_2})^{2N+3} \oplus (\mathfrak{f}_4)^{N+1}$ & $\mathfrak{e_7}\oplus \mathfrak{g}_2$&$\star$\\
$3 \,3 \, A_N \, 4 $ \phantom{$\Bigg|$} & $E_8$  & {\scriptsize$\begin{aligned} 5,1,3,2,2,1,12,( B(E_8),12)^{N},&\\ 1,2,2,3,1,5,1,3,2,1,8&\end{aligned}$} & {\scriptsize$12(N+1)+6$}&  $\mathfrak{e}_7\oplus (\mathfrak{e}_8)^{N+1} \oplus (\mathfrak{sp}_1 \oplus \mathfrak{g_2})^{2N+3} \oplus (\mathfrak{f}_4)^{N+2}$ & $\mathfrak{e_7}\oplus \mathfrak{f}_4$&$\star$\\
$2 \,2 \, 3 \, A_N \, 4 $ \phantom{$\Bigg|$} & $E_7$  & {\scriptsize$2,3,2,1,(8,B(E_7))^{N+1},7$} & {\scriptsize$6(N+1)+5$}& $(\mathfrak{e}_7)^{N+2} \oplus (\mathfrak{so}_7)^{N+2} \oplus (\mathfrak{su}_2)^{2N+4}$ & $\mathfrak{e_7}\oplus \mathfrak{su}_2$&$\spadesuit$\\
$2 \, 2 \,2 \, 3 \, A_N \, 4 $ \phantom{$\Bigg|$} & $E_8$  & {\scriptsize$\begin{aligned} 2,3,1,5,1,3,2,2,1,12,( B(E_8),12)^{N},&\\ 1,2,2,3,1,5,1,3,2,1,8&\end{aligned}$}& {\scriptsize$12(N+1)+9$} &  $\mathfrak{e}_7\oplus (\mathfrak{e}_8)^{N+1} \oplus (\mathfrak{sp}_1 \oplus \mathfrak{g_2})^{2N+4} \oplus (\mathfrak{f}_4)^{N+2}$ & $\mathfrak{e_7}\oplus \mathfrak{su}_2$&$\star\star$\\
$2\, 2 \, 2 \,2  \, 3 \, A_N \, 4 $ \phantom{$\Bigg|$} & $E_8$  & {\scriptsize$\begin{aligned} 2,2,3,1,5,1,3,2,2,1,12,( B(E_8),12)^{N},&\\ 1,2,2,3,1,5,1,3,2,1,8&\end{aligned}$} & {\scriptsize$12(N+1)+10$}&  $\mathfrak{e}_7\oplus (\mathfrak{e}_8)^{N+1} \oplus (\mathfrak{sp}_1 \oplus \mathfrak{g_2})^{2N+4} \oplus (\mathfrak{f}_4)^{N+2}$ & $\mathfrak{e_7} $&$\star$\\
\hline
$7 \, A_N \, 2\,2\,2\,2 $ \phantom{$\Bigg|$} & $E_8$  & {\scriptsize$(12,B(E_8))^N,11,1,2,2,3,1,5,1,3,2,2$}& {\scriptsize$12N+11$} & $ (\mathfrak{e}_8)^{N+1} \oplus (\mathfrak{sp}_1 \oplus \mathfrak{g_2})^{2N+2} \oplus (\mathfrak{f}_4)^{N+1}$ & $\mathfrak{e_8}$ \\
$2\,4 A_N \, 2 \, 2 \, 2 \,2$ \phantom{$\Bigg|$} & $E_8$  & {\scriptsize$3,2,2,1,(12,B(E_8))^N,11,1,2,2,3,1,5,1,3,2,2$}& {\scriptsize$12(N+1)+3$}  & $ (\mathfrak{e}_8)^{N+1} \oplus (\mathfrak{sp}_1 \oplus \mathfrak{g_2})^{2N+3} \oplus (\mathfrak{f}_4)^{N+1}$ & $\mathfrak{g}_2$ \\
$3\,3 \, A_N \, 2\,2\,2\,2 $ \phantom{$\Bigg|$} & $E_8$  & {\scriptsize$\begin{aligned}5,1,3,2,2,(12,B(E_8))^N,&\\ 11,1,2,2,3,1,5,1,3,2,2& \end{aligned}$} & {\scriptsize$12(N+1)+4$}& $ (\mathfrak{e}_8)^{N+1} \oplus (\mathfrak{sp}_1 \oplus \mathfrak{g_2})^{2N+3} \oplus (\mathfrak{f}_4)^{N+2}$ & $\mathfrak{f}_4$ \\
$2 \, 2 \, 3 \, A_N \, 2\,2\,2\,2 $ \phantom{$\Bigg|$} & $E_7$ & {\scriptsize$2,3,2,1,(8,B(E_7))^{N},8,1,2,3,1,5,1,3,2,2$} & {\scriptsize$6N+14$}&$\begin{aligned} (\mathfrak{e}_7)^{N+1} \oplus (\mathfrak{so}_7)^{N+1} \oplus (\mathfrak{su}_2)^{2N+3} &\\ \oplus (\mathfrak{g_2})^2 \oplus \mathfrak{sp}_1 \oplus \mathfrak{f}_4 & \end{aligned}$  & $\mathfrak{su}_2$\\
$2\,2\,2\,3 \, A_N \, 2\,2\,2\,2 $ \phantom{$\Bigg|$} & $E_8$  & {\scriptsize$\begin{aligned}2,3,1,5,1,3,2,2,(12,B(E_8))^N,&\\ 11,1,2,2,3,1,5,1,3,2,2& \end{aligned}$} & {\scriptsize$12(N+1)+7$}& $(\mathfrak{e}_8)^{N+1} \oplus (\mathfrak{sp}_1 \oplus \mathfrak{g_2})^{2N+4} \oplus (\mathfrak{f}_4)^{N+2}$ & $\mathfrak{su}_2$&$\star$ \\
$2\,2\,2\,2\,3 \, A_N \, 2\,2\,2\,2 $ \phantom{$\Bigg|$} & $E_8$  & {\scriptsize$\begin{aligned}2,2,3,1,5,1,3,2,2,(12,B(E_8))^N,&\\ 11,1,2,2,3,1,5,1,3,2,2& \end{aligned}$} & {\scriptsize$12(N+1)+8$}& $ (\mathfrak{e}_8)^{N+1} \oplus (\mathfrak{sp}_1 \oplus \mathfrak{g_2})^{2N+4} \oplus (\mathfrak{f}_4)^{N+2}$ & &\\
\hline
\end{tabular}
}
\end{center}
\end{landscape}

\newpage

\begin{landscape}
\begin{center}
\phantom{ Wozzeck }
\vspace{1.5cm}
Table 3: continued
\vspace{0.3cm}
{\footnotesize
\begin{tabular}{cccccccc}
Series  \phantom{$\Bigg|$}   & Type  & Blow up & $N_T$&Algebra & $W$ & \\
\hline
$7 \, A_N \, 3 $ \phantom{$\Bigg|$} & $E_8$  & {\scriptsize$(12,B(E_8))^N,11,1,2,2,3,1,5$} & {\scriptsize$12 N+7$}& $ (\mathfrak{e}_8)^{N+1} \oplus (\mathfrak{sp}_1 \oplus \mathfrak{g_2})^{2N+1} \oplus (\mathfrak{f}_4)^{N+1}$ & $\mathfrak{e_8} \oplus \mathfrak{f}_4$ \\
$2\,4 \, A_N \, 3 $ \phantom{$\Bigg|$} & $E_8$ & {\scriptsize$3,2,2,1,(12,B(E_8))^N,11,1,2,2,3,1,5$} & {\scriptsize$12 N+11$} & $ (\mathfrak{e}_8)^{N+1} \oplus (\mathfrak{sp}_1 \oplus \mathfrak{g_2})^{2N+2} \oplus (\mathfrak{f}_4)^{N+1}$ & $\mathfrak{g_2} \oplus \mathfrak{f}_4$ \\
$3\,3 \, A_N \, 3 $ \phantom{$\Bigg|$} & $E_8$ & {\scriptsize$5,1,3,2,2,1,(12,B(E_8))^N,11,1,2,2,3,1,5$} & {\scriptsize$12 (N+1)+1$}& $ (\mathfrak{e}_8)^{N+1} \oplus (\mathfrak{sp}_1 \oplus \mathfrak{g_2})^{2N+2} \oplus (\mathfrak{f}_4)^{N+2}$ & $\mathfrak{f_4} \oplus \mathfrak{f}_4$ \\
$2 \, 2 \, 3 \, A_N 3$ \phantom{$\Bigg|$} & $E_7$ & {\scriptsize$2,3,2,1,8,(B(E_7),8)^N,1,2,3,1,5$}& {\scriptsize$6 N+10$} &$(\mathfrak{e}_7)^{N+1} \oplus (\mathfrak{so}_7)^{N+1} \oplus (\mathfrak{su}_2)^{2N+3}\oplus (\mathfrak{g_2}) \oplus \mathfrak{f}_4 $  & $\mathfrak{su}_2 \oplus \mathfrak{f}_4$\\
$2\,2\,2\,3 \, A_N \, 3 $ \phantom{$\Bigg|$} & $E_8$ & {\scriptsize$2,3,1,5,1,3,2,2,1,(12,B(E_8))^N,11,1,2,2,3,1,5$}& {\scriptsize$12 (N+1)+4$} & $ (\mathfrak{e}_8)^{N+1} \oplus (\mathfrak{sp}_1 \oplus \mathfrak{g_2})^{2N+3} \oplus (\mathfrak{f}_4)^{N+2}$ & $\mathfrak{su_2} \oplus \mathfrak{f}_4$ &$\star$\\
$2\,2\,2\,2\,3 \, A_N \, 3 $ \phantom{$\Bigg|$} & $E_8$ & {\scriptsize$2,2,3,1,5,1,3,2,2,1,(12,B(E_8))^N,11,1,2,2,3,1,5$}& {\scriptsize$12 (N+1)+5$} & $ (\mathfrak{e}_8)^{N+1} \oplus (\mathfrak{sp}_1 \oplus \mathfrak{g_2})^{2N+3} \oplus (\mathfrak{f}_4)^{N+2}$ & $ \mathfrak{f}_4$ &\\
\hline
$7 \, A_N \, 3 \,2 $ \phantom{$\Bigg|$} & $E_8$  & {\scriptsize$(12,B(E_8))^N,11,1,2,2,3$}& {\scriptsize$12 N+5$} & $ (\mathfrak{e}_8)^{N+1} \oplus (\mathfrak{sp}_1 \oplus \mathfrak{g_2})^{2N+1} \oplus (\mathfrak{f}_4)^N$ & $\mathfrak{e_8} \oplus \mathfrak{g}_2$ \\
$2 \,4 \, A_N \, 3 \,2 $ \phantom{$\Bigg|$} & $E_8$  & {\scriptsize$3,2,2,1(12,B(E_8))^N,11,1,2,2,3$} & {\scriptsize$12 N+9$}& $ (\mathfrak{e}_8)^{N+1} \oplus (\mathfrak{sp}_1 \oplus \mathfrak{g_2})^{2N+2} \oplus (\mathfrak{f}_4)^N$ & $\mathfrak{g_2} \oplus \mathfrak{g}_2$ \\
$3 \,3 \, A_N \, 3 \,2 $ \phantom{$\Bigg|$} & $E_8$  & {\scriptsize$5,1,3,2,2,1(12,B(E_8))^N,11,1,2,2,3$}& {\scriptsize$12 N+11$} & $ (\mathfrak{e}_8)^{N+1} \oplus (\mathfrak{sp}_1 \oplus \mathfrak{g_2})^{2N+2} \oplus (\mathfrak{f}_4)^{N+1}$ & $\mathfrak{f_4} \oplus \mathfrak{g}_2$ \\
$2 \, 2 \, 3 \, A_N 3\,2 $ \phantom{$\Bigg|$} & $E_7$ & {\scriptsize$2,3,2,1,8,(B(E_7),8)^{N+1},1,2,3$}& {\scriptsize$6(N+1)+8$} &$(\mathfrak{e}_7)^{N+2} \oplus (\mathfrak{so}_7)^{N+2} \oplus (\mathfrak{su}_2)^{2N+5}\oplus (\mathfrak{g_2})$  & $\mathfrak{su}_2 \oplus \mathfrak{g}_2$\\
$2\,2\,2 \,3 \, A_N \, 3 \,2 $ \phantom{$\Bigg|$} & $E_8$  & {\scriptsize$\begin{aligned}2,3,1,5,1,3,2,2,1(12,B(E_8))^N,&\\11,1,2,2,3\end{aligned}$} & {\scriptsize$12 (N+1)+2$}& $ (\mathfrak{e}_8)^{N+1} \oplus (\mathfrak{sp}_1 \oplus \mathfrak{g_2})^{2N+3} \oplus (\mathfrak{f}_4)^{N+1}$ & $\mathfrak{su_2} \oplus \mathfrak{g}_2$&$\star$ \\
$2\, 2\,2\,2 \,3 \, A_N \, 3 \,2 $ \phantom{$\Bigg|$} & $E_8$  & {\scriptsize$\begin{aligned}2,2,3,1,5,1,3,2,2,1(12,B(E_8))^N,&\\11,1,2,2,3\end{aligned}$}& {\scriptsize$12 (N+1)+3$} & $ (\mathfrak{e}_8)^{N+1} \oplus (\mathfrak{sp}_1 \oplus \mathfrak{g_2})^{2N+3} \oplus (\mathfrak{f}_4)^{N+1}$ & $\mathfrak{g}_2$&$\star$ \\
\hline
\end{tabular}
}
\end{center}
\end{landscape}

\newpage

\begin{landscape}
\begin{table}
\begin{center}
\caption{ Minimal resolutions: $(\cJ,\cJ)$ isolated theories }
{\footnotesize
\begin{tabular}{ccccccc}
Theory \phantom{$\Bigg|$}   & Blow up & $N_T$&Algebra & $W$\\
\hline
6 \,3 \phantom{$\Bigg|$}& {\scriptsize 10,1,2,2,3,1,5 } & 7 & $\mathfrak{e}_8 \oplus \mathfrak{sp}_1 \oplus \mathfrak{g}_2 \oplus \mathfrak{f}_4$ & $\mathfrak{e}_8 \oplus \mathfrak{f}_4$  \\
6 \,2\,\phantom{$\Bigg|$} & {\scriptsize 8,1,2,3 } &4& $ \mathfrak{e}_7 \oplus \mathfrak{su}_2 \oplus \mathfrak{g}_2 $ & $\mathfrak{e}_7 \oplus \mathfrak{g}_2$  \\
6 \,2\, 2 \phantom{$\Bigg|$}& {\scriptsize 8,1,2,3,2 } &5& $ \mathfrak{e}_7 \oplus (\mathfrak{su}_2)^2 \oplus \mathfrak{so}_7 $ & $\mathfrak{e}_7 \oplus \mathfrak{su}_2$  \\
6 \,2\, 2 \,2 \phantom{$\Bigg|$}& {\scriptsize 10, 1, 2, 2, 3, 1, 5, 1, 3, 2 } &10& $ \mathfrak{e}_8 \oplus \mathfrak{su}_2 \oplus  (\mathfrak{g}_2)^2 \oplus \mathfrak{sp}_1\oplus \mathfrak{f}_4 $ & $\mathfrak{e}_8 \oplus \mathfrak{su}_2$ \\
6 \,2\, 2 \,2 \,2 \phantom{$\Bigg|$}& {\scriptsize 10, 1, 2, 2, 3, 1, 5, 1, 3, 2, 2 } &11& $ \mathfrak{e}_8 \oplus (\mathfrak{sp}_1 \oplus  \mathfrak{g}_2)^2 \oplus \mathfrak{f}_4 $ & $\mathfrak{e}_8 $  \\
\hline
5 \,2\,\phantom{$\Bigg|$} & {\scriptsize 6,1,3 } &3& $ \mathfrak{e}_6 \oplus \mathfrak{su}_3 $ & $\mathfrak{e}_6 \oplus \mathfrak{su}_3$  \\
5 \,2 \,2  \phantom{$\Bigg|$} & {\scriptsize 7,1,2,3,2 } &5& $ \mathfrak{e}_7 \oplus (\mathfrak{su}_2)^2 \oplus \, \mathfrak{so}_7 $ & $\mathfrak{e}_6 \oplus \,\mathfrak{su}_2$ & $\spadesuit$ \\
5 \,2 \,2 \,2  \phantom{$\Bigg|$} & {\scriptsize 8,1,2,3,1,5,1,3,2 } &9& $ \mathfrak{e}_7 \oplus (\mathfrak{su}_2 \oplus \mathfrak{g}_2)^2 \oplus \, \mathfrak{f}_4 $ & $\mathfrak{e}_7 \oplus \,\mathfrak{su}_2$  \\
\hline
4 \,2\,\phantom{$\Bigg|$} & {\scriptsize 5,1,3 } &3& $ \mathfrak{f}_4 \oplus \mathfrak{su}_3 $ & $\mathfrak{f}_4 \oplus \mathfrak{su}_3$  \\
4 \,2 \, 2 \,\phantom{$\Bigg|$} & {\scriptsize 5,1,3,2 } &4& $ \mathfrak{f}_4 \oplus \mathfrak{su}_2 \oplus \mathfrak{g}_2 $ & $\mathfrak{f}_4 \oplus \mathfrak{su}_2$  \\
4 \,2 \, 2 \, 2 \,\phantom{$\Bigg|$} & {\scriptsize 5,1,3,2,2 } &5& $ \mathfrak{f}_4 \oplus \mathfrak{sp}_1 \oplus \mathfrak{g}_2 $ & $\mathfrak{f}_4 $ \\
\hline
\end{tabular}
}
\end{center}
\end{table}
\end{landscape}

\newpage

\begin{landscape}
\begin{table}
\begin{center}
\caption{ Minimal resolutions: $(\, \cJ , \cJ \,)$ infinite families }
{\footnotesize
\begin{tabular}{cccccccc}
Series  \phantom{$\Bigg|$}   & Type & Blow up & $N_T$ &Algebra & $W$ \\
\hline
$6 \, A_{N} \, 6 $ \phantom{$\Bigg|$} & $E_8$  & {\scriptsize$11,(B(E_8),12)^N,B(E_8),11$} & {\scriptsize$12 (N+1)+1$}& $ (\mathfrak{e}_8)^{N+2} \oplus (\mathfrak{sp}_1 \oplus \mathfrak{g_2})^{2N+2} \oplus (\mathfrak{f}_4)^{N+1}$ &$\mathfrak{e}_8\oplus \mathfrak{e}_8$&\\
$6 \, A_{N} \, 5 $ \phantom{$\Bigg|$} & $E_8$  & {\scriptsize$11,(B(E_8),12)^N,B(E_8),10$}& {\scriptsize$12 (N+1)+1$} & $ (\mathfrak{e}_8)^{N+2} \oplus (\mathfrak{sp}_1 \oplus \mathfrak{g_2})^{2N+2} \oplus (\mathfrak{f}_4)^{N+1}$ &$\mathfrak{e}_8\oplus \mathfrak{e}_8$&\\
$6 \, A_{N} \, 4 $ \phantom{$\Bigg|$} & $E_8$  & {\scriptsize$11,(B(E_8),12)^N,1,2,2,3,1,5,1,3,2,1,8$}& {\scriptsize$12 (N+1)$} & $ \mathfrak{e}_7 \oplus(\mathfrak{e}_8)^{N+1} \oplus (\mathfrak{sp}_1 \oplus \mathfrak{g_2})^{2N+2} \oplus (\mathfrak{f}_4)^{N+1}$ &$\mathfrak{e}_8\oplus \mathfrak{e}_7$&$\star$\\
$6 \, A_{N} \, 3 \, 2$ \phantom{$\Bigg|$} & $E_8$  &  {\scriptsize$ 11,(B(E_8),12)^N,B(E_8),11,1,2,2,3$}& {\scriptsize$12 (N+1)+5$} & $ (\mathfrak{e}_8)^{N+2} \oplus (\mathfrak{sp}_1 \oplus \mathfrak{g_2})^{2N+3} \oplus (\mathfrak{f}_4)^{N+1}$ &$\mathfrak{e}_8\oplus \mathfrak{g}_2$&\\
$6 \, A_{N} \,2 \, 3 $ \phantom{$\Bigg|$} & $E_8$  & {\scriptsize$ 11,(B(E_8),12)^N,B(E_8),11,1,2,2,3,1,5$}& {\scriptsize$12 (N+1)+7$} & $ (\mathfrak{e}_8)^{N+2} \oplus (\mathfrak{sp}_1 \oplus \mathfrak{g_2})^{2N+3} \oplus (\mathfrak{f}_4)^{N+2}$ &$\mathfrak{e}_8\oplus \mathfrak{f}_4$&\\
$6 \, A_{N} \, 2 \,2 \, 2 \, 2 \, 2 $ \phantom{$\Bigg|$} & $E_8$  & {\scriptsize$11,(B(E_8),12)^N,B(E_8),11,1,2,2,3,1,5,1,3,2,2$}& {\scriptsize$12 (N+1)+11$} & $ (\mathfrak{e}_8)^{N+2} \oplus (\mathfrak{sp}_1 \oplus \mathfrak{g_2})^{2N+4} \oplus (\mathfrak{f}_4)^{N+2}$ &$\mathfrak{e}_8$&\\
\hline
$5 \, A_N \, 5$ \phantom{$\Bigg|$} & $E_7$  & {\scriptsize$(8,B(E_7))^{N+1},8$} & {\scriptsize$6(N+1)+1$}& $ (\mathfrak{e}_7)^{N+2} \oplus (\mathfrak{su}_2)^{2N+2} \oplus (\mathfrak{so}_7)^{N+1}$ &$\mathfrak{e}_7\oplus \mathfrak{e}_7$\\
$5 \, A_N \, 4$ \phantom{$\Bigg|$} & $E_7$  & {\scriptsize$(8,B(E_7))^{N+1},7$} & {\scriptsize$6(N+1)+1$}& $ (\mathfrak{e}_7)^{N+2} \oplus (\mathfrak{su}_2)^{2N+2} \oplus (\mathfrak{so}_7)^{N+1}$ &$\mathfrak{e}_7\oplus \mathfrak{e}_7$&$\spadesuit$\\
$5 \, A_N \, 3$ \phantom{$\Bigg|$} & $E_7$  & {\scriptsize$(8,B(E_7))^N,8,1,2,3,1,5$} & {\scriptsize$6 N+6$}& $ (\mathfrak{e}_7)^{N+1} \oplus (\mathfrak{su}_2)^{2N+1} \oplus (\mathfrak{so}_7)^N \oplus \mathfrak{g}_2 \oplus \mathfrak{f}_4$ &$\mathfrak{e}_7\oplus \mathfrak{f}_4$&\\
$5 \, A_N \, 2 \, 2 \,2 \,2$ \phantom{$\Bigg|$} & $E_7$  & {\scriptsize$(8,B(E_7))^N,8,1,2,3,1,5,1,3,2,2$} & {\scriptsize$6(N+2)+4$}& $ (\mathfrak{e}_7)^{N+1} \oplus (\mathfrak{su}_2)^{2N+1} \oplus (\mathfrak{so}_7)^N \oplus (\mathfrak{g}_2)^2 \oplus \mathfrak{f}_4 \oplus \mathfrak{sp}_1$ &$\mathfrak{e}_7$&\\
$5 \, A_N \, 3\,2$ \phantom{$\Bigg|$} & $E_7$  & {\scriptsize$(8,B(E_7))^{N+1},8,1,2,3$}& {\scriptsize$6(N+1)+4$} & $ (\mathfrak{e}_7)^{N+2} \oplus (\mathfrak{su}_2)^{2N +3} \oplus (\mathfrak{so}_7)^{N+1} \oplus \mathfrak{g}_2$ &$\mathfrak{e}_7\oplus \mathfrak{g}_2$&\\
\hline
\end{tabular}
}
\end{center}
\end{table}
\end{landscape}

\newpage

\begin{landscape}

\begin{center}

\phantom{ Wozzeck }
\vspace{1.5cm}
Table 5: continued
\vspace{0.3cm}

{\footnotesize
\begin{tabular}{cccccccc}
Series  \phantom{$\Bigg|$}   & Type  & Blow up & $N_T$ & Algebra & $W$ \\
\hline
$4 \, A_N \, 4$ \phantom{$\Bigg|$} & $E_6$ & {\scriptsize$(6,B(E_6))^{N+1},6$} & {\scriptsize$4(N+1)+1$}& $ (\mathfrak{e}_6)^{N+2} \oplus (\mathfrak{su}_3)^{N+1}$ &$\mathfrak{e}_6\oplus \mathfrak{e}_6$ &\\
$4 \, A_N \, 3$ \phantom{$\Bigg|$} & $E_6$  & {\scriptsize$(6,B(E_6))^{N+1},5$}& {\scriptsize$4(N+1)+1$} & $ (\mathfrak{e}_6 \oplus \mathfrak{su}_3)^{N+1} \oplus \mathfrak{f}_4$ &$\mathfrak{e}_6\oplus \mathfrak{f}_4$ &\\
$4 \, A_N \, 3\,2$ \phantom{$\Bigg|$} &$E_6$ & {\scriptsize$(6,B(E_6))^{N+1},6,1,3$} & {\scriptsize$4(N+1)+3$}& $ (\mathfrak{e}_6 \oplus \mathfrak{su}_3)^{N+2} $ &$\mathfrak{e}_6\oplus \,\mathfrak{su}_3$ &\\
$4 \, A_N \, 2\, 2\,2\,2$  \phantom{$\Bigg|$} & $E_6$ & {\scriptsize$(6,B(E_6))^{N+1},5,1,3,2,2$} & {\scriptsize$4(N+2)+1$}& $ (\mathfrak{e}_6 \oplus \mathfrak{su}_3)^{N} \oplus \mathfrak{f}_4 \oplus \mathfrak{sp}_1 \oplus \mathfrak{g}_2$ &$\mathfrak{e}_6 $ &\\
\hline
$3 \, A_N \, 3$ \phantom{$\Bigg|$} &$D_4$ & {\scriptsize$4,(1,4)^{N+1}$} & {\scriptsize$2N+5$}& $ (\mathfrak{so}_8)^{N+2}$ &$\mathfrak{so}_8\oplus \mathfrak{so}_8$&\\
$3 \, A_N \, 3\,2 $ \phantom{$\Bigg|$} & $E_6$  & {\scriptsize$5,(B(E_6),6)^{N+1},1,3$} & {\scriptsize$4(N+1)+3$}& $\mathfrak{f}_4 \oplus (\mathfrak{su_3})^{N+2} \oplus (\mathfrak{e}_6)^{N+1}$ &$\mathfrak{f}_4 \oplus \mathfrak{su}_3$&\\
$3 \, A_N \, 2\,2\,2$ \phantom{$\Bigg|$} &$D_4$ & {\scriptsize$(4,1)^{N+1},3,2,2$}& {\scriptsize$2(N+2)+1$} & $ (\mathfrak{so}_8)^{N+1} \oplus \, \mathfrak{sp_1}\oplus \mathfrak{g}_2$ &$\mathfrak{so}_8$&\\
\hline
$2\,3\, A_N \, 3 \,2 $ \phantom{$\Bigg|$} & $E_6$ & {\scriptsize$3,1,6,(B(E_6),6)^{N+1},1,3$}& {\scriptsize$4(N+1)+5$} & $ (\mathfrak{e}_6)^{N+2} \oplus (\mathfrak{su}_3)^{N+3}$ &$\mathfrak{su}_3\oplus \mathfrak{su}_3$&\\
$2\,3\, A_N \, 2 \,2 \,2 \, 2 $ \phantom{$\Bigg|$} &$E_6$ & {\scriptsize$3,1,(6,B(E_6))^{N+1},5,1,3,2,2$} & {\scriptsize$4(N+1)+7$}& $ (\mathfrak{e}_6)^{N+1} \oplus (\mathfrak{su}_3)^{N+2} \oplus \, \mathfrak{su}_2 \oplus \mathfrak{g}_2 \oplus \mathfrak{f}_4$ &$\mathfrak{su}_3$&\\
\hline
\end{tabular}
}
\end{center}
\end{landscape}

\bibliographystyle{utphys}
\bibliography{REVfractionalFINALv4}

\end{document}